\documentclass[aps,notitlepage,floatfix, onecolumn,
superscriptaddress,nofootinbib,11pt]{revtex4-1}
\usepackage[utf8]{inputenc}
\usepackage{graphicx}
\graphicspath{ {./Figures/} }
\usepackage{amsmath,amssymb,color}
\usepackage{appendix}
\usepackage{feynmf}
\usepackage{xcolor}
\usepackage{hyperref}
\hypersetup{colorlinks = true, allcolors=blue}

\begin{document}
\title{Information scrambling at finite temperature in local quantum systems}
\author{Subhayan Sahu}
\affiliation{Condensed Matter Theory Center and Department of Physics, University of Maryland, College Park, MD 20742, USA}
\author{Brian Swingle}
\affiliation{Condensed Matter Theory Center and Department of Physics, University of Maryland, College Park, MD 20742, USA}
\affiliation{Maryland Center for Fundamental Physics and Joint Center for Quantum Information and Computer Science,  University of Maryland, College Park, MD 20742, USA}


\begin{abstract}

    This paper investigates the temperature dependence of quantum information scrambling in local systems with an energy gap, $m$, above the ground state. We study the speed and shape of growing Heisenberg operators as quantified by out-of-time-order correlators, with particular attention paid to so-called contour dependence, i.e. dependence on the way operators are distributed around the thermal circle. We report large scale tensor network numerics on a gapped chaotic spin chain down to temperatures comparable to the gap which show that the speed of operator growth is strongly contour dependent. The numerics also show a characteristic broadening of the operator wavefront at finite temperature $T$. To study the behavior at temperatures much below the gap, we perform a perturbative calculation in the paramagnetic phase of a 2+1D O($N$) non-linear sigma model, which is analytically tractable at large $N$. Using the ladder diagram technique, we find that operators spread at a speed $\sqrt{T/m}$ at low temperatures, $T\ll m$. In contrast to the numerical findings of spin chain, the large $N$ computation is insensitive to the contour dependence and does not show broadening of operator front. We discuss these results in the context of a recently proposed state-dependent bound on scrambling.
\end{abstract}

\maketitle
\tableofcontents

\section{Introduction}	
	
Quantum information scrambling has emerged as an important dynamical feature of interacting quantum systems ranging from tabletop atomic systems to toy models of black holes~\cite{Sekino2008,Hayden2007,brown2012scrambling,Lashkari2013,Shenker2014,Shenker2014stringy,Hosur2016}. Scrambling refers to the way a closed chaotic quantum system delocalizes initially simple information such that it becomes inaccessible to all local measurements. Scrambling can be identified as a quantum analogue of the classical butterfly effect, as first discussed in a condensed matter context~\cite{Larkin1969}, and more recently explored in the context of holographic field theories and many-body systems such as the SYK model \cite{Kitaev2015,Sachdev2015,Maldacena2016b,Kitaev2018}. Scrambling can be studied for generic quantum systems by calculating out-of-time-ordered correlation (OTOC) functions, which, for geometrically local systems, gives rise to a state dependent velocity of information propagation---the butterfly velocity \cite{Roberts2016,Chowdhury2017,Xu2018}. OTOC functions can be measured for engineered quantum many body systems in the lab, with many proposals \cite{Swingle_2016,Zhu_2016,Yao_2016,Yunger_Halpern_2017,Yunger_Halpern_2018,Campisi_2017,Yoshida_2017,Vermersch_2019,Qi_2019} and subsequent experiments \cite{Garttner_2017,Wei_2018,Li_2017,Meier_2019,Landsman_2019,Wei_2019,Nie_2019}.

For quantum systems at the semiclassical limit, the deviation of an OTOC function from its initial value grows exponentially with time, with an exponent that can be viewed as a quantum analogue of the classical Lyapunov exponent $\lambda_L$ \cite{Kitaev2015}, although the connection to classical chaos is subtle \cite{Rozenbaum_2017,Xu_2020}. Deforming the contour along which path integrals are evaluated is a general technique one can use to regulate quantities in field theory and it leads to different choices of OTOCs at finite temperature, based on the contour on the thermal circle used to define it. One particular choice of contour leads to a well-behaved version of the OTOC that obeys a bound \cite{Maldacena2016}, $\lambda_L \leq 2\pi/\beta$, where $\beta$ is the inverse temperature. This bound was later understood in the more general context of the growth of operator complexity and thermalization~\cite{Parker_2019,Murthy_2019}. However, exponents arising from other versions of OTOCs can have a strong dependence on the choice of contour \cite{Liao2018,Romero-Bermudez2019}. 

In this work, we systematically study the temperature and contour dependence of OTOCs in generic quantum systems with spatial locality and a mass gap. Our motivation for this study comes from two directions. First, we want to understand possible contour dependence of OTOCs in a non-perturbative calculation. Second, we want to understand the temperature dependence of various characteristics of scrambling as a system is cooled below its mass gap. At high temperature, we indeed find contour dependence of the OTOC. At low temperature, where our expectation is that the physics is that of a weakly interacting dilute gas of quasiparticle excitations, we find that the rate of growth of scrambling is exponentially suppressed while the butterfly velocity is of order the sound speed. Technically, these results are obtained by studying a gapped spin chain at large size numerically and a field theory model analytically. The remainder of the introduction provides neccessary background material for our study.

\subsection{Squared commutators}

Consider a local quantum system, where the dynamical degrees of freedom are operators supported on local subsystems labelled by their positions in real space, $\mathbf{x}$. An operator $W_{\mathbf{0}}$ originally localized at position $\mathbf{0}$ can spread in real space under a Heisenberg time evolution that generates $W_{\mathbf{0}}(t)$. The extent of its physical spreading can be diagnosed by taking its commutator with another local operator $V_{\mathbf{x}}$, i.e. $[W_{\mathbf{0}}(t),V_{\mathbf{x}}]$. The squared commutator, evaluated on a particular choice of initial state, can quantify the extent of operator growth, as it is a valid norm of the commutator. 
	
However, in a quantum system at a finite temperature, T, this norm can be evaluated in several ways. Let us denote $\rho = e^{-\beta H}/Tr(e^{-\beta H})$ as the thermal density matrix ($\beta = 1/T$ is the inverse temperature). For any $0\leq\alpha\leq1$, 
\begin{equation}\label{eq:alpha_square_norm}
\mathcal{C}_{(\alpha)}(t,{\mathbf{x}})= Tr\left(\rho^{\alpha}[W_\mathbf{0}(t),V_\mathbf{x}]^{\dagger}\rho^{(1-\alpha)}[W_\mathbf{0}(t),V_\mathbf{x}]\right),
\end{equation}
is a Frobenius norm of the thermally smeared commutator $\rho^{(1-\alpha)/2}[W_\mathbf{0}(t),V_\mathbf{x}]\rho^{\alpha/2}$, which encodes a notion of the size of operator spreading.

\begin{figure}
    \centering
    \includegraphics[width=\columnwidth]{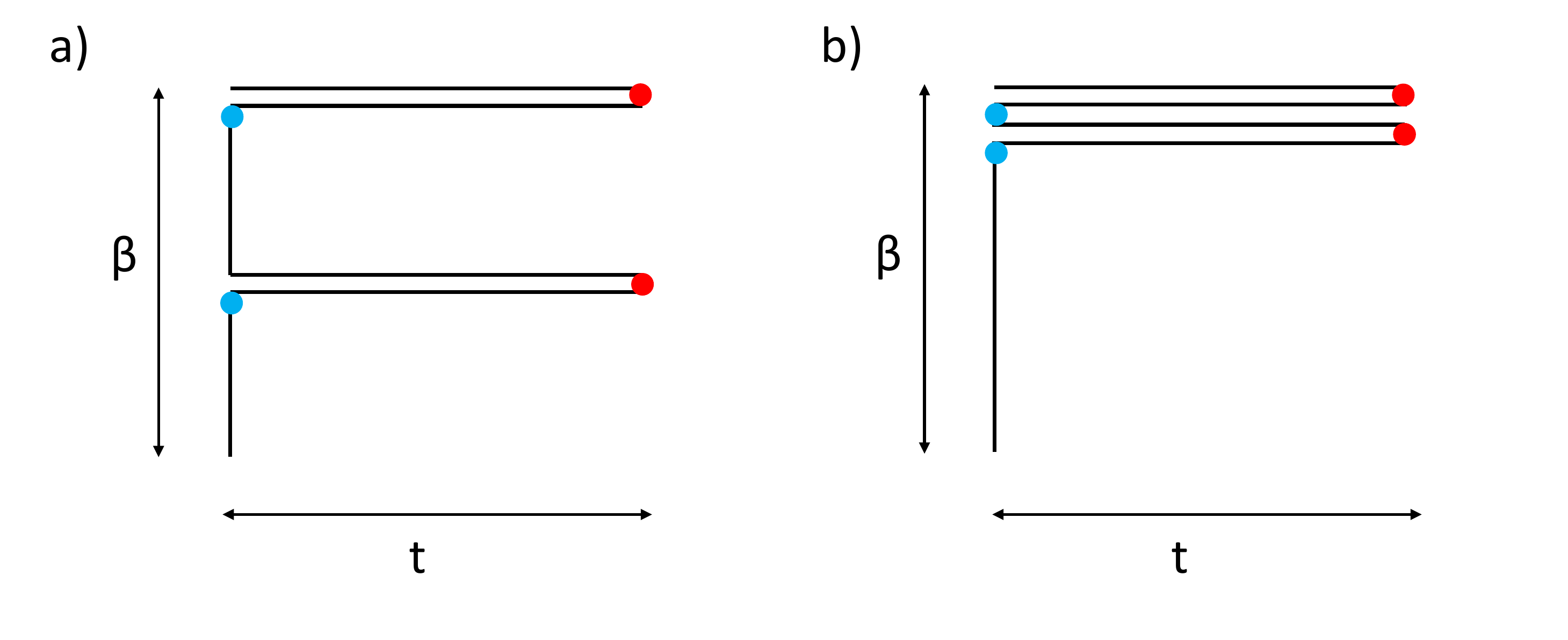}
    \caption{Contour for the (a) regulated and (b) unregulated out of time ordered correlators. The red points refer to the time evolved operators $W_{\mathbf{0}}(t)$, and the blue points refer to the probe operators $V_{\mathbf{x}}$. The regulated and the unregulated correlators are distributed in distinct ways along the thermal circle.}
    \label{fig:regulated_contour}
\end{figure}

Two choices of the squared commutator which have been studied in the literature, are the `regulated' squared commutator, $C_{r}(t,\mathbf{x})=\mathcal{C}_{1/2}(t,\mathbf{x})$, and the `unregulated' squared commutator, $C_{u}(t,\mathbf{x})=\mathcal{C}_{1}(t,\mathbf{x})$. When the expressions of the regulated and unregulated squared commutators are expanded, they contain terms which are thermally smeared versions of out of time ordered four point correlators of the form $W_{\mathbf{0}}(t)V_{\mathbf{x}}W_{\mathbf{0}}(t)V_{\mathbf{x}}$, evaluated on two distinct thermal contours, as shown in Fig. \ref{fig:regulated_contour} a and b. In this work, we study these two squared commutators, and explore the difference in the physics that they capture \cite{Liao2018,Romero-Bermudez2019}.

\subsection{Lyapunov exponent, butterfly velocity, and wavefront broadening}
The squared commutator in holographic models, or in quantum systems with a semiclassical limit, grows exponentially at early times with a `Lyapunov exponent' $\lambda_{L}$,
$\mathcal{C}(t)\sim e^{\lambda_{L}t}$. In spatially local systems, the time argument can be replaced by the appropriate $t \to t-x/v_B$, where $v_B$ is a velocity determining the speed of information scrambling, called the `Butterfly velocity' \cite{Roberts2016,Aleiner2016,Patel2017,Chowdhury2017}. The butterfly velocity is state dependent analogue of the microscopic Lieb Robinson velocity \cite{Lieb1972}.

However, interacting local quantum systems which are not in a semi-classical limit (that is, the number of local degrees of freedom is finite, and not large as in the case for systems with a semi-classical limit), show a qualitatively different behavior. As studies of random unitary circuits \cite{Nahum2018,VonKeyserlingk2018}, stochastic local Hamiltonian spin models \cite{Xu2019}, and numerical studies on deterministic quantum spin models \cite{Xu2018,Khemani_2018,Sahu2019,Han2019} have shown, the near wave-front behavior of the squared commutator is,

\begin{equation}
    \mathcal{C}(t,\mathbf{x}) \sim \exp \left(-\lambda\frac{ \left(x/v_{B}-t\right)^{1+p}}{t^{p}}\right)\text{, for }x\gtrsim v_{B}t.
\end{equation}

This behavior satisfies a ballistically growing and a broadening operator wavefront, $x\sim v_{B}t + \# t^{p/(1+p)}$, where $v_{B}$ is the Butterfly velocity and $p$ is the broadening coefficient. For $p=1$, the broadening is diffusive, which is observed in the case of random unitary circuits \cite{Nahum2018,VonKeyserlingk2018}. This ballistic-diffusive form doesn't exhibit an exponential `chaotic' behavior. Until now, most studies of broadening were done at infinite temperature. However, unlike the `Lieb Robinson velocity' of local quantum systems, the `Butterfly velocity' is a state dependent information spreading velocity, and hence is a temperature dependent quantity. Furthermore the Lyapunov exponent and butterfly velocity could depend non-trivially on the choice of the contour. In this paper we explore these questions through a combination of numerical studies on quantum spin systems and analytical studies of tractable semi-classical field theory models.

\subsection{Summary of our results}
In this work we use a combination of numerical and analytical techniques to study the temperature and contour dependence of squared commutator in strongly interacting, gapped, local quantum systems. We do this firstly using a novel numerical technique based on matrix product operator (MPO) representation of Heisenberg operators to study scrambling in 1D quantum spin chains. We can access both the regulated and unregulated squared commutators in the early growth regime for a gapped, local Hamiltonian for large spin chains of $\mathcal{O}(200)$ spins upto long times $t \sim 100 J^{-1}$, where $J^{-1}$ is the interaction scale of the Hamiltonian. Next, we study the low temperature behavior of the squared commutator in the paramagnetic phase of the $2+1 D$ non-linear $O(N)$ model using perturbative calculation of the ladder-sum for the OTOC functions. We first list out the important results and the structure of the paper,

\textbf{1.} In Sec. \ref{sec:mpo_spins}, we introduce the MPO numerical technique and apply it to calculate both the regulated and unregulated squared commutators in 1D mixed field Ising Hamiltonian. We observe a broadening of the expanding operator wave-front at all temperatures. This broadening behavior had been previously observed for the infinite $T$ ensemble \cite{Nahum2018,VonKeyserlingk2018,Xu2018,Khemani_2018}; but here we confirm the persistence of the broadening behavior even at low temperatures.

For the regulated squared commutator we notice a strong temperature dependence of the broadening coefficient and butterfly velocity. We observe that at temperatures lower than the gap, $\beta > m^{-1}$, the butterfly velocity is consistent with a power-law ($(\beta m)^{-a}$ with $a>0$) behavior. 

For the unregulated squared commutator, on the other hand, we observe that the butterfly velocity and the broadening coefficient have no observable temperature dependence, and in fact remain constant even as the temperature is tuned from $\beta=0$ to $\beta>m^{-1}$. This confirms a strong contour dependence of the OTOC \cite{Liao2018,Romero-Bermudez2019}. We also numerically study the contour dependence of $\partial_{t}C_{(\alpha)}(t,\mathbf{x})$ and make a comparison with the chaos bound to demonstrate that the bound doesn't apply to these squared commutators.

\textbf{2.} While the MPO technique can access temperatures below the gap, it is challenging to access very low temperatures. In order to calculate the temperature dependence at low temperatures, in Sec. \ref{sec:field_theory}, we calculate the behavior of the regulated and unregulated squared commutator in the paramagnetic phase of the $2+1 D$ non-linear $O(N)$ model. This is a gapped strongly interacting theory for which we can analytically calculate the scrambling behavior at large $N$ using a diagrammatic ladder technique. We find that the Lyapunov exponent is $\lambda_{L}\sim e^{-\beta m}/\beta$, and the butterfly velocity is $v_{B}\sim (\beta m)^{-1/2}$ at low temperatures such that $\beta >> m^{-1}$. This shows that the butterfly velocity has the same scaling as the speed of sound of semiclassical massive particles. The field theory calculation can't, however, reproduce the broadening behavior or the contour dependence, indicating that finite N corrections need to be taken into account for those features. 

\textbf{3.} In Sec. \ref{sec:discussion}, we summarize our results and compare the numerical and analytical approaches. We discuss the relation between the temperature dependence of butterfly velocity obtained in this paper with a recently derived temperature dependent bound on butterfly velocity \cite{Han2019}. The bound is not sensitive to the contour dependence, and we show that it is consistent with temperature dependence of the butterfly velocities observed in Sec. \ref{sec:mpo_spins} and \ref{sec:field_theory}. 

\section{Matrix product operator method for numerical calculation of scrambling}\label{sec:mpo_spins}

We now numerically study scrambling in a spatially local quantum system, consisting of tensor product of finite dimensional local Hilbert spaces, like spins on a lattice. The Hamiltonian is assumed to be a sum of geometrically local terms, and the lattice has a well defined position label.

Operators acting on vectors in a Hilbert space $\mathcal{H}$ can be viewed as vectors on a `doubled' Hilbert space $\mathcal{H}_{L}\otimes\mathcal{H}_{R}$. Here the tensor product structure refers to the two copies - `left' and `right' - of the state Hilbert spaces. We introduce the notation $|..)$ to denote the operator as a vector. A local operator acting on the $\mathbf{0}$ position in the lattice, $|W_{\mathbf{0}})$, can be time evolved in the Heisenberg picture, 
\begin{equation}\label{eq:mpo_tebd}
|W_{\mathbf{0}}(t))=|U_{t}W_{\mathbf{0}}U_{t}^{\dagger})=e^{it(H_{L}\otimes I-I\otimes H^{*}_{R})}|W_{\mathbf{0}}).
\end{equation}
One can now probe the evolved operator using a second local operator at a position $\mathbf{x}$ by constructing its commutator,
\begin{equation}
|O(\mathbf{x},t))=|[W_{\mathbf{0}}(t),V_{\mathbf{x}}])=(1\otimes V_{\mathbf{x}}^{T}-V_{\mathbf{x}}\otimes I)|W_{\mathbf{0}}(t)),
\end{equation}

The squared commutator can be obtained by squaring this operator which measures the extent of quantum information scrambling in the system. The $\alpha$ dependent squared commutator defined in Eq. \ref{eq:alpha_square_norm} can be expressed as a norm of an operator state, $\mathcal{C}_{(\alpha)}=(O_{\alpha}(\mathbf{x},t,\beta)|O_{\alpha}(\mathbf{x},t,\beta))$, where, 
\begin{equation} \label{eq:op_state_time_finiteT}
|O_{\alpha}(\mathbf{x},t,\beta))=|\rho^{(1-\alpha)/2}O(\mathbf{x},t)\rho^{\alpha/2}).
\end{equation}

\subsection{Model and numerical method}
We consider the mixed field quantum Ising model, 
\begin{equation}\label{eq:MFI}
H=-\frac{1}{E_{0}}\left(J\sum_{i=1}^{L-1} Z_{i}Z_{i+1}+h_{x}\sum_{i=1}^{L}X_{i}+h_{z}\sum_{i=1}^{L}Z_{i}\right)
\end{equation}
with $E_{0}=\sqrt{4J^{2}+2h_{x}^{2}+2h_{z}^{2}}$, on a one dimensional lattice. The $X$ and $Z$ matrices are the usual Pauli matrices. The parameters chosen are, $J=1, h_{x}=1.05, h_{z}=0.5$. Time is measured in the units of $J^{-1}=1$. This is a gapped system, and the spectral gap between the ground state and the first excited state is $\sim 1.13$ as extracted from small size exact diagonalization.

We want to calculate $C_{u,r}(t,x)$ for large system sizes and upto long times, and we employ the Matrix product operators (MPO) based technique to time evolve operator states which extends the time dependent density matrix renormalization group (t-DMRG) technique \cite{Vidal2003,Vidal2004,Daley2004,White2004} to super-operators \cite{Xu2018}. We first time evolve the local operator $W$ by doing time evolution using super-operator $H\otimes I-I\otimes H^{*}$ on the operator state, following Eq. \ref{eq:mpo_tebd}. We also obtain $|\rho)$ by evolving the identity $|I)$ operator state in imaginary time. Now, we can construct the operator state $|O_{\alpha}(t,\mathbf{x},\beta))$ as defined in Eq. \ref{eq:op_state_time_finiteT}, for $\alpha = 1/2 (1)$, and its norm squared is the required (un)regulated squared commutator.

In the MPO based method, at each Trotter step, we must truncate the MPO to a fixed bond dimension, thereby introducing errors. However, we will demonstrate that our numerical procedure converges (for small values of the squared commutator) at large system sizes ($L\sim200$) and upto long times $t\sim 100$, even at low temperatures, which makes it a powerful method to study the temperature and contour dependence of quantum information scrambling.
\begin{figure}
	\includegraphics[width=0.505\columnwidth]{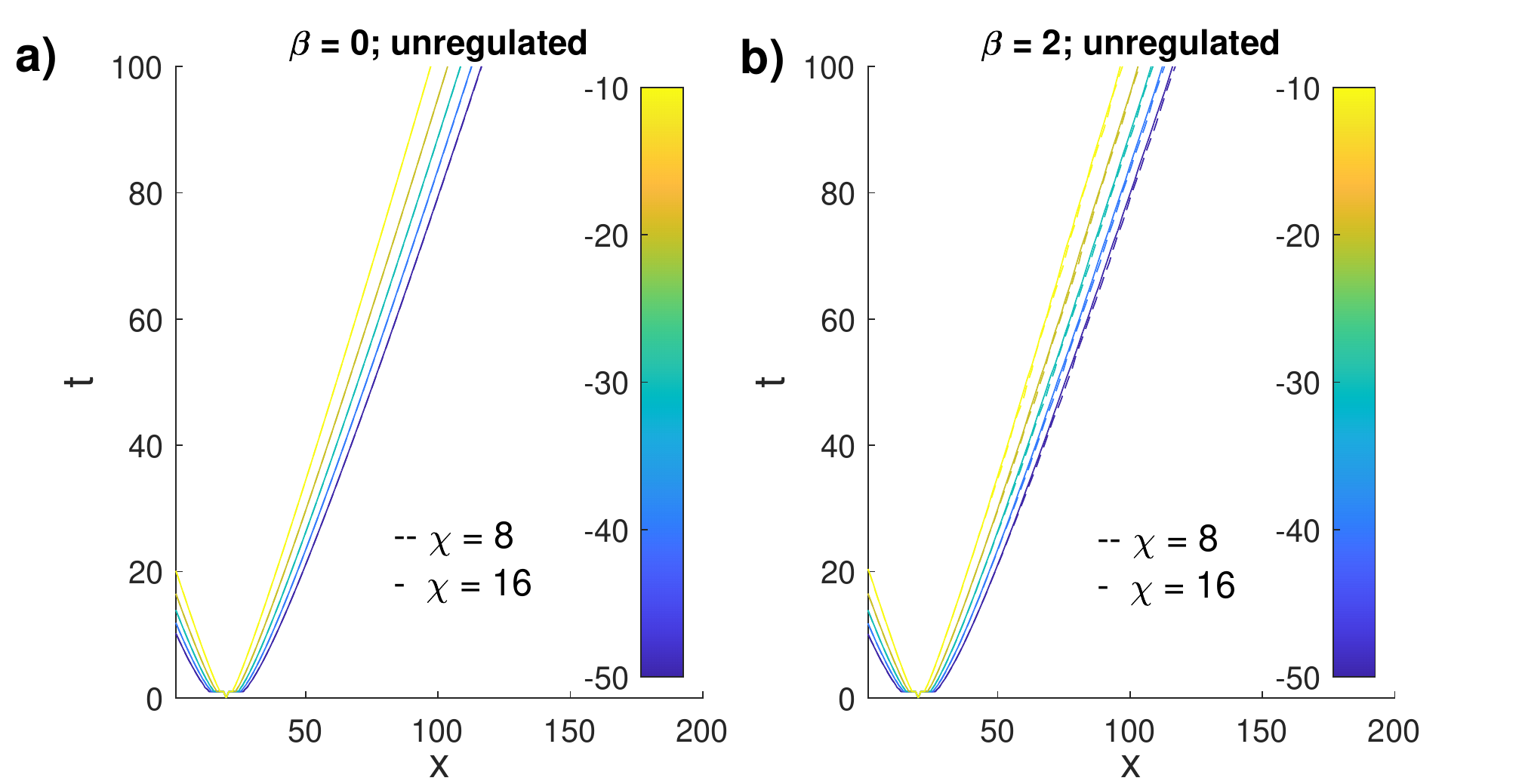}
	\includegraphics[width=0.485\columnwidth]{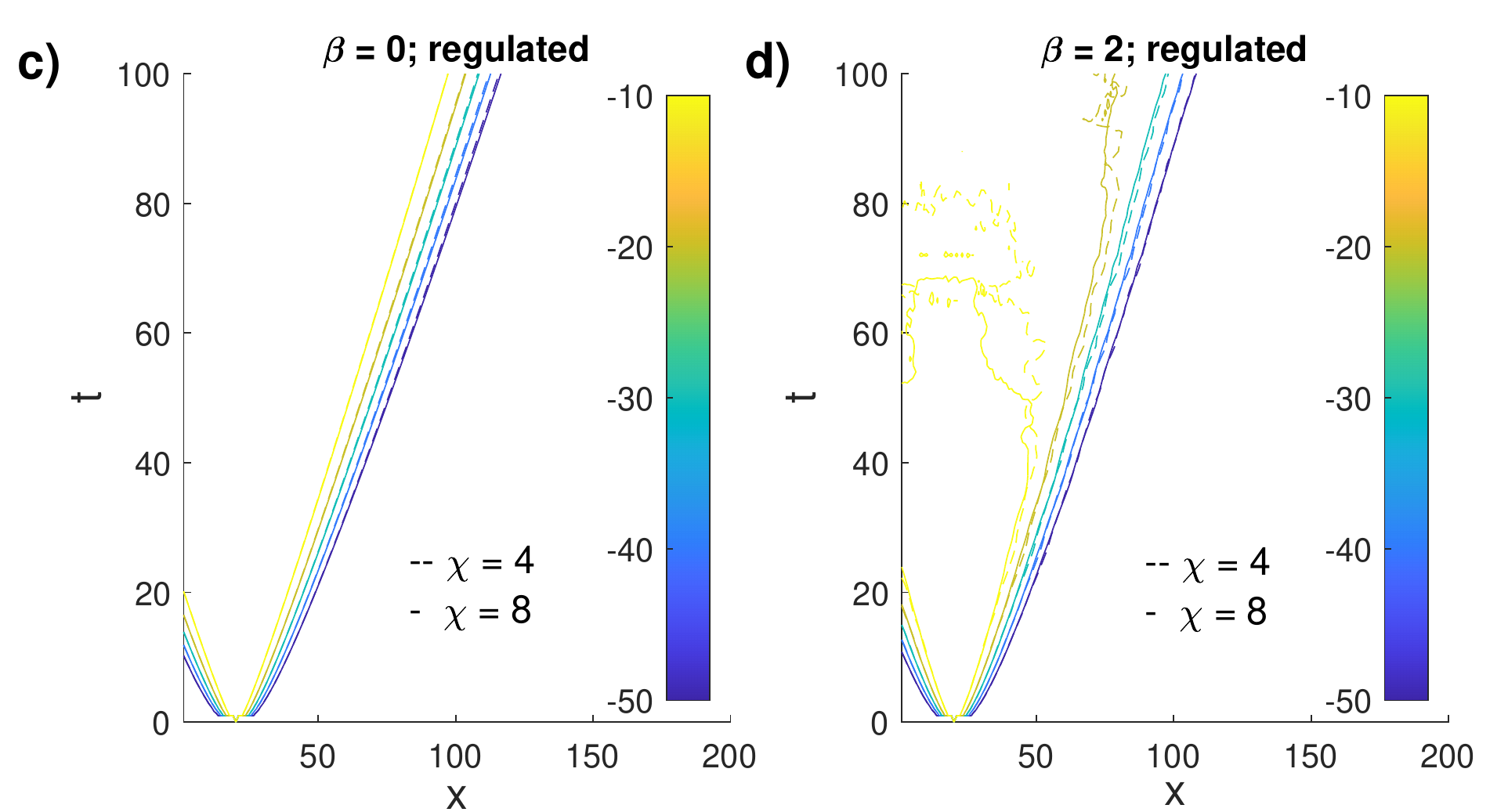}
	\caption{The contours of the logarithm of the regulated and unregulated squared commutator at different temperatures - a) $\beta = 0$ (unregulated), b) $\beta = 2$ (unregulated), c) $\beta = 0$ (regulated) and d) $\beta = 2$ (regulated) are shown. For the unregulated case bond dimensions, $\chi = 8$ and $\chi = 16$, and for the regulated case bond dimensions, $\chi = 4$ and $\chi = 8$ are considered. The data shows convergence even at low temperatures for $\log C_{r}<-30$, and for $\log C_{u}<-20$.}\label{Fig:contour}
\end{figure}

We consider a $L=200$ spin chain with the mixed field Ising Hamiltonian as in Eq. \ref{eq:MFI}. We start with an operator $X_{20}$, a Pauli $X$ operator localized at the site $20$, and construct the squared commutator with $Z$ operators at all sites of the chain. We perform the MPO-TEBD method with Trotter steps, $\delta t = 0.005$ for time evolution (to generate $X(t)$) and $\delta \beta = 0.05$ for imaginary time evolution (to generate $\rho$), for bond dimensions $\chi = 4,8$ (regulated) and $\chi=8,16$ (unregulated). To calculate the regulated and unregulated squared commutators, we need to construct the MPOs $|O_{1/2}(t,x,\beta))$ and $|O_{1}(t,x,\beta))$, as defined in Eq. \ref{eq:op_state_time_finiteT}, respectively. For $|O_{1/2})$ we need to perform two MPO multiplications, $\rho^{1/4}\to [X_{20}(t),Z_{x}]\rho^{1/4} \to \rho^{1/4}[X_{20}(t),Z_{x}]\rho^{1/4}$, while for $|O_{1})$, we need to perform one MPO multiplication, $\rho^{1/2}\to [X_{20}(t),Z_{x}]\rho^{1/2}$. The details of the numerical implementation, which include a comparison to exact diagonalization, discussions on convergence with bond dimension, and the fitting procedure, are provided in App. \ref{appsec:numerics_mpo}.

A heuristic justification of why the MPO approximation works is as follows - it was shown in \cite{Xu2018} that the commutator $[X(t),Z_{x}]$ has a small operator entanglement outside the light-cone. It is also well understood that the thermal density matrix $\rho$ satisfies an area law in mutual information \cite{Wolf2008}, and hence is expected to be reliably approximated by a low bond dimension matrix product operator. These two arguments imply that the operator $|O_{\alpha}(t,\mathbf{x},\beta))$ as defined in Eq. \ref{eq:op_state_time_finiteT}, which is an MPO multiplication of powers of $\rho$ and the commutator $[X(t),Z_{x}]$, should have a small operator entanglement outside the light-cone (i.e. when the squared commutator is small), and hence can be well approximated by a low bond dimension MPO.

As has been pointed out previously, in \cite{Xu2018,Hemery2019,Sahu2019}, the MPO-TEBD method can capture the qualitative features of scrambling only if the scrambling data has converged with bond dimension. We ensure that all our further analysis is done on scrambling data only in the spatio-temporal domain where it has converged with bond dimension. We plot the contours of the squared commutator in Fig. \ref{Fig:contour}, and demonstrate that the contours converge very well for small values of the squared commutator. The shape of the contours, where the data has converged, show that the wavefront propagates ballistically with a velocity.

\subsection{Broadening of the wavefront}
\begin{figure}
	\includegraphics[width=0.485\columnwidth]{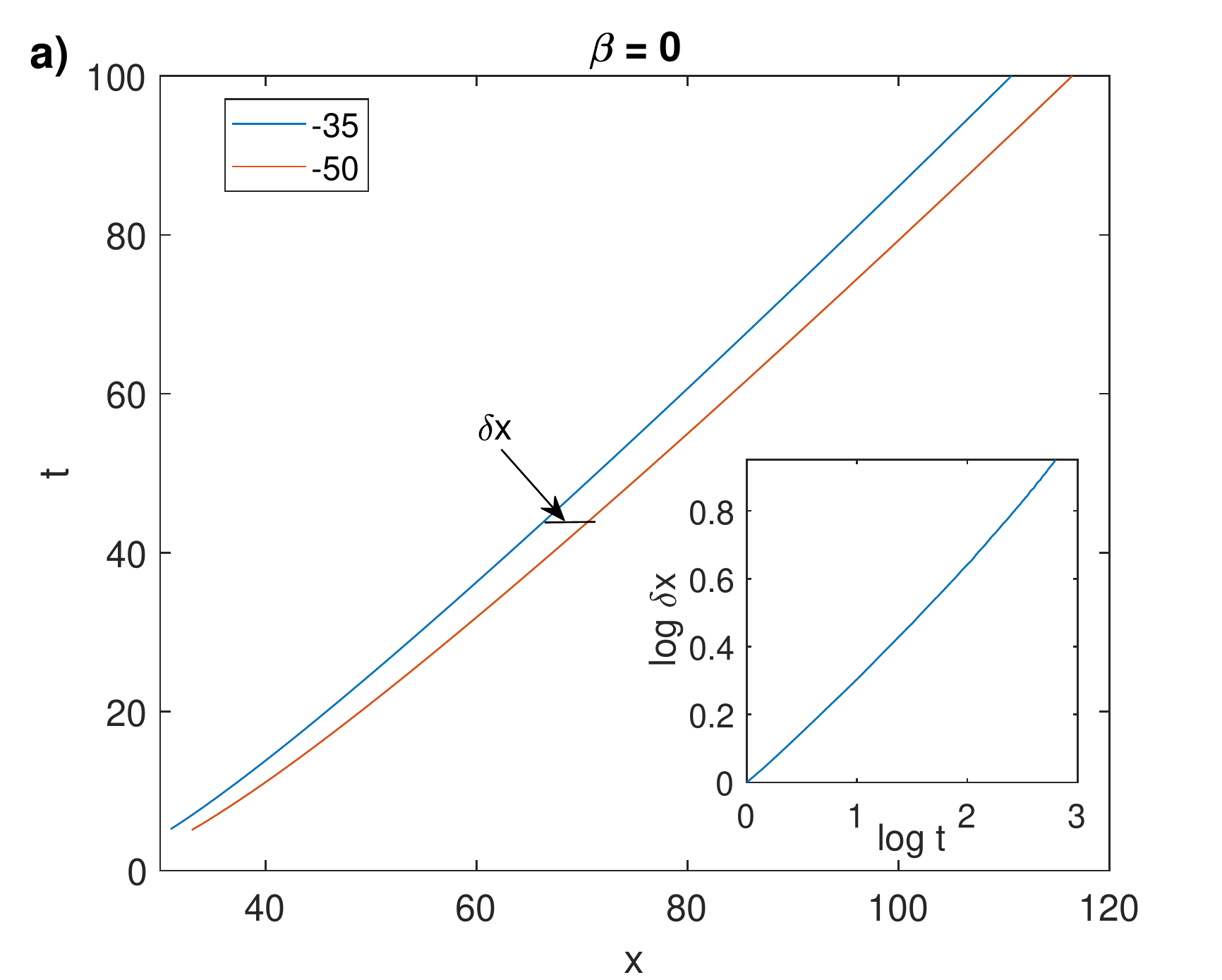}
	\includegraphics[width=0.485\columnwidth]{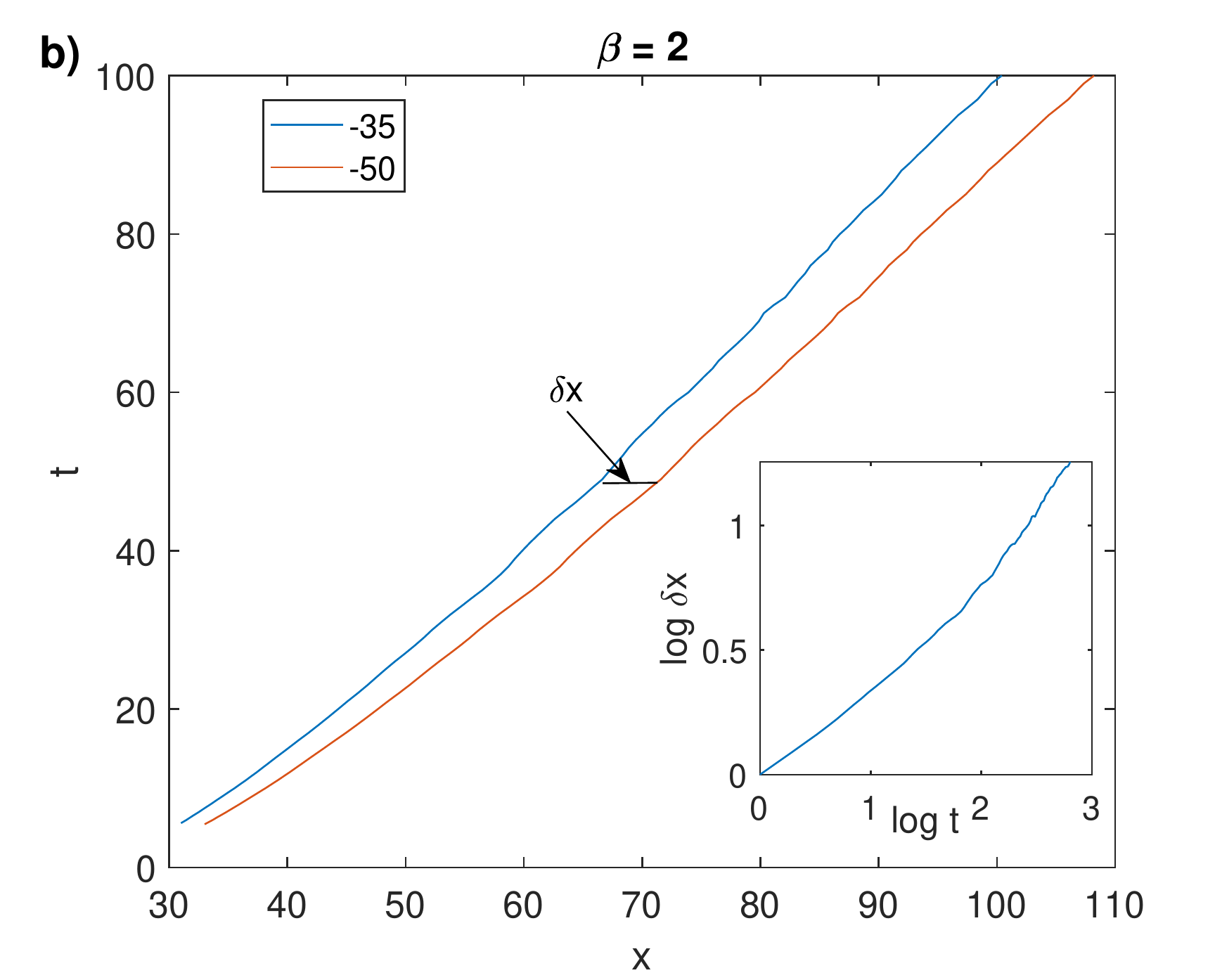}
	\caption{We extract the contours of $\log C_{r} = -35$ and $-50$ at different temperatures, for the data with $\chi = 8$. From the contours we extract $\delta x$, which is the spatial distance between the two contours. The time dependence of $\delta x$ is shown in the inset; the fact that it is increasing with time demonstrates a broadening of the wavefront. The broadening persists even at a) high temperature $\beta = 0$ and b) low temperature $\beta = 2$.}\label{Fig:reg_broadening}
\end{figure}
Without any numerical fitting, we demonstrate the broadening behavior of the operator wavefront even at low temperatures in the Fig. \ref{Fig:reg_broadening}. We extract the spatial separation $\delta x$ between two chosen contours of the $\log C_{r}$, and plot its time dependence in the insets of Fig. \ref{Fig:reg_broadening}. A positive (and an increasing) slope implies a broadening behavior. In Fig. \ref{Fig:reg_broadening}, we show data for the regulated case, but a similar study for the unregulated squared commutator also demonstrates a broadening behavior. Thus, the Figs. \ref{Fig:contour} and \ref{Fig:reg_broadening} together show that the early time (before the light-cone is reached) behavior of the squared commutator has a ballistic growth and a broadening wavefront.

In \cite{Xu2018,Xu2019,Khemani2018}, it was argued that the squared commutator, near the wavefront, when $C(x,t)<<1$, can be captured by the following ansatz, 
\begin{equation} \label{eq:scrambling_ansatz}
C(x,t)\sim exp\left(-\lambda_{p}\frac{\left((x-x_{0})/v_{B}-t\right)^{1+p}}{t^{p}}\right).
\end{equation}
One can identify the broadening coefficient $p$ as, 
\begin{equation}
\frac{\delta \log \delta x}{\delta t}\sim \frac{p}{p+1}.
\end{equation}
We now fit our data to the ansatz in Eq. \ref{eq:scrambling_ansatz} to extract the Lyapunov exponent, butterfly velocity and broadening coefficient.

\subsection{Temperature dependence of butterfly velocity}
\begin{figure}
\includegraphics[width=0.48\columnwidth]{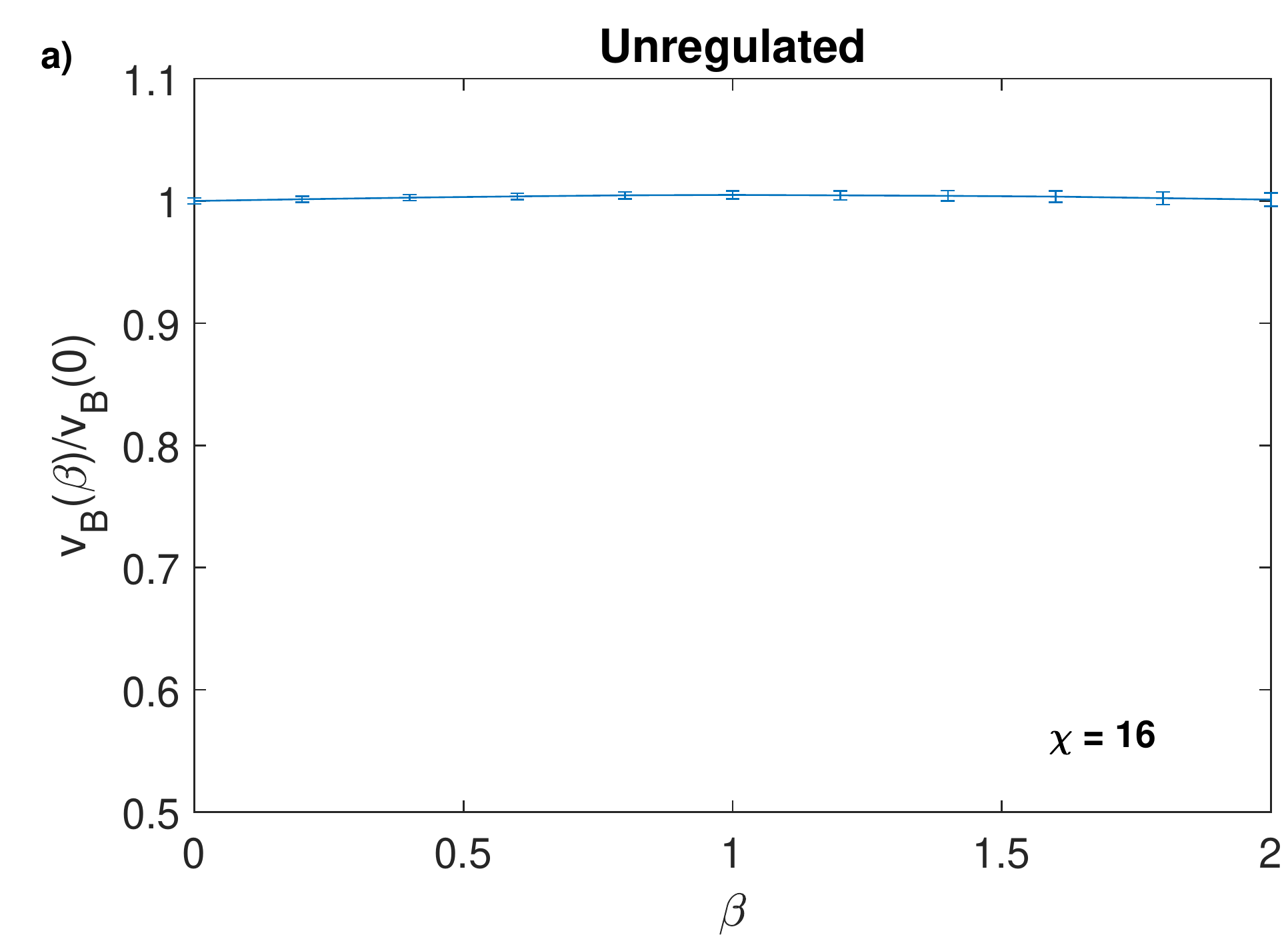}
\includegraphics[width=0.48\columnwidth]{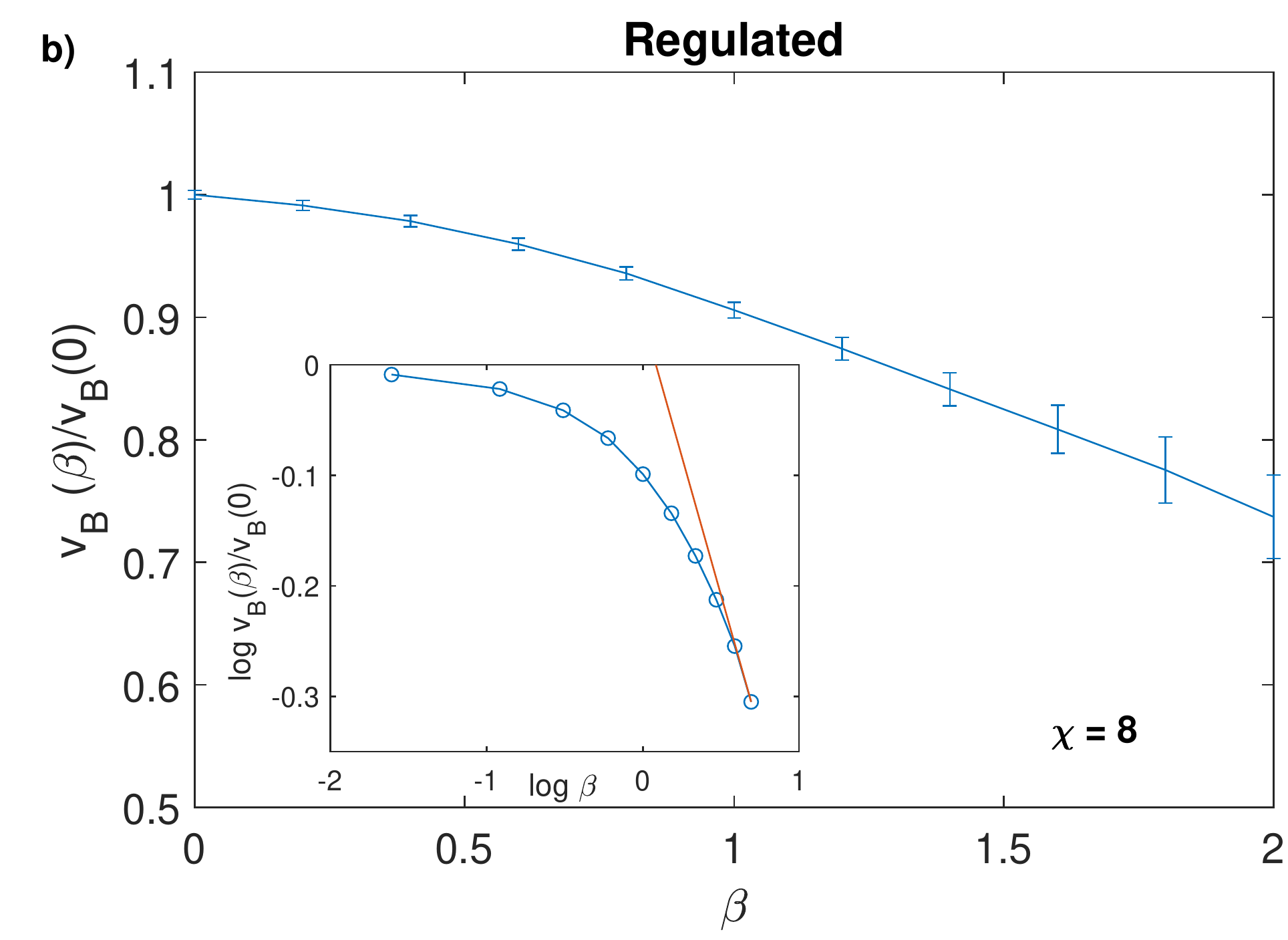}
	\caption{\textbf{a)} We plot the extracted $v_{B}(\beta)/v_{B}(0)$ for the unregulated case, as a function of $\beta$. The data is for $\chi=16$ bond dimension. The butterfly velocity is practically constant at all temperatures. \textbf{b)} For the regulated case, we plot the normalized $v_B$, (i.e. $v_{B}(\beta)/v_{B}(0)$), extracted from the $\chi=8$ data, as a function of $\beta$. In the inset, in the log-log scale, we demonstrate that the low temperature behavior of $v_B$ is consistent with $\beta^{-1/2}$ (which is the slope of the red line plotted.). }\label{Fig:reg_fit_velo}
\end{figure}
We extract the butterfly velocity, velocity dependent Lyapunov exponent and the broadening coefficient from the obtained numerical data by fitting them to the near wave-front ansatz in Eq. \ref{eq:scrambling_ansatz}. In Fig. \ref{Fig:reg_fit_velo}a, we plot fitted $v_{B}(\beta)/v_{B}(0)$ as a function of $\beta$ for the unregulated case, and see that the fitted butterfly velocity has almost no discernible temperature dependence. In Fig. \ref{Fig:reg_fit_velo}b, we plot the same for the regulated case, and notice a strong temperature dependence. The low temperature behavior is consistent with a power law decrease in the butterfly velocity as a function of $\beta$, as is shown in the inset of Fig. \ref{Fig:reg_fit_velo}b. In Sec. \ref{sec:field_theory}, we show that at the low temperature limit of an analytically tractable field theory model with a mass gap $m$, the butterfly velocity has a temperature scaling which is the same as the equipartition behavior - $\sqrt{1/\beta m}$. The asymptotic low temperature behavior in the MPO calculation (even though the temperatures we access here are not very low compared to the spectral gap) is close to the $\sqrt{1/\beta m}$ behavior, as is demonstrated in Fig. \ref{Fig:reg_fit_velo}b.  

In App. \ref{appsec:numerics_mpo}, we also study the temperature dependence of the broadening coefficient $p$. In Fig. \ref{Fig:p_reg_unreg_fit}, we show that $p$ for the unregulated case has a very weak dependence on temperature and remains practically constant as the temperature is lowered. The regulated case, however, has an increasing trend for $p$ with decreasing temperature. 

\subsection{Contour dependence and chaos bound}\label{sec:chaosbound}
\begin{figure}
	\includegraphics[width=0.8\columnwidth]{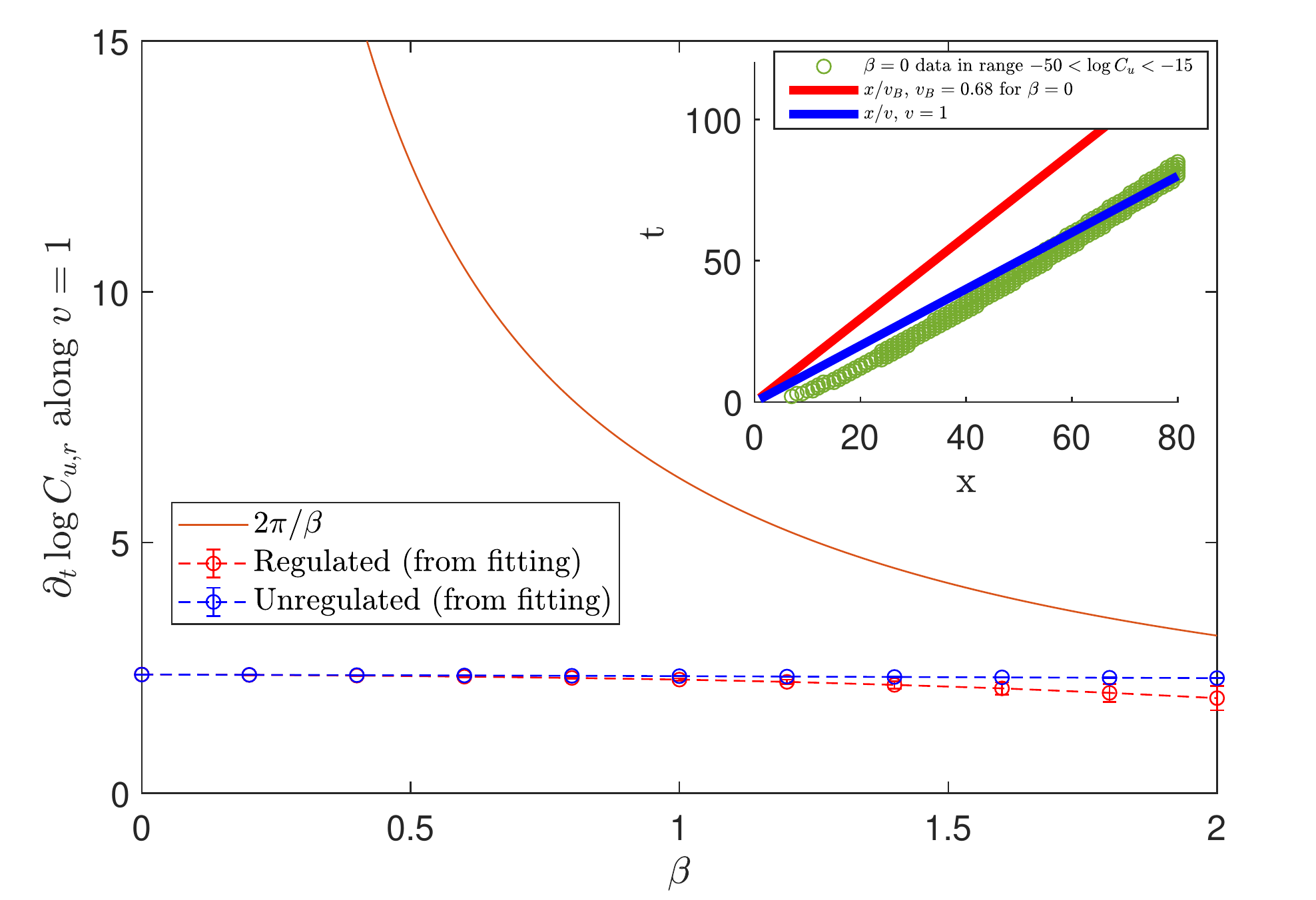}
	\caption{From the fitting of the obtained data of the regulated and unregulated squared commutators, we obtain the $\partial_{t} C_{u,r}$ from the near wavefront ansatz, along a `ray' $x = t$ and compare it against the `bound on chaos' $2\pi/\beta$. In the inset, we show the `ray' $x=vt$ at $v=1$, and compare that to the butterfly velocity $v_{B} = 0.68$ at $\beta=0$ for $C_{u}$.  }\label{Fig:lambda_reg_unreg}
\end{figure}
For a symmetrically defined out of time ordered correlation function, there exists the Maldacena-Shenker-Stanford (MSS) chaos bound $\lambda_{L}\leq 2\pi/\beta$ \cite{Maldacena2016}. The symmetric OTOC is defined as, 
\begin{equation}
    F(t,\mathbf{x}) = Tr\left(\rho^{1/4}V_{\mathbf{x}}\rho^{1/4}W_{0}(t)\rho^{1/4}V_{\mathbf{x}}\rho^{1/4}W_{0}(t)\right).
\end{equation}
This is related to the regulated squared commutator, as the $C_{r}(t,\mathbf{x})$, when expanded, 
\begin{equation}
    C_{r}(t,\mathbf{x}) = 2\left(Tr\left(\rho^{1/2}V_{\mathbf{x}}W_{0}(t)\rho^{1/2}W_{0}(t)V_{\mathbf{x}}\right)-Re F(t+i\beta/4,\mathbf{x})\right).
\end{equation}
Let's introduce a related quantity $F_{d}(t,\mathbf{x}) = Tr\left(\rho^{1/2}V_{\mathbf{x}}\rho^{1/2}V_{\mathbf{x}}\right)Tr\left(\rho^{1/2}W_{0}(t\rho^{1/2}W_{0}(t\right)$. In \cite{Maldacena2016}, it was proven that the following bound exists,
\begin{equation}
    \frac{\partial\log \left(F_{d}(t,\mathbf{x})-F(t,\mathbf{x})\right)}{\partial t}\leq \frac{2\pi}{\beta}.
\end{equation}

Given this result, one might conjecture that the related quantity $\partial_{t}\log C_{r}(t,\mathbf{x})$ also satisfies the same bound. To study this, we can calculate $\partial_{t}\log C_{r}(t,\mathbf{x})$ along different `rays' $x=vt$ \cite{Khemani_2018}; if the near wavefront scrambling ansatz (Eq. \ref{eq:scrambling_ansatz}) is satisfied, then $\partial_{t}\log C_{u,r}$ along a ray of velocity $v$ is given by $\lambda_{p}(v/v_{B}-1)^{p}(1+pv/v_{B})$. At sufficiently large $v$, this will violate the chaos bound. In Fig. \ref{Fig:lambda_reg_unreg}, we plot the $\partial_{t}\log C_{r,u}(t,\mathbf{x})$, for a fixed `ray' $x = t$, obtained from fitting of the unregulated and regulated cases to the ansatz, as a function of $\beta$ and notice that the unregulated case is practically constant, and can violate the bound at lower temperatures. We confirm this without numerical fitting, in App. \ref{appsec:chaosbound}, Fig. \ref{fig:chaos_bound_data}. In App. \ref{appsec:chaosbound} we also study $\partial_{t}\log C_{r,u}(t,x=vt)$, as a function of `ray' velocity $v$. We find that at high ray velocities $v$, both $\partial_{t}\log C_{r}(t,vt)$ and $\partial_{t}\log C_{u}(t,vt)$ violate the bound. This shows that the MSS bound doesn't hold for the squared commutators we considered.

\subsection{Summary of findings from the MPO numerics}
By studying squared commutators for large-sized, gapped spin chain which is spatially local, and has finite dimensional local Hilbert spaces, we got three distinctive features. First, the spatial locality leads to a ballistic wavefront propagating at the butterfly velocity, which has distinct temperature scaling for the regulated and unregulated cases. In the unregulated case the velocity is constant, while for the regulated case, the velocity decreases with temperature. Second, the wavefront broadens with time for both contours, and thus the squared commutator doesn't have pure exponential growth. Third, there are numerical indications that the chaos bound is not satisfied for these squared commutators. 

Can we explain these behaviors using an analytically tractable model? In particular, can we understand the low temperature limit which is not accessible in the spin chain numerics?  We explore that in the next section, where we consider a non-linear $O(N)$ model in $2+1 D$, which is spatially local, and solvable at large $N$. We study the scrambling behavior at low temperatures for the gapped phase of the model, and find that the butterfly velocity indeed varies as $\sqrt{T/m}$ at low temperatures. However, we will find that the field theory calculation doesn't show contour dependence or wavefront broadening. 

\begin{fmffile}{whole_file}
\section{Scrambling in the paramagnetic phase of the non-linear $O(N)$ model}\label{sec:field_theory}
\begin{figure}
    \centering
    \includegraphics[height=0.6\columnwidth]{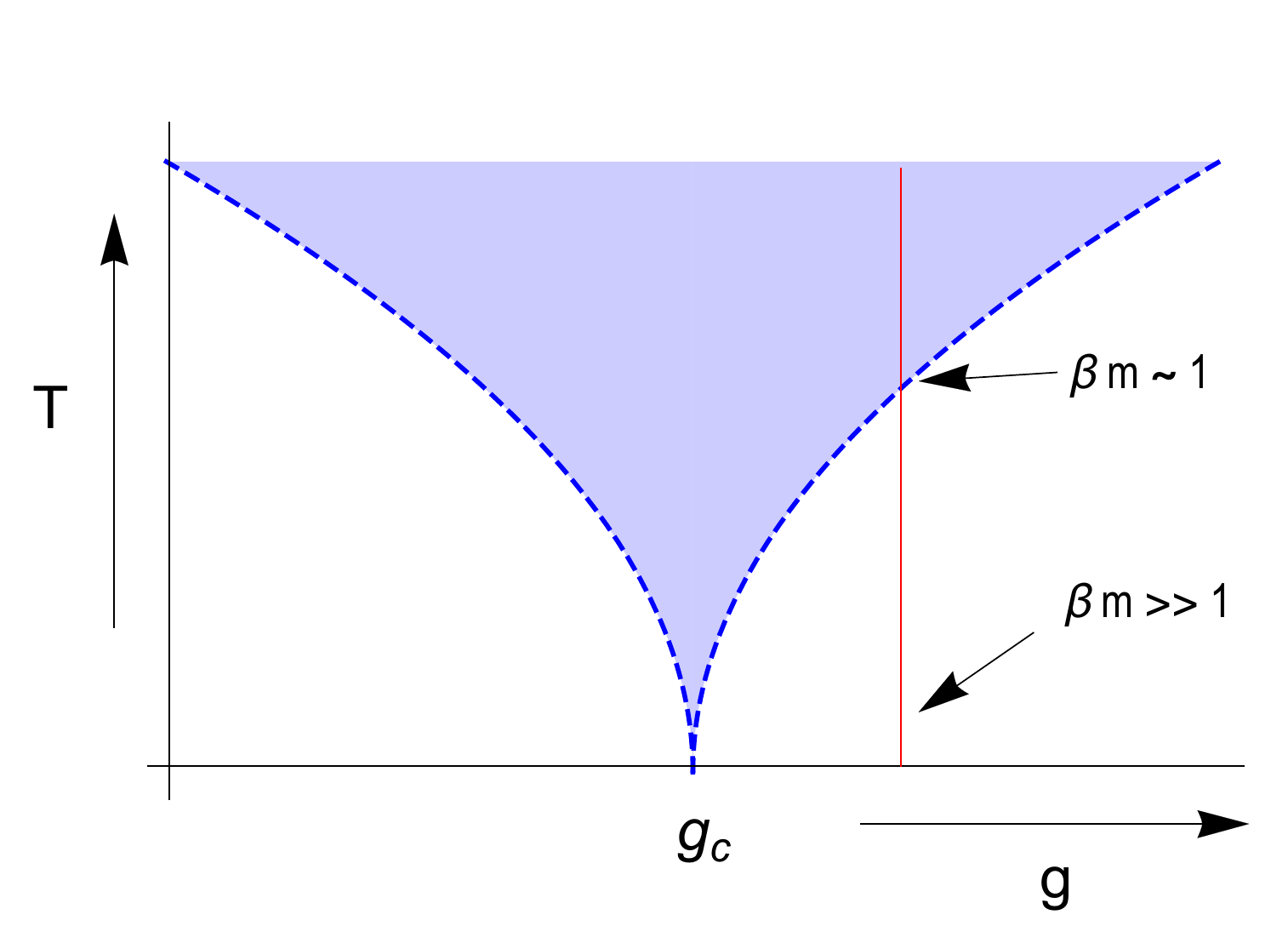}
    \caption{This is the critical phase diagram of the non-linear $O(N)$ model. The blue shaded region is controlled by the critical theory around the quantum critical point at $T=0$ and $g=g_{c}$, while the dashed lines indicate a cross-over to the phases controlled by the symmetry of the zero temperature phases away from the critical point. We focus on the low temperature behavior of the symmetry unbroken paramagnetic phase $g>g_{c}$.}
    \label{fig:crit_diag}
\end{figure}

The non-linear $O(N)$ model is a spatially local field theory of an $O(N)$ symmetric vector field $\phi_{a}$, with $a=1,..,N$. The theory is solvable at large $N$, and in this limit this model differs from the spin chain in the fact that the local Hilbert space is not finite. Furthermore, to avoid complications in the field theory at $1+1D$ due to scattering, we study this model at $2+1D$, and we expect that dimensionality will not affect qualitative features of the temperature and contour dependence. The critical phase diagram \cite{Sachdev2011} of this model is shown in Fig. \ref{fig:crit_diag}. We analyse this model using Ladder sum techniques developed in \cite{Stanford2016,Chowdhury2017} (see also \cite{Steinberg_2019,cheng2019chaos,Alavirad_2019,Gu_2019}), and study both the temperature and contour dependence of the squared commutators. 

The real time lagrangian for this theory is given by,
\begin{equation}
    \mathcal{L}=\frac{1}{2}\left[\sum_{a}(\partial \phi_{a})^{2}-\frac{v}{2N}\left(\phi_{a}^{2}-\frac{N}{g}\right)^{2}\right]
\end{equation}
The action is given by $\int_{x}\mathcal{L}$, where the space-time integration $\int_{x}$ is over $2+1D$. We have set the speed of light $c$ and $\hbar$ to 1.  The parameter $g$ (which determines the bare mass) can be tuned across a quantum critical point that occurs at $g=g_{c}$, and $v$ is the self-interaction coupling constant. We consider the strong coupling (large $v$) and large-$N$ limit. In \cite{Chowdhury2017}, scrambling behavior was studied at the critical point $g_{c}$, by evaluating the regulated squared commutator using a perturbative ladder sum calculation with $1/N$ as the small parameter \cite{Stanford2016,Chowdhury2017}. Following the diagrammatic techniques used in these studies, we study scrambling on the paramagnetic phase of the model at $g>g_{c}$, where there are quasiparticle-like excitations with finite bare mass $m$. We study the temperature dependence of the scrambling in the low temperature limit $\beta m>>1$.

The main goal of this section is to analytically obtain temperature dependence of the butterfly velocity at low temperatures. We didn't have access to very low temperatures in Sec. \ref{sec:mpo_spins}, and we intend to explore the regime $\beta m >>1 $ using this field theory model. 

The generalized squared commutator in different contours given in Fig. \ref{fig:regulated_contour} is given by,
\begin{equation}\label{eq:o_n_reg_sq_comm}
C_{\alpha}(t,\mathbf{x})=-\frac{1}{N^{2}}\sum_{ab}Tr\left(\rho^{\alpha}[\phi_{a}(t,\mathbf{0}),\phi_{b}(0,\mathbf{x})]\rho^{1-\alpha}[\phi_{a}(t,\mathbf{0}),\phi_{b}(0,\mathbf{x})]\right).
\end{equation}
The regulated and the unregulated squared commutators are given by $C_{r} = C_{1/2}$, and $C_{u} = C_{1}$, respectively.

We summarize the results of this section before showing the explicit calculations. Using the ladder-sum calculation, we find that both the regulated and unregulated squared commutators have the following early time behavior,
\begin{equation}
    C_{r,u}(t,\mathbf{x})\sim\frac{1}{N}e^{\lambda_{0}\left(t-\frac{x^{2}}{v_{B}t}\right)},
\end{equation}
where the `Lyapunov' exponent, $\lambda_{0}\sim e^{-\beta m}/\beta$, and the butterfly velocity, $v_{B}\sim (\beta m)^{-1/2}$. This implies that at low temperatures, the butterfly velocity has the same temperature scaling as the speed of sound (which also scales as $(\beta m)^{-1/2}$) of the semi-classical gas of dilute quasiparticle excitations of the paramagnetic phase of the $O(N)$ model at low temperature.

\subsection{Basic diagrammatics and low temperature relaxation rate}

We introduce auxiliary Hubbard Stratonovich (HS) field $\lambda$ to solve the interacting problem. The Euclidean Lagrangian we consider is 
\begin{equation}
    \mathcal{L}_{E}=\frac{1}{2}\left[\sum_{a}(\partial \phi_{a})^{2}-\frac{\lambda}{\sqrt{N}}\left(\sum_{a}\phi_{a}^{2}-\frac{N}{g}\right)-\frac{\lambda^{2}}{4v}\right]
\end{equation}
The HS field $\lambda$ is chosen so that it generates a zero temperature mass, $m$, such that, $\frac{-\langle\lambda\rangle}{\sqrt{N}}=m^{2}$. The HS field also acts as a Lagrange multiplier, fixing (at large N), $\langle\sum\phi_{a}^{2}\rangle=\frac{N}{g}$. At finite temperature $T$, the constraint imposed by the HS field is 
\begin{equation}
    \begin{split}
        & NT\sum_{i\omega_{n}}\int_{\mathbf{k}}^{\Lambda}\frac{1}{\omega_{n}^{2}+\epsilon_{\mathbf{k}}^{2}} = \frac{N}{g}, \text{ where } \epsilon_{\mathbf{k}}=\sqrt{\mathbf{k}^{2}+m^{2}}. 
    \end{split}
\end{equation}
Here, and in the rest of the paper, $\int_{\mathbf{p}}$ stands for $\int \frac{d^{2}\mathbf{p}}{(2\pi)^{2}}$. At $\beta=\frac{1}{T}=\infty$, this fixes $g$ in terms of $m$ and $\Lambda$, 
\begin{equation}
    \frac{1}{4\pi}(\Lambda-m)=\frac{1}{g}
\end{equation}
At finite temperature, the mass will be modified, as a function $m(\beta)$. We restrict ourselves to low temperature, assuming the hierarchy of scales $\Lambda>>m>>\beta^{-1}$. This implies $m(\beta)\approx m$, i.e., the thermal mass is approximately the same as the bare mass.

The perturbative calculation of the squared commutator can be set up using the basic ingredients - the real time retarded and Wightman propagators of the fields $\phi_{a}$ and the HS field $\lambda$. The retarded propagators are identified as horizontal lines, while the Wightman propagators are denoted as the vertical lines in the diagrams (in momentum space). 

For the $\phi$ field, bare Euclidean propagator in imaginary time $\tau$ is $\mathcal{G}(\tau,\mathbf{x})=Tr\left(\rho\phi_{a}(\tau,\mathbf{x})\phi_{b}(0,\mathbf{0})\right)$, where, $\rho$ is the thermal density matrix, $\rho = e^{-\beta H}/(Z=Tr(e^{-\beta H}))$. The retarded propagator is defined as $\mathcal{G}_{R}(t,\mathbf{x})\delta_{ab} = -i Tr\left(\rho[\phi_{a}(t,\mathbf{x}),\phi_{b}(0,\mathbf{0})]\right) \theta(t)$. In the Fourier space, they are related by analytic continuation of the Matsubara frequencies, $\mathcal{G}_{R}(\omega,\mathbf{k})=-\mathcal{G}(i\omega_{n}\to\omega,\mathbf{k})$. We can calculate and denote the retarded bare propagator as,
\begin{equation}
\begin{split}
    \parbox{20mm}{\begin{fmfgraph}(40,15)
    \fmfleft{i} \fmfright{o} \fmf{plain,width=0.5pt}{i,o}
    \end{fmfgraph}} &:= \quad \mathcal{G}^{(0)}_{R}(\omega,\mathbf{k}) =  \frac{1}{(\omega+i0^{+})^{2}-\epsilon_{\mathbf{k}}^{2}}.
\end{split}
\end{equation}
The spectral function is defined as $A(\omega,\mathbf{k})=-2Im[\mathcal{G}_{R}(\omega,\mathbf{k})]$. The bare $\phi$ spectral function is given by,
\begin{equation}\label{eq:bare_phi_spec}
    A^{(0)}(\omega,\mathbf{k})=\frac{\pi}{\epsilon_{\mathbf{k}}}[\delta(\omega-\epsilon_{\mathbf{k}})-\delta(\omega+\epsilon_{\mathbf{k}})].
\end{equation}

The generalized Wightman function is defined as,
\begin{equation}\label{eq:gen_wightman}
    \mathcal{G}^{(\alpha)}_{W}(t,\mathbf{x})\delta_{ab} := Tr\left(\rho^{\alpha}\phi_{a}(t,\mathbf{x})\rho^{1-\alpha}\phi_{b}(0,\mathbf{0})\right). 
\end{equation} 
By going to the spectral representation, we show in App. \ref{appsec:spec_wightman},
\begin{equation}\label{eq:wightman_spectral}
    \mathcal{G}^{(\alpha)}_{W}(\omega,\mathbf{k}) =  \frac{A(\omega,\mathbf{k})}{2\sinh \frac{\beta\omega}{2}}e^{(\alpha-1/2)\beta \omega}.
\end{equation}
For the $\lambda$ field, the bare Euclidean propagator is $\mathcal{G}_{\lambda}^{(0)}(i\omega_{n},\mathbf{k})=-4v$. At infinite $v$, one can dress the $\lambda$ propagators as shown in Fig. \ref{fig:resum_lambda}. In that case,
\begin{equation}
    \mathcal{G}_{\lambda}(i\omega_{n},\mathbf{k})=\frac{\mathcal{G}_{\lambda}^{0}}{1-\Pi\mathcal{G}_{\lambda}^{0}}\underbrace{\longrightarrow}_{v\to\infty}-\frac{1}{\Pi(i\omega_{n},\mathbf{k})},
\end{equation}
where $\Pi$ is the one loop $\phi$ bubble,
\begin{align}
    \Pi(i\nu_{n},\mathbf{k})&=\frac{T}{2}\sum_{i\omega_{n}}\int_{\mathbf{q}}^{\Lambda}\frac{1}{(\omega_{n}+\nu_{n})^{2}+\epsilon_{\mathbf{q}+\mathbf{k}}^{2}}\frac{1}{\omega_{n}^{2}+\epsilon_{\mathbf{q}}^{2}}.
\end{align}
The retarded polarization bubble is given by analytic continuation, $\Pi_{R}(\omega,\mathbf{k})=\Pi(i\omega_{n}\to\omega,\mathbf{k})$. The resummed retarded $\lambda$ propagator is then denoted as,
\begin{equation}
    \parbox{20mm}{\begin{fmfgraph}(40,15)
    \fmfleft{i} \fmfright{o} \fmf{dashes,width=2pt}{i,o}
    \end{fmfgraph}} :=\quad \mathcal{G}_{R,\lambda}(\omega,\mathbf{k}) = \frac{1}{\Pi_{R}(\omega,\mathbf{k})}.
\end{equation}
From the $\lambda$ spectral function, $A_{\lambda}(\omega,\mathbf{k})=-2Im[\mathcal{G}_{R}(\omega,\mathbf{k})]$, we can define the generalized $\lambda$ Wightman function,
\begin{equation}
    \mathcal{G}_{W,\lambda}^{(\alpha)}(\omega,\mathbf{k})=\frac{A_{\lambda}(\omega,\mathbf{k})}{2\sinh{\frac{\beta \omega}{2}}}e^{(\alpha-1/2)\beta \omega}.
\end{equation}
\begin{figure}
    \centering
\begin{equation*}   
  \begin{gathered} \begin{fmfgraph*}(60,30)
        \fmfleft{l1}
        \fmfright{r1}
        \fmf{dashes,width=2pt}{l1,r1}
    \end{fmfgraph*}
  \end{gathered} \quad = \quad
    \begin{gathered} \begin{fmfgraph*}(60,30)
        \fmfleft{l1}
        \fmfright{r1}
        \fmf{dashes,width=0.5pt}{l1,r1}
    \end{fmfgraph*}
  \end{gathered} \quad + \quad
  \begin{gathered}  \begin{fmfgraph*}(90,30)
        \fmfleft{l1}
        \fmfright{r1}
        \fmf{dashes,width=0.5pt}{l1,v1}
        \fmf{dashes,width=2pt}{v2,r1}
        \fmf{plain,width=0.5pt,tension=0.5,left}{v1,v2,v1} 
    \end{fmfgraph*}
    \end{gathered}
\end{equation*}
    \caption{The resummed $\lambda$ propagator}
    \label{fig:resum_lambda}
\end{figure}
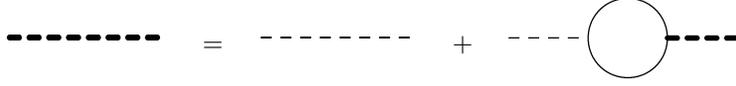
\begin{figure}
    \centering
\begin{equation*}   
\Sigma(i\omega_{n},\mathbf{q})=
    \begin{fmfgraph*}(120,80)
        \fmfleft{i1,i2,i3,i4,i5,i6}
        \fmfright{o1,o2,o3,o4,o5,o6}
        \fmf{plain,width=0.5pt}{i1,v1,v2,o1}
        \fmf{dashes,width=2pt,tension=0,left}{v1,v2}
    \end{fmfgraph*} +
    \begin{fmfgraph*}(120,100)
        \fmfleft{i1,i2,i3,i4,i5,i6}
        \fmfright{o1,o2,o3,o4,o5,o6}
        \fmf{plain,width=0.5pt}{i1,v1,o1}
        \fmffreeze
        \fmf{dashes,width=2pt}{v1,v2}
        \fmf{phantom}{i2,v2}
        \fmf{phantom}{o2,v2}
        \fmf{phantom}{i5,v3,o5}
        \fmf{phantom}{i5,v4}
        \fmf{phantom}{v4,i2}
        \fmf{phantom}{v4,o4}
        \fmf{phantom}{o5,v5}
        \fmf{phantom}{v5,o2}
        \fmf{phantom}{v5,i4}
        \fmf{plain,width=0.5pt,left=0.8}{v2,v4}
        \fmf{plain,width=0.5pt,left=0.25}{v4,v5}
        \fmf{dashes,width=2pt,left=1.6}{v4,v5}
        \fmf{plain,width=0.5pt,left=0.8}{v5,v2}
    \end{fmfgraph*}
\end{equation*}
    \caption{The $\phi$ self energy}
    \label{fig:self_energy}
\end{figure}
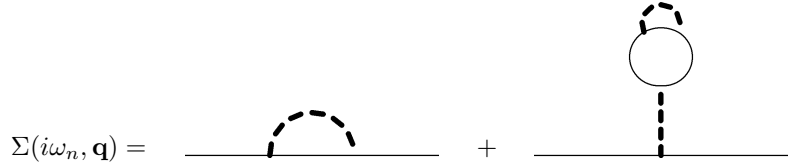
We need to dress the bare $\phi$ propagator, for which we need to calculate the self energy as given in Fig. \ref{fig:self_energy}, from which one can obtain the retarded self energy by analytic continuation. The resummed retarded propagator is denoted by a thick line,
\begin{equation}
\begin{split}
    \parbox{20mm}{\begin{fmfgraph}(40,15)
    \fmfleft{i} \fmfright{o} \fmf{plain,width=2pt}{i,o}
    \end{fmfgraph}} &:= \quad \mathcal{G}_{R}(\omega,\mathbf{k}) =  \frac{1}{(\omega+i0^{+})^{2}-\epsilon_{\mathbf{k}}^{2}+\Sigma_{R}(\omega,\mathbf{k})},
\end{split}
\end{equation}
where $\Sigma_{R}$ is the retarded self energy. In App. \ref{appsec:polarization} and App. \ref{appsec:self_energy} we calculate the polarization bubble (Fig. \ref{fig:resum_lambda}) and the self energy (Fig. \ref{fig:self_energy}) respectively, in the low temperature regime, $\beta m >> 1$.

From the self energy, we can obtain the relaxation rate of $\phi$ quasiparticles at momentum $\mathbf{q}$, which is defined as, 
\begin{equation}
        \Gamma_{\mathbf{q}}=\frac{Im[\Sigma_{R}(\epsilon_{\mathbf{q}},\mathbf{q})]}{2\epsilon_{q}}.
\end{equation}
In App. \ref{appsec:self_energy}, we demonstrate that at $\mathbf{q}=0$, the inverse lifetime $\tau_{\phi}^{-1}=\Gamma_{\mathbf{q}=\mathbf{0}}$ \cite{Chubukov1994}, can be evaluated at low temperature,
\begin{align} \label{eq:gamma_0}
    \Gamma_{0} = \frac{1}{\tau_{\phi}}\approx \frac{2\pi}{N\beta}e^{-\beta m}.
\end{align}

For general $\mathbf{q}$, we have, \begin{equation}\label{eq:gamma_q}
    \Gamma_{\mathbf{q}} \approx \frac{1}{2N}e^{\beta \epsilon_{\mathbf{q}}/2}\int_{\mathbf{k}}e^{-\beta \epsilon_{\mathbf{k}}/2}\mathcal{R}^{(1/2)}_{1+}(\mathbf{k},\mathbf{q}),
\end{equation}
where, $\mathcal{R}^{(1/2)}_{1+}(\mathbf{k},\mathbf{q})$ is given in Eq. \ref{eq:r1_p} in App. \ref{appsec:self_energy}.

\subsection{Ladder sum calculation}\label{sec:ladder_sum}
\begin{figure}
    \centering
\begin{align*}
    \begin{gathered}
        \begin{fmfgraph*}(120,80)
        \fmfstraight
            \fmfleft{i1,i2,i3,i4,i5}
            \fmfright{o1,o2,o3,o4,o5}
            \fmfpoly{shaded,tension=0}{v2,v4,v3,v1}
            \fmf{plain}{i2,v1,v2,o2}
            \fmf{plain}{i4,v3,v4,o4}
        \end{fmfgraph*}
    \end{gathered} &=
        \begin{gathered}
        \begin{fmfgraph*}(60,80)
        \fmfstraight
            \fmfleft{i1,i2,i3,i4,i5}
            \fmfright{o1,o2,o3,o4,o5}
            \fmf{plain}{i2,v1,v2,o2}
            \fmf{plain}{i4,v3,v4,o4}
        \end{fmfgraph*}
    \end{gathered} +
    \begin{gathered}
        \begin{fmfgraph*}(120,80)
        \fmfstraight
            \fmfleft{i1,i2,i3,i4,i5}
            \fmfright{o1,o2,o3,o4,o5}
            \fmfpoly{shaded,tension=0}{v2,v4,v3,v1}
            \fmf{plain}{i2,v1,v2,o2}
            \fmf{plain}{i4,v3,v4,o4}
            \fmffreeze
            \fmf{phantom,tension=20}{i2,d1,v1}
            \fmf{phantom,tension=20}{i4,d2,v3}
            \fmf{dashes}{d1,d2}
        \end{fmfgraph*}
    \end{gathered} + 
        \begin{gathered}
        \begin{fmfgraph*}(120,80)
        \fmfstraight
            \fmfleft{i1,i2,i3,i4,i5}
            \fmfright{o1,o2,o3,o4,o5}
            \fmfpoly{shaded,tension=0}{v2,v4,v3,v1}
            \fmf{plain}{i2,v1,v2,o2}
            \fmf{plain}{i4,v3,v4,o4}
            \fmffreeze
            \fmf{phantom,tension=20}{i2,d1,v1}
            \fmf{phantom,tension=20}{i4,d2,v3}
            \fmf{photon}{d1,d2}
        \end{fmfgraph*}
    \end{gathered}\\
    \begin{gathered}
        \begin{fmfgraph*}(120,80)
        \fmfstraight
            \fmfleft{i1,i2,i3,i4,i5}
            \fmfright{o1,o2,o3,o4,o5}
            \fmf{phantom}{i2,v1,o2}
            \fmf{phantom}{i4,v2,o4}
            \fmffreeze
            \fmf{photon}{v1,v2}
        \end{fmfgraph*}
    \end{gathered} &= 
        \begin{gathered}
        \begin{fmfgraph*}(120,80)
        \fmfstraight
            \fmfleft{i1,i2,i3,i4,i5}
            \fmfright{o1,o2,o3,o4,o5}
            \fmf{phantom}{i2,v1,o2}
            \fmf{phantom}{i4,v2,o4}
            \fmffreeze
            \fmf{phantom,tension=20}{i2,d1,v1}
            \fmf{phantom,tension=20}{i4,d2,v2}
            \fmf{phantom,tension=20}{v1,d3,o2}
            \fmf{phantom,tension=20}{v2,d4,o4}
            \fmf{dashes}{d2,d4}
            \fmf{dashes}{d1,d3}
            \fmf{plain}{d1,d2}
            \fmf{plain}{d3,d4}
        \end{fmfgraph*}
    \end{gathered}
\end{align*}
    \caption{Bethe Saltpeter equation for the out of time ordered correlation function. In the diagram, all horizontal lines are retarded propagators, while the vertical lines are the Wightman propagators.}
    \label{fig:bs_eq}
\end{figure}
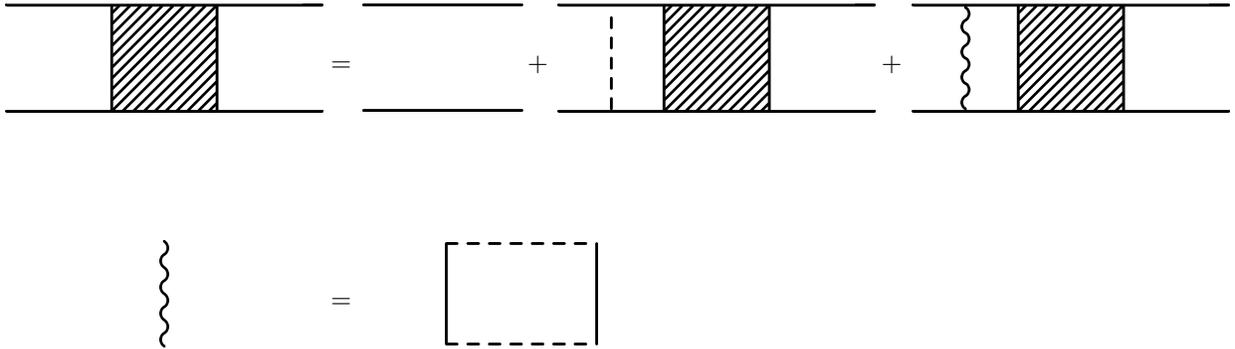
We finally calculate the regulated squared commutator, given in Eq. \ref{eq:o_n_reg_sq_comm} perturbatively in $1/N$, using the ladder-sum rules described in \cite{Chowdhury2017}, which we will extensively use. The calculation boils down to solving a Bethe Saltpeter equation in momentum space for the out of time ordered 4 point function, as shown in Fig. \ref{fig:bs_eq}. There are two sides of the ladder, which are connected by `rungs' - which are the Wightman functions. The first diagram on the RHS of Fig. \ref{fig:bs_eq} is the `free' term $\frac{1}{N}[\mathcal{G}_{R}(t,\mathbf{x})]^{2}$, which doesn't have any exponential in time behavior, hence is not important for diagnosing chaos. There are two types of rungs - the Type I and Type II rungs correspond to the second and third diagram on the RHS of the top line in Fig. \ref{fig:bs_eq} respectively. The expressions for the two rung contributions can be easily written down from the diagram; for example, the Type I rung can be expressed as,
\begin{equation}
    \begin{split}
        C_{\alpha,\text{Type I}}(\nu,\mathbf{k})=\frac{1}{N}\int \frac{d\omega}{2\pi}\int_{\mathbf{p}}\int \frac{d\omega^{\prime}}{2\pi}\int_{\mathbf{p^{\prime}}}\mathcal{G}_{R}(\nu-\omega,\mathbf{k}-\mathbf{p})\mathcal{G}_{R}(\omega,\mathbf{p})\\
        \mathcal{G}^{(\alpha)}_{W,\lambda}(\omega^{\prime}-\omega,\mathbf{p^{\prime}}-\mathbf{p})\mathcal{G}_{R}(\nu-\omega^{\prime},\mathbf{k}-\mathbf{p^{\prime}})\mathcal{G}_{R}(\omega,\mathbf{p^{\prime}}).
    \end{split}
\end{equation}
The result for the Type II rung is very similar, with the replacement $\mathcal{G}^{(\alpha)}_{W,\lambda}(\omega^{\prime}-\omega,\mathbf{p^{\prime}}-\mathbf{p})\to\mathcal{G}^{(\alpha)}_{\text{eff}}(\omega^{\prime},\omega,\mathbf{p^{\prime}},\mathbf{p})$, where,
\begin{equation}\label{eq:g_eff}
    \begin{split}
        \mathcal{G}^{(\alpha)}_{\text{eff}}(\omega^{\prime},\omega,\mathbf{p^{\prime}},\mathbf{p})=\int \frac{d\omega^{\prime\prime}}{2\pi}\int_{\mathbf{p^{\prime\prime}}} \mathcal{G}^{(\alpha)}_{W}(\omega^{\prime\prime}-\omega,\mathbf{p^{\prime\prime}}-\mathbf{p})\mathcal{G}^{(\alpha)}_{W}(\omega^{\prime}-\omega^{\prime\prime},\mathbf{p^{\prime}}-\mathbf{p^{\prime\prime}})\\
        \mathcal{G}_{R,\lambda}(\nu-\omega^{\prime\prime},-\mathbf{p^{\prime\prime}})\mathcal{G}_{R,\lambda}(\omega^{\prime\prime},\mathbf{p^{\prime\prime}}).
    \end{split}
\end{equation}
We set up the Bethe Saltpeter equation by defining a function $f(\nu,\mathbf{k};\omega,\mathbf{p})$, such that,
\begin{equation}
    C_{(\alpha)}(\nu,\mathbf{k})=\frac{1}{N}\int \frac{d\omega}{2\pi}\int_{\mathbf{p}}f^{(\alpha)}(\nu,\mathbf{k};\omega,\mathbf{p}).
\end{equation}
As was shown in \cite{Chowdhury2017}, it is convenient to consider the ``on-shell" ansatz for $f(\nu,\mathbf{k};\omega,\mathbf{p})$,
\begin{equation}
    f^{(\alpha)}(\nu,\mathbf{k};\omega,\mathbf{p})=\frac{f^{(\alpha)}_{+}(\nu,\mathbf{k};\mathbf{p})}{2\epsilon_{\mathbf{p}}}\delta(\omega-\epsilon_{\mathbf{p}})+\frac{f^{(\alpha)}_{-}(\nu,\mathbf{k};\mathbf{p})}{2\epsilon_{\mathbf{p}}}\delta(\omega+\epsilon_{\mathbf{p}}).
\end{equation}
We can approximate the product of the retarded Green functions by their most singular (in $\nu$) terms (for small $k$, such that $\Gamma_{\mathbf{k}-\mathbf{p}}\approx\Gamma_{\mathbf{p}}$),
\begin{equation}
    {G}_{R}(\nu-\omega,\mathbf{k}-\mathbf{p})\mathcal{G}_{R}(\omega,\mathbf{p}) \to \frac{\pi i}{2\epsilon_{\mathbf{p}}\epsilon_{\mathbf{k}-\mathbf{p}}}\left[\frac{\delta(\omega-\epsilon_{\mathbf{p}})}{\nu-(\epsilon_{\mathbf{p}}-\epsilon_{\mathbf{k}-\mathbf{p}})+2i\Gamma_{\mathbf{p}}}+\frac{\delta(\omega+\epsilon_{\mathbf{p}})}{\nu+(\epsilon_{\mathbf{p}}-\epsilon_{\mathbf{k}-\mathbf{p}})+2i\Gamma_{\mathbf{p}}}\right].
\end{equation}
Further, we have, $\epsilon_{\mathbf{p}}-\epsilon_{\mathbf{k}-\mathbf{p}}\approx\mathbf{k}.\nabla_{\mathbf{p}}\epsilon_{\mathbf{p}}$, and for small $\mathbf{p}$, $\nabla_{\mathbf{p}}\epsilon_{\mathbf{p}}\approx \mathbf{p}/m$.
The Bethe Saltpeter equation can now be written as \cite{Chowdhury2017}, 
\begin{equation}\label{eq:bs_eq}
    (-i\nu \pm i\frac{\mathbf{k}.\mathbf{p}}{m}) f^{(\alpha)}_{\pm}(\nu,\mathbf{k};\mathbf{p}) = \frac{1}{N}\int_{\mathbf{p^{\prime}}}\hat{\mathcal{K}^{(\alpha)}}(\mathbf{p^{\prime}},\mathbf{p})f^{(\alpha)}_{\pm}(\nu,\mathbf{k};\mathbf{p^{\prime}}),
\end{equation}
where,
\begin{equation}
    \begin{split}
        &\hat{\mathcal{K}^{(\alpha)}}(\mathbf{p^{\prime}},\mathbf{p})=\mathcal{R}^{(\alpha)}_{1}(\mathbf{p^{\prime}},\mathbf{p})+\mathcal{R}^{(\alpha)}_{2}(\mathbf{-p^{\prime}},\mathbf{p})-2N\Gamma_{\mathbf{p}}(2\pi)^{2}\delta^{(2)}(\mathbf{p^{\prime
        }}-\mathbf{p}), \text{ and,}\\
        &\mathcal{R}^{(\alpha)}_{1,2}(\mathbf{p^{\prime}},\mathbf{p}):=\mathcal{R}^{(\alpha)}_{1,2+}(\mathbf{p^{\prime}},\mathbf{p})+\mathcal{R}^{(\alpha)}_{1,2-}(\mathbf{p^{\prime}},\mathbf{p}), \text{ where,}\\&\mathcal{R}^{(\alpha)}_{1\pm}(\mathbf{p^{\prime}},\mathbf{p}):=\frac{1}{4\epsilon_{\mathbf{p^{\prime}}}\epsilon_{\mathbf{p}}}\mathcal{G}^{(\alpha)}_{W,\lambda}(\pm\epsilon_{\mathbf{p^{\prime}}}-\epsilon_{\mathbf{p}},\mathbf{p^{\prime}}-\mathbf{p})\text{ and }\mathcal{R}^{(\alpha)}_{2\pm}(\mathbf{p^{\prime}},\mathbf{p}):=\frac{1}{4\epsilon_{\mathbf{p^{\prime}}}\epsilon_{\mathbf{p}}}\mathcal{G}^{(\alpha)}_{\text{eff}}(\pm\epsilon_{\mathbf{p^{\prime}}},\epsilon_{\mathbf{p}},\mathbf{p^{\prime}},\mathbf{p}).
    \end{split}
\end{equation}
The inverse life-time $\Gamma_{\mathbf{p}}$ was defined in Eq. \ref{eq:gamma_q}. Recall $\alpha = 1/2$ refers to the regulated case, while, $\alpha = 1$ refers to the unregulated case. Because of the spectral relation in Eq. \ref{eq:wightman_spectral}, we have, $\mathcal{G}^{(1)}_{W}(\omega) = e^{\beta \omega/2}\mathcal{G}^{(1/2)}_{W}(\omega)$. Thus, the kernel functions are also related simply as, $\mathcal{R}^{(1)}_{1,2}(\mathbf{p^{\prime}},\mathbf{p}) = e^{\beta\left(\epsilon_{\mathbf{p^{\prime}}}-\epsilon_{\mathbf{p}}\right)/2}\mathcal{R}^{(1/2)}_{1,2}(\mathbf{p^{\prime}},\mathbf{p})$.  We calculate the kernel functions from the Type I and Type II rungs, $\mathcal{R}^{(1/2)}_{1,2\pm}$, at low temperature, in App. \ref{appsec:kernel}.

\subsubsection{Kernel functions at low temperature}
From the expressions for the kernel functions $\mathcal{R}^{(1/2)}_{1,2}(\mathbf{p^{\prime}},\mathbf{p})$, obtained in Eqs. \ref{eq:r1_p}, \ref{eq:r1_m}, \ref{eq:r2_p} and \ref{eq:r2_m} in App. \ref{appsec:kernel}, it becomes clear that the kernel functions are exponentially suppressed as $\exp{\left(-\beta(\epsilon_{\mathbf{p^{\prime}}}-\epsilon_{\mathbf{p}})/2\right)}$. Expanding in terms of the small parameter $|\mathbf{p^{\prime}}-\mathbf{p}|$ in the kernel functions, we get the following low temperature approximation, 
\begin{equation}\label{eq:r_approx1}
    \mathcal{R}^{(1/2)}_{1}(\mathbf{p^{\prime}},\mathbf{p}) = \mathcal{R}^{(1/2)}_{2}(\mathbf{p^{\prime}},\mathbf{p}) = e^{-\beta m}\frac{8\pi\sqrt{2\pi }}{\sqrt{\beta m}|\mathbf{p^{\prime}}-\mathbf{p}|m^{2}}e^{-\beta( |\mathbf{p^{\prime}}-\mathbf{p}|^{2}/8m )}.
\end{equation}

We can extract the temperature scaling of the kernel integration, by rescaling $\mathbf{p},\mathbf{p^{\prime}}\to\mathbf{p}\sqrt{m}/\sqrt{\beta},\mathbf{p^{\prime}}\sqrt{m}/\sqrt{\beta}$. Furthermore, to solve the Bethe Saltpeter equation numerically, we need to create a discrete 2D grid of momenta, with momentum spacing $\Delta p$. We can thus replace the integral in Eq. \ref{eq:bs_eq} with a discrete sum,
\begin{equation}\label{eq:bs_eq_mat}
    (-i\nu \pm \frac{i\mathbf{k}.\mathbf{p}}{\sqrt{\beta m}})f^{(\alpha)}_{\pm}(\nu,\mathbf{k};\mathbf{p}) = \frac{(\Delta p)^{2}}{4\pi^{2}N}\frac{e^{-\beta m}}{\beta} \sum_{\mathbf{p}^{\prime}} \hat{K}^{(\alpha)}_{\mathbf{p}^{\prime}\mathbf{p}}f^{(\alpha)}_{\pm}(\nu,\mathbf{k};\mathbf{p}^{\prime}),
\end{equation}
with the kernel matrix defined as,
\begin{equation}
    \begin{split}
        &\hat{K}^{(\alpha)}_{\mathbf{p}^{\prime}\mathbf{p}} = \left[\hat{R}^{(\alpha)}_{1}\right]_{\mathbf{p}^{\prime}\mathbf{p}}+\left[\hat{R}^{(\alpha)}_{2}\right]_{\mathbf{p}^{\prime}\mathbf{p}}-2\hat{\Gamma}_{\mathbf{p}}\delta_{\mathbf{p}^{\prime}\mathbf{p}} \text{ and, } \hat{\Gamma}_{\mathbf{p}} = \frac{1}{2}e^{\left(\mathbf{p}^{2}-\mathbf{p^{\prime}}^{2}\right)/4}(\Delta p)^{2}\left[\hat{R}^{(1/2)}_{1}\right]_{\mathbf{p}^{\prime}\mathbf{p}}\text{ where,}\\
        &\left[\hat{R}^{(\alpha)}_{1}\right]_{\mathbf{p}^{\prime}\mathbf{p}} = \left[\hat{R}^{(\alpha)}_{2}\right]_{\mathbf{p}^{\prime}\mathbf{p}} = \frac{8\pi\sqrt{2\pi}}{|\mathbf{p^{\prime}}-\mathbf{p}|}e^{-( |\mathbf{p^{\prime}}-\mathbf{p}|^{2}/8 )}e^{\left(\alpha-1/2\right)\left(\mathbf{p^{\prime}}^{2}-\mathbf{p}^{2}\right)/2}.
    \end{split}
\end{equation}

 We create a discrete 2D grid of rescaled non-dimensionalized momenta, with a hard cutoff of $\Lambda = 1$. This is justified as the kernel matrix is exponentially suppressed in $|\mathbf{p^{\prime}}-\mathbf{p}|^{2}$.  

We want to find the temporal behavior of $C_{r,u}(t,\mathbf{x})$. We can thus replace $-i\nu\to\partial_t$ in Eq. \ref{eq:bs_eq_mat} and solve the matrix equation for its eigenvalues. If there are real positive eigenvalues, we can infer that there is an exponential growth in the regulated squared commutator. We denote the leading eigenvalue as $\lambda^{r,u}_{L}(\mathbf{k})$.

\begin{figure}
	\includegraphics[width=0.7\columnwidth]{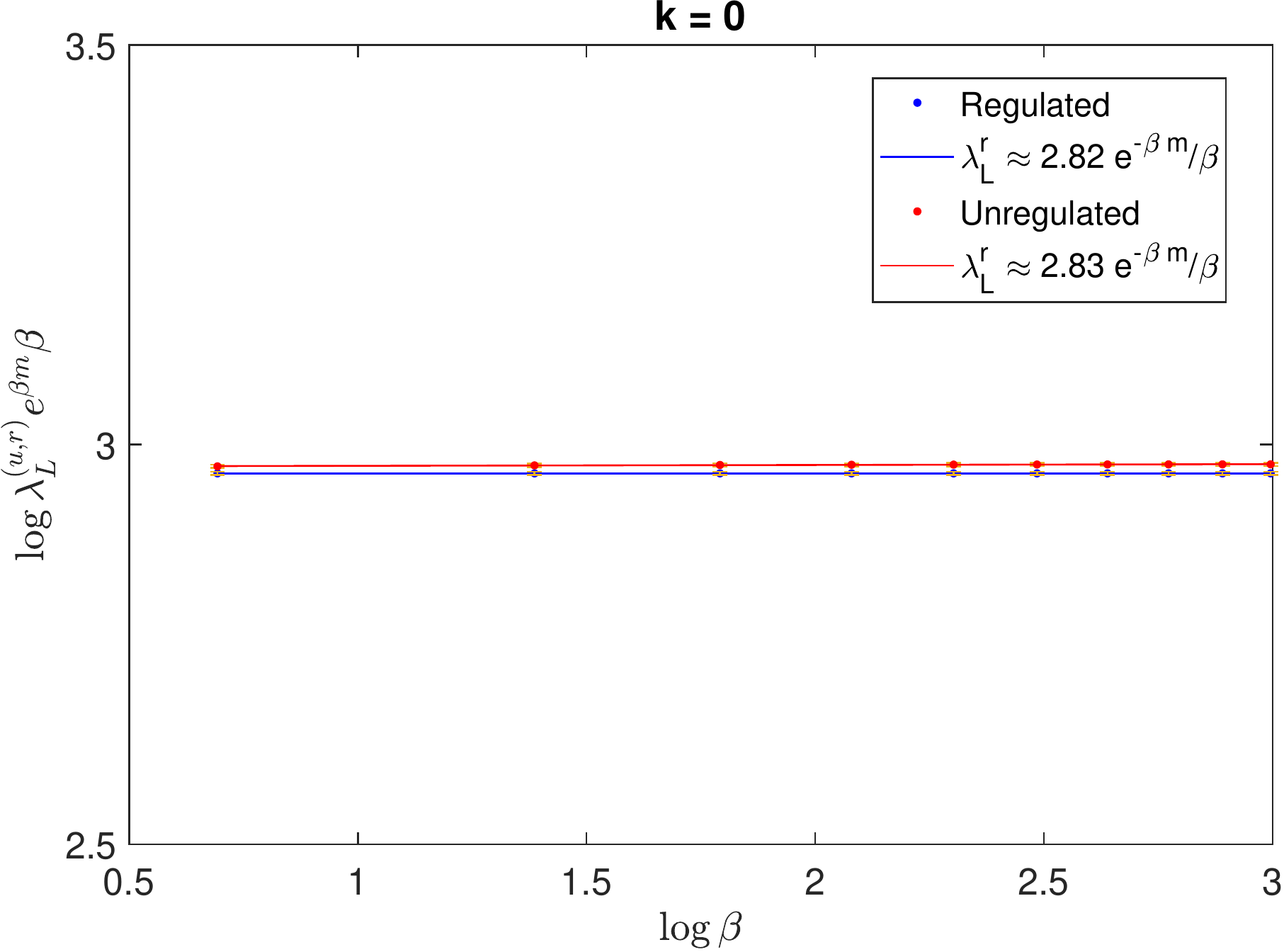}
	\caption{Scaled maximal eigenvalue of the Eq. \ref{eq:bs_eq_mat} at $\mathbf{k}=0$, $\lambda_{L}(k=0)e^{\beta m}\beta N$, is plotted as a function of inverse temperature $\beta$ in the log-log scale (we rescaled factors of N in the numerics). The errorbars are estimated from the uncertainty of extrapolating the eigenvalues to the continuous limit $dp \to 0$. The behaviour is constant with temperature, confirming $\lambda_{L} \sim e^{-\beta m}/\beta N$. Also, the result is same for both the regulated and unregulated cases showing that the ladder method is contour-independent.}\label{Fig:k0_ladder}
\end{figure}

\subsubsection{Temperature scaling of the butterfly velocity}
First, let us restrict to $k=0$. From Eq. \ref{eq:bs_eq_mat}, we have, $\lambda^{r,u}_{L}(k=0)\sim e^{-\beta m}/\beta N$.  By numerically finding the largest eigenvalue of the matrix equation we assert that the leading eigenvalue is always real and positive, leading to an exponential growth in the squared commutator. The details of the numerical computation are given in Appendix \ref{appsec:numerics_ladder}, and the results for both the regulated and the unregulated cases are demonstrated in Fig. \ref{Fig:k0_ladder}. Furthermore, the relevant inverse time-scale is also given by $\Gamma_{0} = e^{-\beta m}/\beta N$, (Eq. \ref{eq:gamma_0}). Hence, we can rescale the Bethe Saltpeter equation by this scale, and introduce a rescaled external momentum, $\mathbf{u} = \mathbf{k}/\left(\sqrt{\beta m}\Gamma_{0}\right)$, and a rescaled time $\Tilde{t} = \Gamma_{0} t$. 



The matrix equation can be now recast as,
\begin{equation}
    \left(\partial_{\Tilde{t}}\pm i \mathbf{u}.\mathbf{p}\right)f^{(\alpha)}_{\pm}(\nu,\mathbf{k};\mathbf{p}) \sim \sum_{\mathbf{p^{\prime}}} \hat{K}_{\mathbf{p}^{\prime}\mathbf{p}}f^{(\alpha)}_{\pm}(\nu,\mathbf{k};\mathbf{p}^{\prime}).
\end{equation}

\begin{figure}
	\includegraphics[width=0.70\columnwidth]{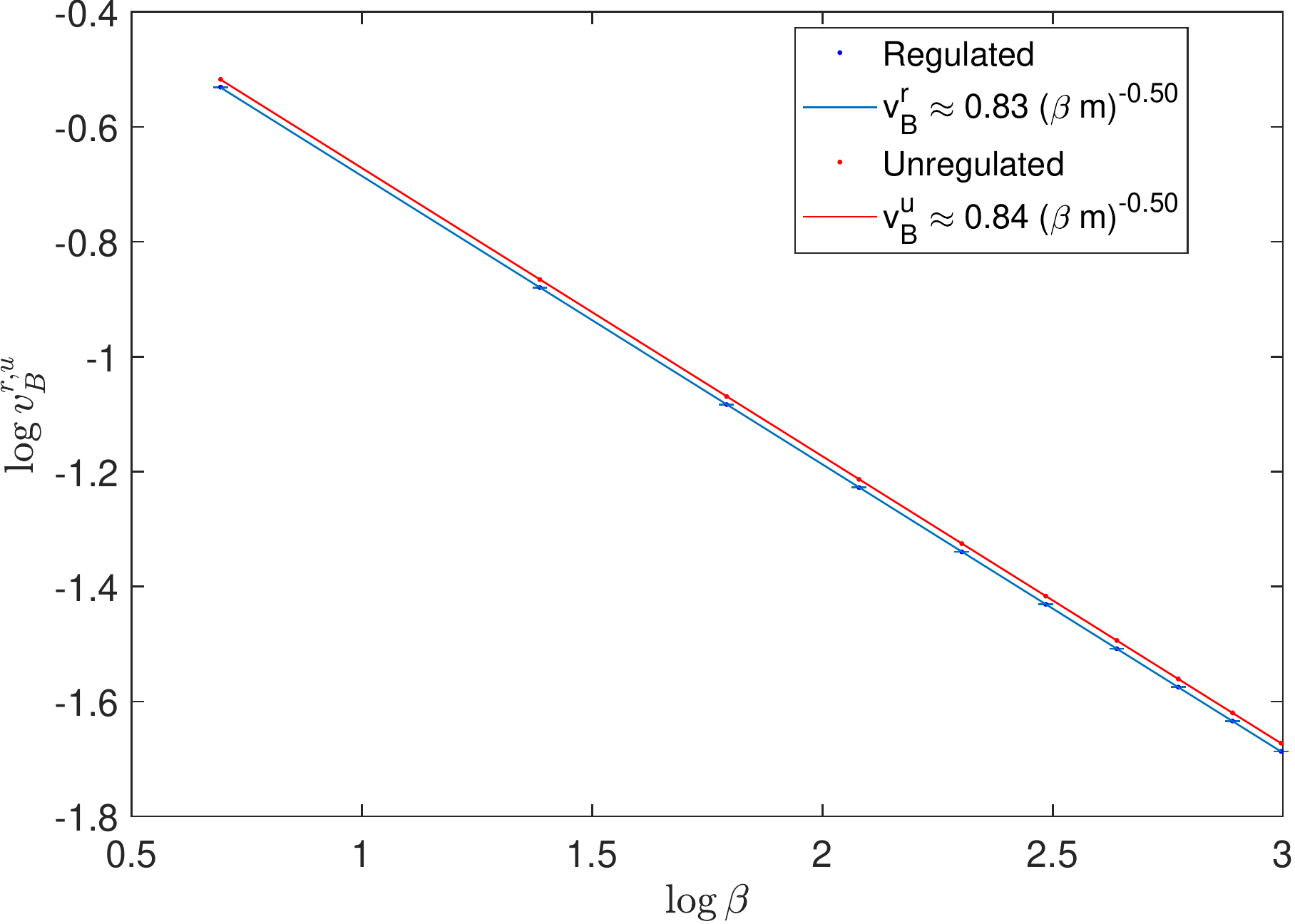}
	\caption{Using the fitted $\lambda_{0}$, $\lambda_{2}$ and $\lambda_{i}$, the butterfly velocity $v_{B}$ is calculated from Eq. \ref{eq:vb_ladder}, and plotted against $\beta$ in a log-log scale. The low temperature behavior of $v_{B}$ is $v_{B}\approx \frac{0.83}{\sqrt{\beta m}}$ - for both the regulated and the unregulated cases.}\label{Fig:knon0_ladder}
\end{figure}

For small $u$, the eigenvalues of this matrix equation can be approximated by 
\begin{equation}
    \Tilde{\lambda}_{L}(u) \approx \Tilde{\lambda}_{0}-\Tilde{\lambda}_{2}u^{2}\pm i\Tilde{\lambda}_{i}u ,
\end{equation}
because of the spherical symmetry of the leading eigenvector at $\mathbf{k}=0$. Here, $\Tilde{\lambda}_{0,2,i}\sim \mathcal{O}(1)$, and by rescaling back, $\lambda_{0,2,i}\sim e^{-\beta m}/\beta N$. The quadratic form of the real part and the linear form for the imaginary part have been verified numerically in Fig. \ref{Fig:knon0_quadfit} in App. \ref{appsec:numerics_ladder}. Now, the regulated and unregulated squared commutator can be evaluated as,
\begin{equation}\label{eq:kernel_int}
    \begin{split}
        C_{r,u}(t,\mathbf{x}) &= \frac{1}{N}\int_{\nu}\int_{\mathbf{k}}\int_{\mathbf{p}}e^{i\mathbf{k}.\mathbf{x}-i\nu t}\left(\frac{f^{r,u}_{+}(\nu,\mathbf{k};\mathbf{p})}{2\epsilon_{\mathbf{p}}}+\frac{f^{r,u}_{-}(\nu,\mathbf{k};\mathbf{p})}{2\epsilon_{\mathbf{p}}}\right)\\
        &=\frac{1}{N}\int_{\mathbf{k}}e^{i\mathbf{k}.\mathbf{x}+\lambda_{0}t-\lambda_{2}u^{2}t}\left(e^{i\lambda_{i}ut}\chi^{r,u,+}_{\mathbf{k}}+e^{-i\lambda_{i}ut}\chi^{r,u,-}_{\mathbf{k}}\right),
    \end{split}
\end{equation}
where, $\chi^{r,u,\pm}_{\mathbf{k}}$ is the eigenvector of the matrix eigenvalue in Eq. \ref{eq:bs_eq_mat}. If there are no singularities in $\chi^{r,u,\pm}_{\mathbf{k}}$, we can assume the two terms in the integral depends only on the saddle points of the exponents. Recalling $u=k/\left(\sqrt{\beta m}\Gamma_{0}\right)$, the two saddle points are given by,
\begin{equation}
    \mathbf{k}^{*}_{\pm} = (\beta m \Gamma_{o}^{2})\frac{i\left(\mathbf{x}\pm\frac{\lambda_{i}t}{\sqrt{\beta m}\Gamma_{0}}\right)}{2\lambda_{2}t}.
\end{equation}
When $C_{r,u}(t,\mathbf{x})$ is evaluated, one of the terms will be exponentially suppressed in $x$ compared to the other. Keeping only the leading term, we get,
\begin{equation}\label{eq:largeN_scrambling}
    C_{r,u}(t,\mathbf{x})\sim\frac{1}{N}\exp\left[\lambda_{0}t-\frac{\beta m \Gamma_{0}^{2}\left(x-\frac{\lambda_{i}t}{\sqrt{\beta m}\Gamma_{0}}\right)^{2}}{4\lambda_{2}t}\right].
\end{equation}
The first term comes from the pure exponential growth that was present for the $u=0$ case, and the second term is reminiscent of the broadening form of the squared commutator in Eq. \ref{eq:scrambling_ansatz}. By finding the level sets of the exponential for the ballistic condition $x\sim v_{B} t$, we have the following expression for the butterfly velocity $v_{B}$,
\begin{equation}\label{eq:vb_ladder}
    \begin{split}
        v_{B}^{r,u} = \sqrt{\frac{4\lambda_{0}\lambda_{2}}{\beta m \Gamma_{0}^{2}}}+\frac{\lambda_{i}}{\sqrt{\beta m \Gamma_{0}^{2}}}.
    \end{split}
\end{equation}
Since $\lambda_{0,2,i}\sim\Gamma_{0}$, we get the following temperature dependence of the butterfly velocity,
\begin{equation}
    v_{B}^{r,u}\sim\sqrt{\frac{1}{\beta m}}.
\end{equation}
Note that this is the same scale as the speed of sound of the ideal classical gas at finite temperature. Hence the butterfly velocity from the regulated squared commutator of this essentially classical gas has the same temperature scaling as the speed of sound. Furthermore, the particular temperature scaling $\sqrt{1/\beta m}$ of the butterfly velocity arises because the thermal scale is the appropriate scale to non-dimensionalize the momenta, and doesn't depend on the exact form of $\Tilde{\lambda}_{L}(u)$.

From the numerically obtained eigenvalues, we can see from Fig. \ref{Fig:knon0_ladder}, that the butterfly velocity from regulated and unregulated squared commutators are the same at low temperatures,
\begin{equation}
    v_{B} \approx \frac{0.83}{\sqrt{\beta m}}.
\end{equation}
This shows that the ladder calculation is insensitive to contour dependence.

At fixed $t$, for a fixed difference of $C_{r,u}(t,\mathbf{x})$, one finds from Eq. \ref{eq:largeN_scrambling} that the spread $\epsilon = x-v_{B}t \sim $ constant. This implies that this form of the squared commutator doesn't have a broadening behavior. A similar exercise for the spin chain result in Eq. \ref{eq:scrambling_ansatz}, would show a time dependent spread, $\epsilon \sim t^{p/(p+1)}$, implying broadening.

In deriving these results, we assumed that the integral expression of the squared commutator in Eq. \ref{eq:kernel_int} is dominated by the saddle point contribution. In \cite{Gu_2019}, it was noted that OTOCs obtained from ladder sum calculations generically have a pole in momentum space wherever $\lambda_{L}(k)=2\pi/\beta$,
\begin{equation}
    C(t,\mathbf{x}) \sim \frac{1}{N}\int_{\mathbf{k}}\frac{e^{i\mathbf{k}.\mathbf{x}+\lambda_{L}(k)t}}{\cos \frac{\lambda_{L}(k)\beta}{4}}.
\end{equation}
However, in the $O(N)$ theory, the chaos exponent $\lambda_{L}(k)\sim 1/N$ is $N$ suppressed, hence these poles occur at parametrically large values of the momentum. Provided that $x/t$ is $N$-independent, the saddle point momentum is always closer to the real axis than the pole and hence controls the integral. For example, as we have seen from the $k$ dependence of $\lambda_{L}(k)$ in Fig. \ref{Fig:extrapolate} in Appendix. \ref{appsec:numerics_ladder}, if $\lambda_L(k = i |k|) \sim \lambda_0 \beta |k|^2/m$ at large imaginary $k$, then the closest pole in the upper half plane would be at $|k| \sim \sqrt{\frac{m}{\beta}\frac{N \lambda_{\text{max}}}{\lambda_0}} $. This momentum is very large due to large $N$ and the large ratio $\lambda_{\text{max}}/\lambda_0$.

\subsection{Summary of findings from the field theory calculation}

In this section, we studied the temperature and contour dependence of squared commutator in a solvable large $N$ local model using the ladder technique. We find that our analysis can describe the temperature scaling of the butterfly velocity. However, it is insensitive to the contour of thermal ordering. This is not unexpected, as the ladder method is not expected to exhibit contour dependence \cite{Kobrin2020}. It also doesn't capture the broadening behavior that was observed in Sec. \ref{sec:mpo_spins}.

The field theory model differs from the spin chain numerics in two ways - the number of spatial dimensions, and in the fact that the spin chain has finite local Hilbert space unlike the field theory model, which is solvable at large $N$ - an effectively classical description. It is thus likely that the broadening and the contour dependence are sourced by quantum fluctuations due to the finiteness of the local Hilbert space \cite{Xu2019}, which is not captured in this calculation. 

\end{fmffile}

\section{Discussions}\label{sec:discussion}

In this paper we have studied the temperature and contour dependence of quantum information scrambling for local gapped interacting systems in two different models and for a wide range of temperatures. 

We first introduced a tensor network based technique to calculate both regulated and unregulated squared commutators in quantum spin chains at temperatures ranging across the spectral gap. For the regulated case, the butterfly velocity decreases with lowering temperature, and is consistent with a power law $v_B \sim \beta^{-a}$ for $a>0$ at intermediate-to-large $\beta$. We also observe a strong contour dependence, and point out that the butterfly velocity obtained from the unregulated squared commutator remains insensitive to the temperature variation. In fact, a careful study of $\partial_{t}C(t,\mathbf{x})$ shows that the chaos bound cannot be generalized away from the special contour ordering used to prove it.

To get an analytical handle on local gapped systems at temperatures lower than what can be accessed in the spin chain numerics, we use a perturbative field theoretic ladder sum technique, and calculate the temperature dependence of the squared commutator in the paramagnetic phase of the $O(N)$ model. There we confirmed that the characteristic speed of information scrambling at low temperature is proportional to the speed of sound of a classical gas, i.e. $v_B \sim \beta^{-1/2}$, confirming the intuition that the low temperature state can be accurately modeled as a weakly interacting dilute gas of massive quasiparticles. However, the scrambling in this model is insensitive to the contour, and also doesn't have the broadening feature.

The strong contour dependence we observe in our spin-chain numerics is in the spirit of the results from previous Schwinger-Keldysh calculations in \cite{Liao2018,Romero-Bermudez2019}, which showed similar contour dependence. Our result for the strongly interacting quantum spin chain compliments their perturbative arguments. These results taken together suggest that the unregulated case accesses high energy modes even at low temperatures, thereby remaining insensitive to the effects of temperature. Although we did not find such behavior in the $O(N)$ model at leading order in $1/N$, we expect higher order corrections will modify this conclusion since there are multiple energy scales in the problem in addition to temperature.

The numerical study also reveals the existence of a wave-front broadening effect that persists even at low temperatures. This feature is not captured in the field theory calculations, and remains an interesting theoretical challenge for the future. As was suggested in \cite{Xu2019}, quantum fluctuations due to the finiteness of the local Hilbert spaces will play a significant role in the broadening behavior.

Using Lieb Robinson \cite{Lieb1972} bounds, it has recently been demonstrated \cite{Han2019} that locality and short ranged correlations imply temperature dependent bounds on the butterfly velocity defined from the unregulated squared commutator. In App.~\ref{appsec:scrambling_bound}, we review the derivation of this bound and extend it to the regulated case. In particular, it can be shown that the butterfly velocity (obtained from either unregulated or regulated cases) obeys the bound,
\begin{equation}
    \partial_{\beta}v_{B}\to 0 \text{, as  } \beta \to \infty.    
\end{equation}
These bounds are consistent with a constant butterfly velocity at low temperatures $v_{B}\sim \text{constant}$ (unregulated case from spin chain numerics) and with a butterfly velocity proportional to a power of temperature $v_{B}\sim\beta^{-a}$ for $a>0$ (regulated case from the spin chain dynamics and field theory calculation, with $a=1/2$). The existing bounds are contour independent and hence cannot constrain the contour dependence.

The strong contour dependence that we observe has non-trivial implications for temperature dependent scrambling studies in future experiments. Our work shows that the regulated and the unregulated cases capture different physics, thus enriching the large set of phenomena falling under the umbrella of scrambling.

\section{Acknowledgement}
The authors would like to thank Shenglong Xu for useful discussions, and Shenglong Xu and Gregory Bentsen for comments on the manuscript. The authors acknowledge the University of Maryland supercomputing resources (http://hpcc.umd.edu) made available for conducting the research reported in this paper. This material is based upon work supported by the Air Force Office of Scientific Research under award number FA9550-17-1-0180.

\bibliography{References}

\begin{appendices}
\section{Details of MPO numerics}\label{appsec:numerics_mpo}
\begin{figure}
	\includegraphics[width=0.47\columnwidth]{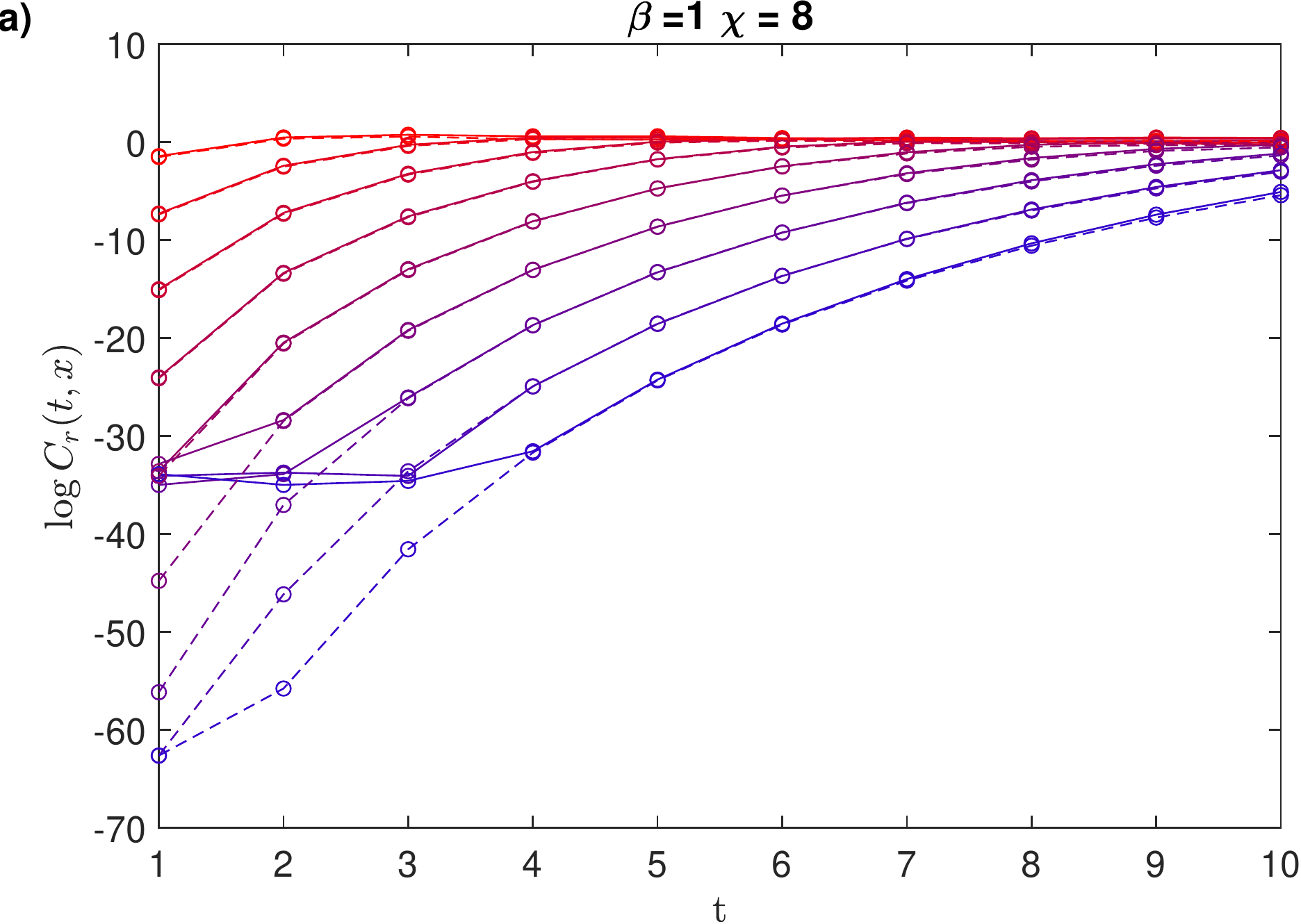}
	\includegraphics[width=0.47\columnwidth]{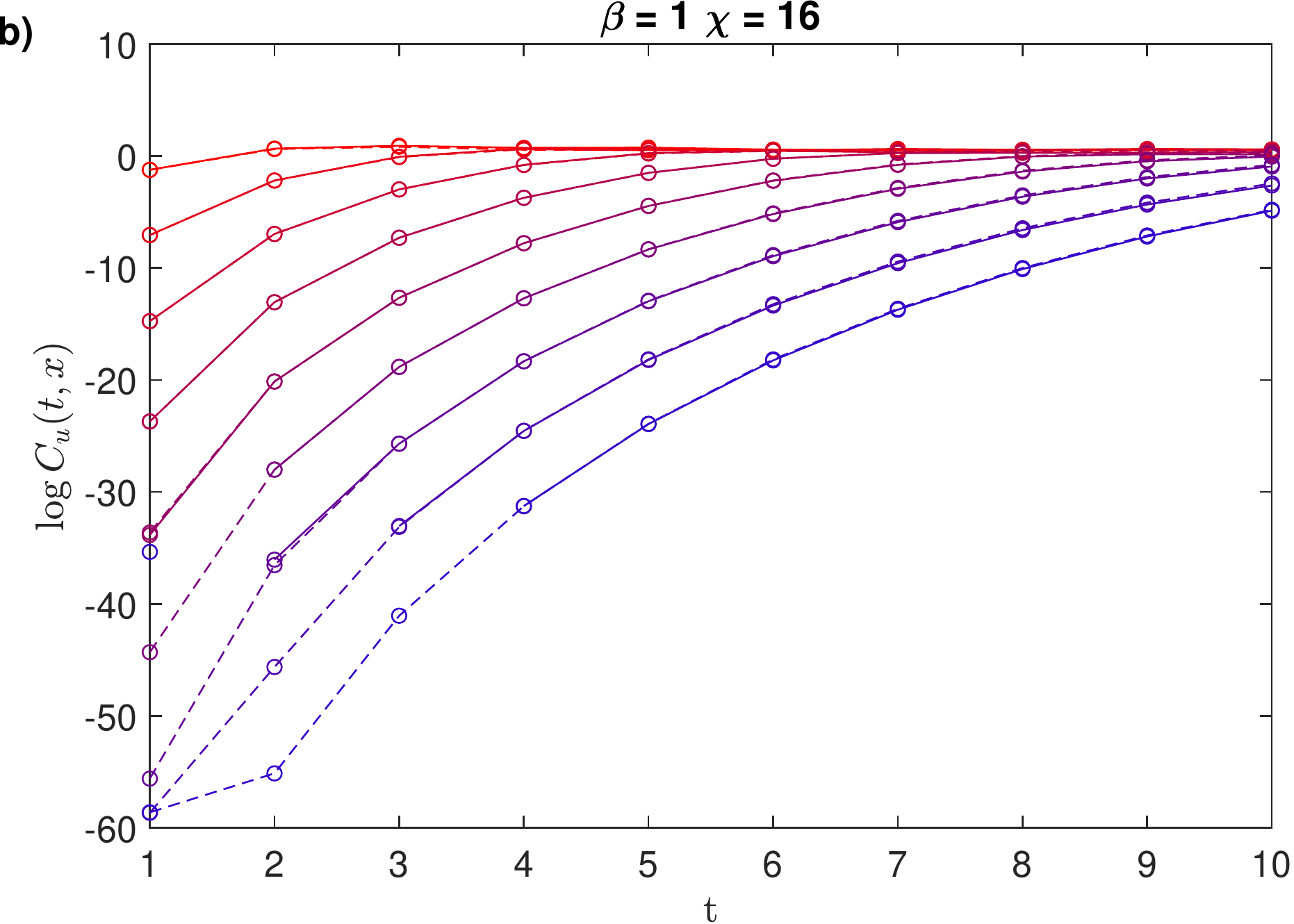}
	\caption{We demonstrate the convergence of our numerical method with exact diagonalization for small systems. a) For the regulated case, the $\chi = 8$ result has excellent agreement with exact diagonalization for $L=10$ spin chain at $\beta =1$. b) For the unregulated case, the same agreement is demonstrated for $\chi = 16$ result at $\beta = 1$. }\label{Fig:comp_ed}
\end{figure}
We first check the MPO TEBD numerical technique against exact diagonalization. In Fig. \ref{Fig:comp_ed}, we show the comparison of the MPO method to the results of exact diagonalization for a $L=10$ sized spin chain. The machine precision of MATLAB being $\sim e^{-36}$, accuracy of $\log C$ from exact diagonalization is $\sim -30$. However, in our MPO numerical method, we express the squared commutator as the square of a norm, hence the precision is squared, with reliable numerical data of $C$ down to $\sim e^{-60}$.

\begin{figure}
	\includegraphics[width=0.495\columnwidth]{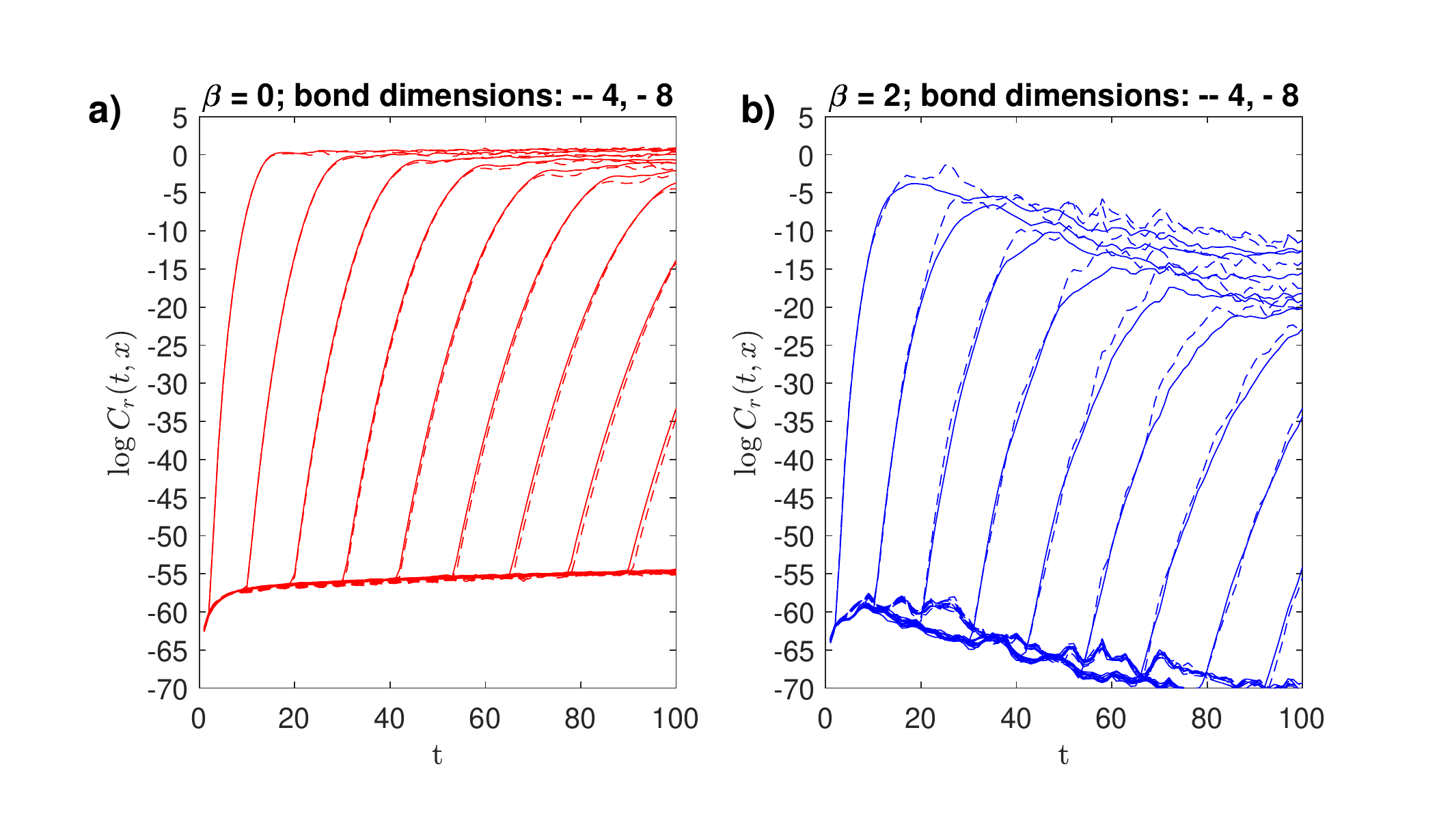}
	\includegraphics[width=0.495\columnwidth]{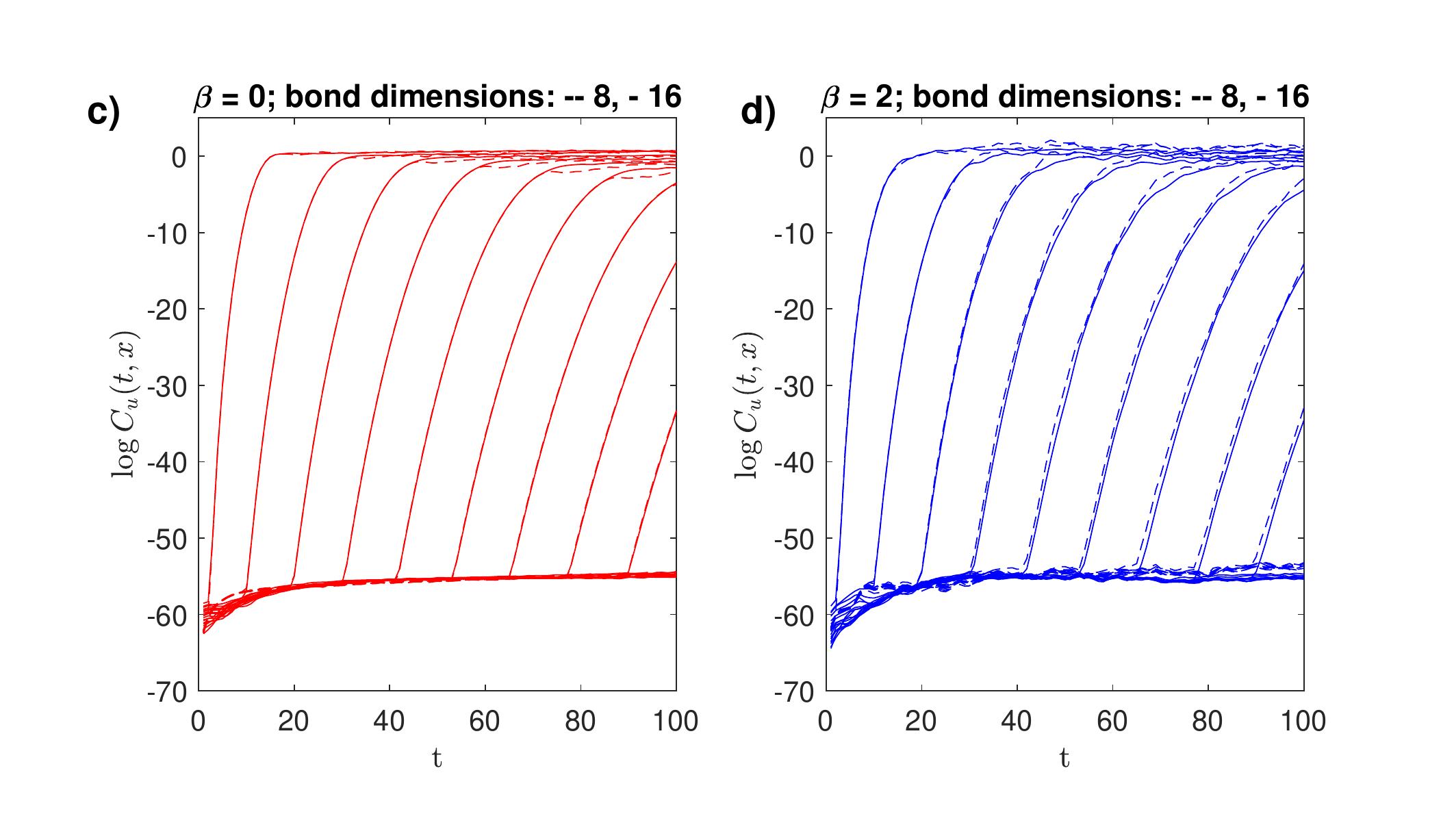}
	\caption{\textbf{a, b)} The log of the regulated squared commutator is plotted as a function of time, for the case of an operator $X_{20}(t)$ and $Z_{r}$, with $r=30,40,..,200$, for bond dimensions $\chi = 4$ (dotted) and $\chi = 8$. The left and the right figures correspond to $\beta = 0$ (a) and $\beta = 2$ (b) respectively. Even at the low temperature, the data is seen to be converged for the range $-50<\log C_{r} <-35$. Note we are able to access such small values accurately because we have expressed the regulated squared commutator as a square of a norm, and the norm can be estimated upto the numerical precision of MATLAB which is $\sim e^{-36}$, allowing us to push to around $e^{-60}$ in precision. \textbf{c, d)} The log of the unregulated squared commutator is plotted as a function of time, for the case of an operator $X_{20}(t)$ and $Z_{r}$, with $r=30,40,..,200$, for bond dimensions $\chi = 8$ (dotted) and $\chi = 16$. The left and the right figures correspond to $\beta = 0$ (c) and $\beta = 2$ (d) respectively. Even at the low temperature, the data is seen to be converged for the range $-50<\log C_{u} <-15$.}\label{Fig:reg_unreg_logc}
\end{figure}
In order to demonstrate the convergence of the obtained squared commutator with bond dimension, we plot the log of the regulated and unregulated squared commutators as a function of time for different spatial differences in Fig. \ref{Fig:reg_unreg_logc}. Even without numerical fitting, it is clear from Fig. \ref{Fig:reg_unreg_logc} that the regulated squared commutator has a strong temperature dependence, while the unregulated squared commutator is much less sensitive to temperature even when the temperature is tuned from $\beta=0$ to $\beta = 2 >m^{-1}$, where the mass is the spectral gap $\sim 1.13$.

It is also seen that the early time data converges well with bond dimension. As has been noted before in \cite{Hemery2019}, the qualitative lightcone behavior of the unconverged data obtained from the MPO method can be qualitatively different; hence for all our analysis and fitting we only use numerical data which are shown to converge.

\begin{figure}
	\includegraphics[width=0.8\columnwidth]{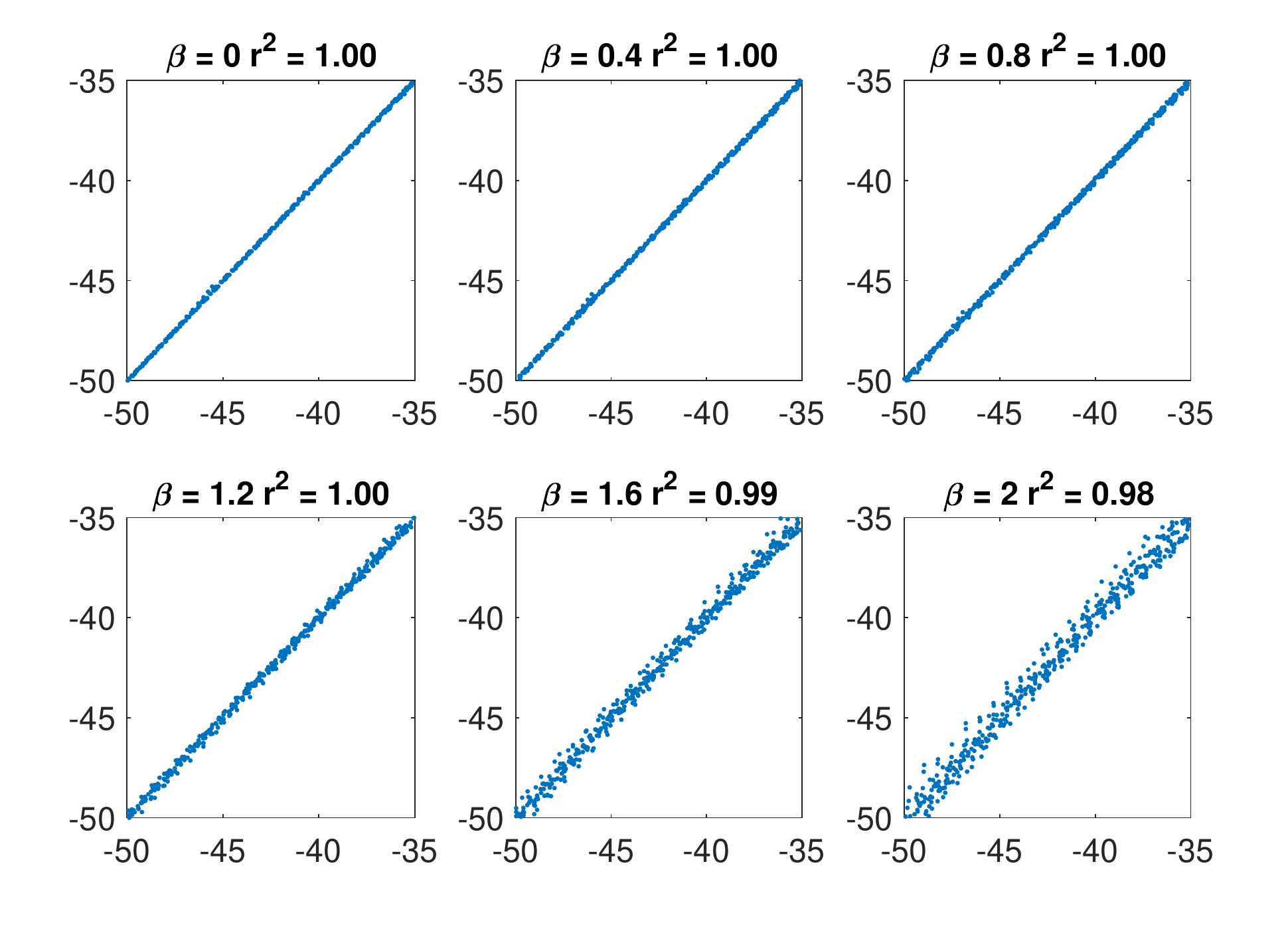}
	\caption{The collapse of the obtained regulated squared commutator for the data range $-50<\log C_{r}<-35$, $20<x<200$ and $20<t<100$, to the near wave-front ansatz by least squared method. We have chosen this data range as we have confirmed the convergence of our numerical procedure in this range.}\label{Fig:reg_goodness_fit}
\end{figure}

\begin{figure}
	\includegraphics[width=0.47\columnwidth]{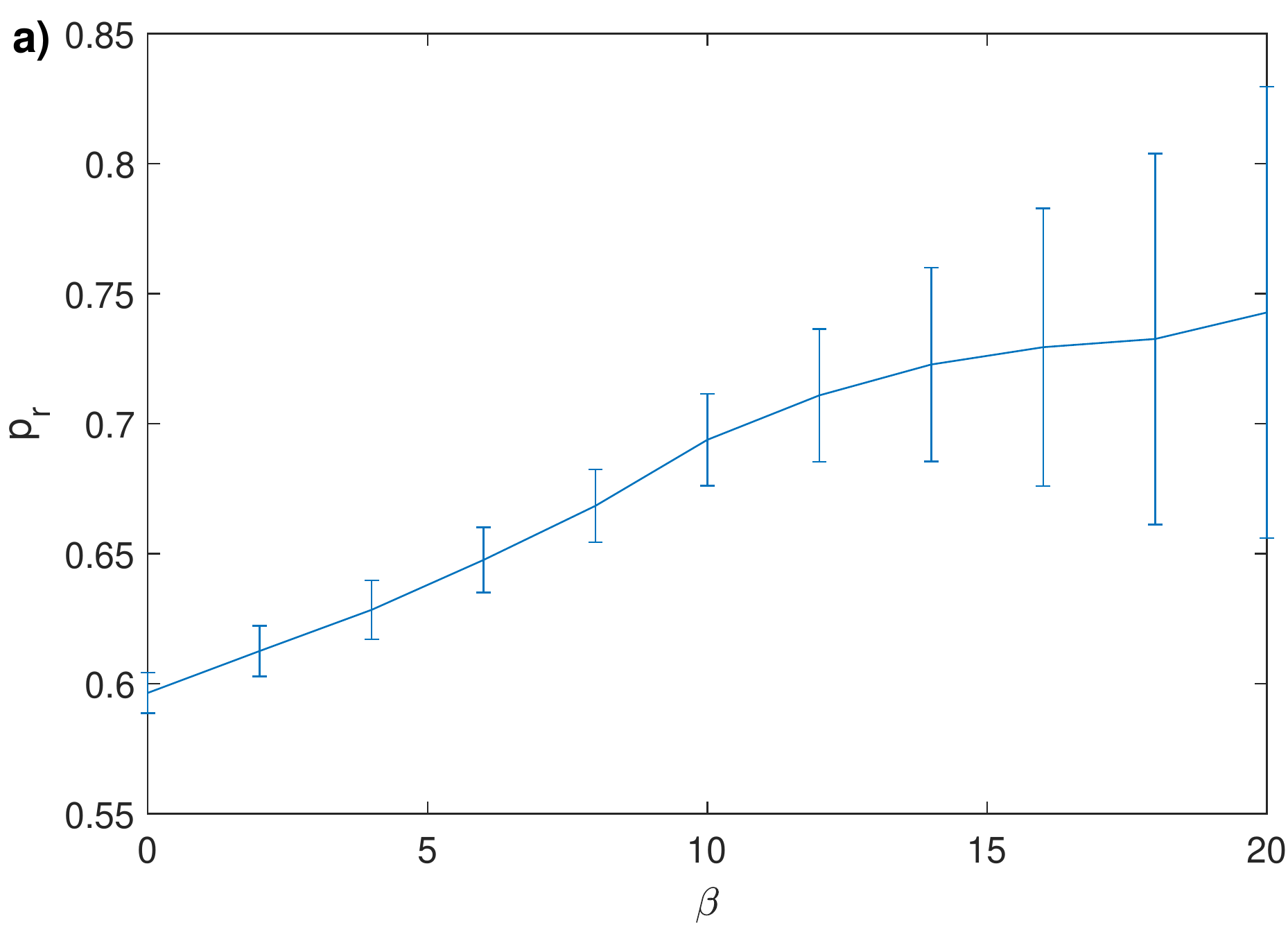}
	\includegraphics[width=0.47\columnwidth]{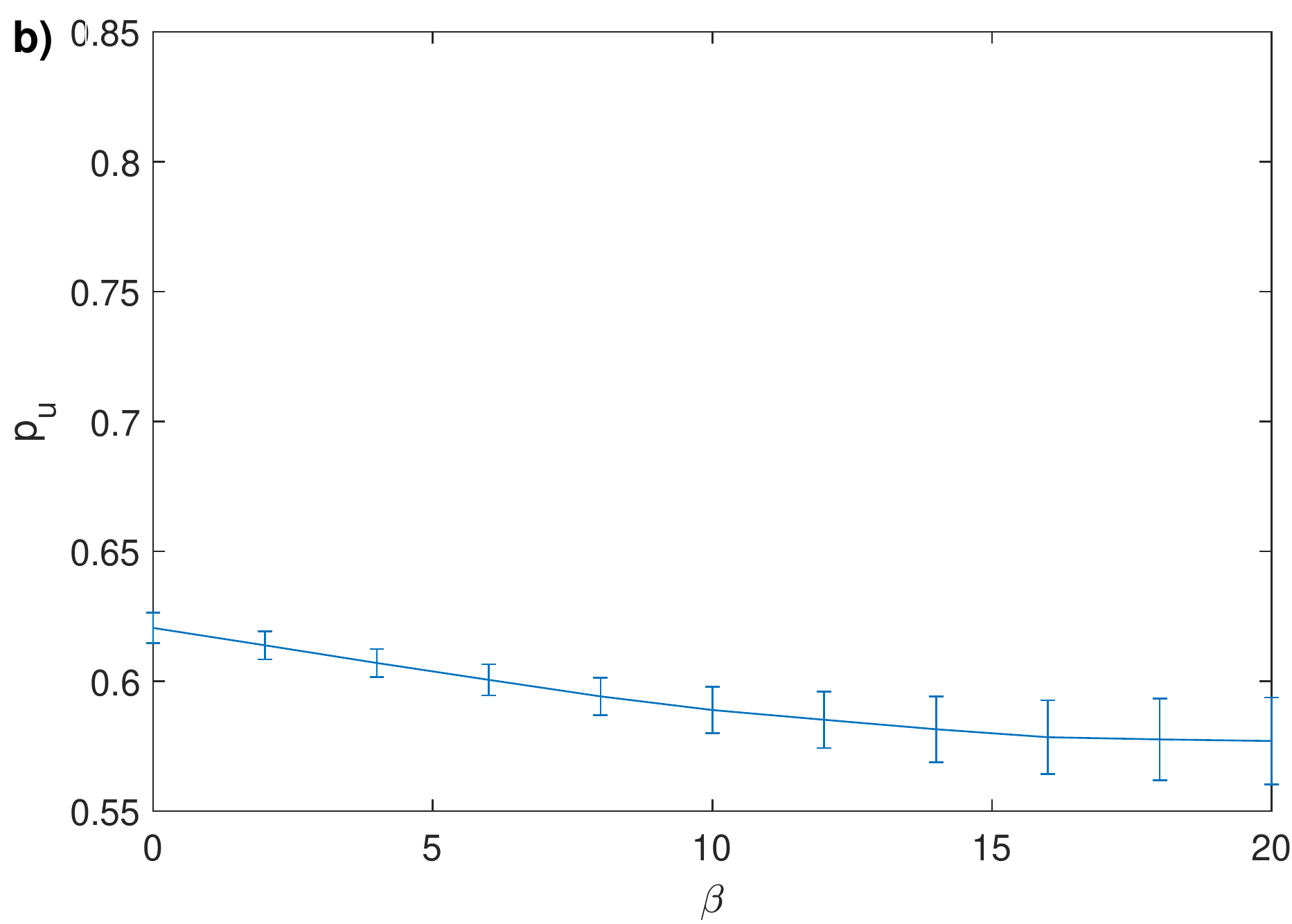}
	\caption{\textbf{a)} Broadening coefficient $p$ obtained from the numerical fitting of regulated squared commutator is plotted as a function of $\beta$. \textbf{b)} $p$ from fitting of the unregulated squared commutator is plotted as a function of $\beta$. The errorbars are from the 95\% confidence intervals of the fit. To compare the regulated and the unregulated cases we have fixed the y-axis scales to be the same in the two plots.}\label{Fig:p_reg_unreg_fit}
\end{figure}

We fit the converged data using least squared error method to the near wave-front ansatz of Eq. \ref{eq:scrambling_ansatz}. The goodness of fit is studied in Fig. \ref{Fig:reg_goodness_fit}, where the data collapse to the fitted model is shown at different temperatures.

The unregulated squared commutator was studied using a similar numerical technique in \cite{Han2019}. Our results indicate that the butterfly velocity obtained from the unregulated squared commutator is constant as function of temperature, even at temperatures lower than the gap, in contradiction with the indicated result from \cite{Han2019}. We checked the case for the $[Z(t),Z]$ type squared commutators as well, and our results are the same for both cases. In \cite{Han2019}, the fitting was done for a much smaller spatio-temporal region $20<x<45$ and $1<t<5$ (in our units), and for a much smaller range $\log C_{u}>-22$, as compared to the situation considered here.

We also study the temperature dependence of the broadening coefficient obtained from the fitting in Fig.~\ref{Fig:p_reg_unreg_fit}\textbf{a} (regulated) and Fig.~\ref{Fig:p_reg_unreg_fit}\textbf{b}(unregulated). For the unregulated case, we see a fairly constant $p$ which is insensitive to decreasing temperature. The regulated case shows an increasing trend with decreasing temperature.

\section{Contour dependence and chaos bound}\label{appsec:chaosbound}
\begin{figure}
	\includegraphics[width=0.48\columnwidth]{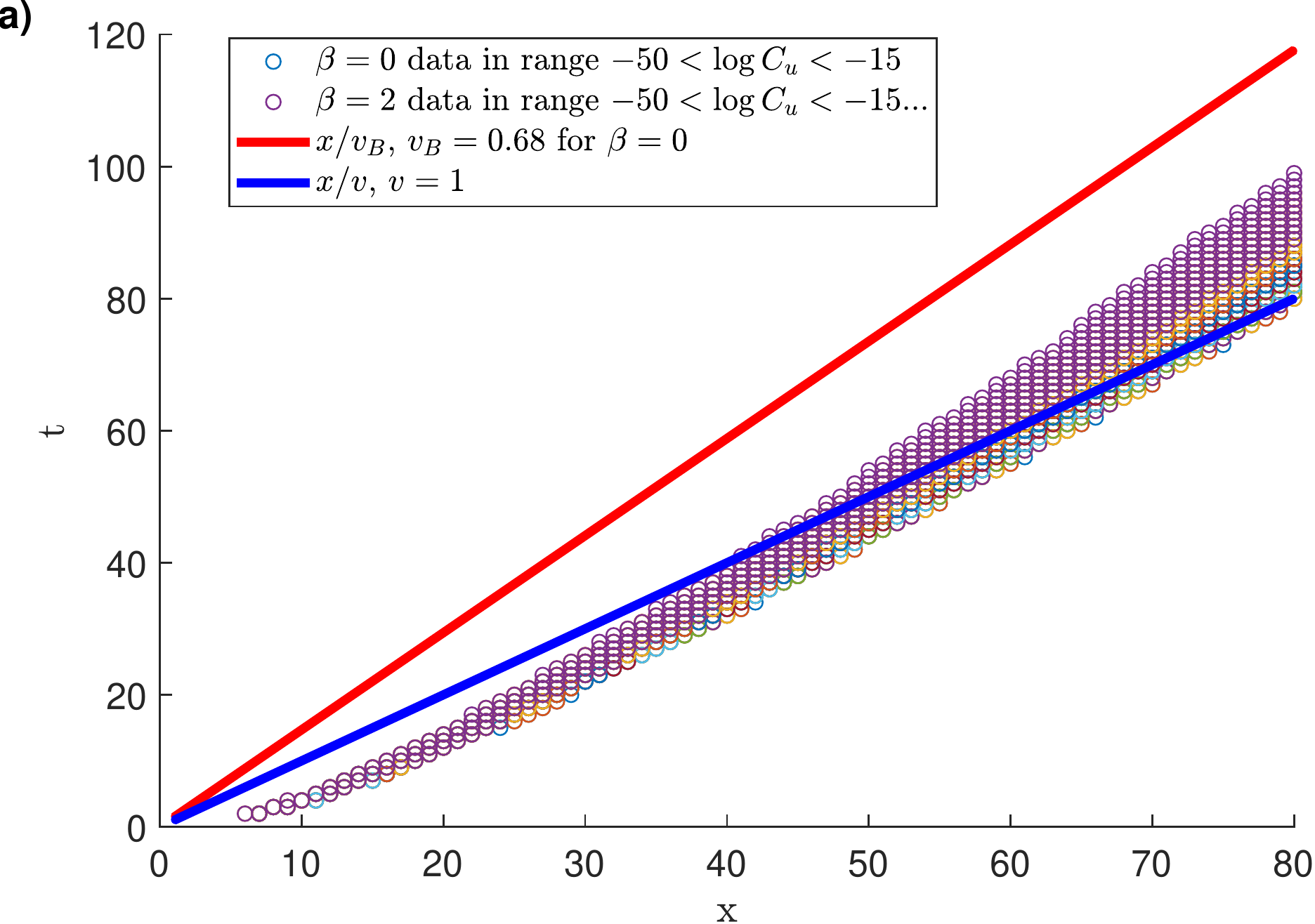}
	\includegraphics[width=0.48\columnwidth]{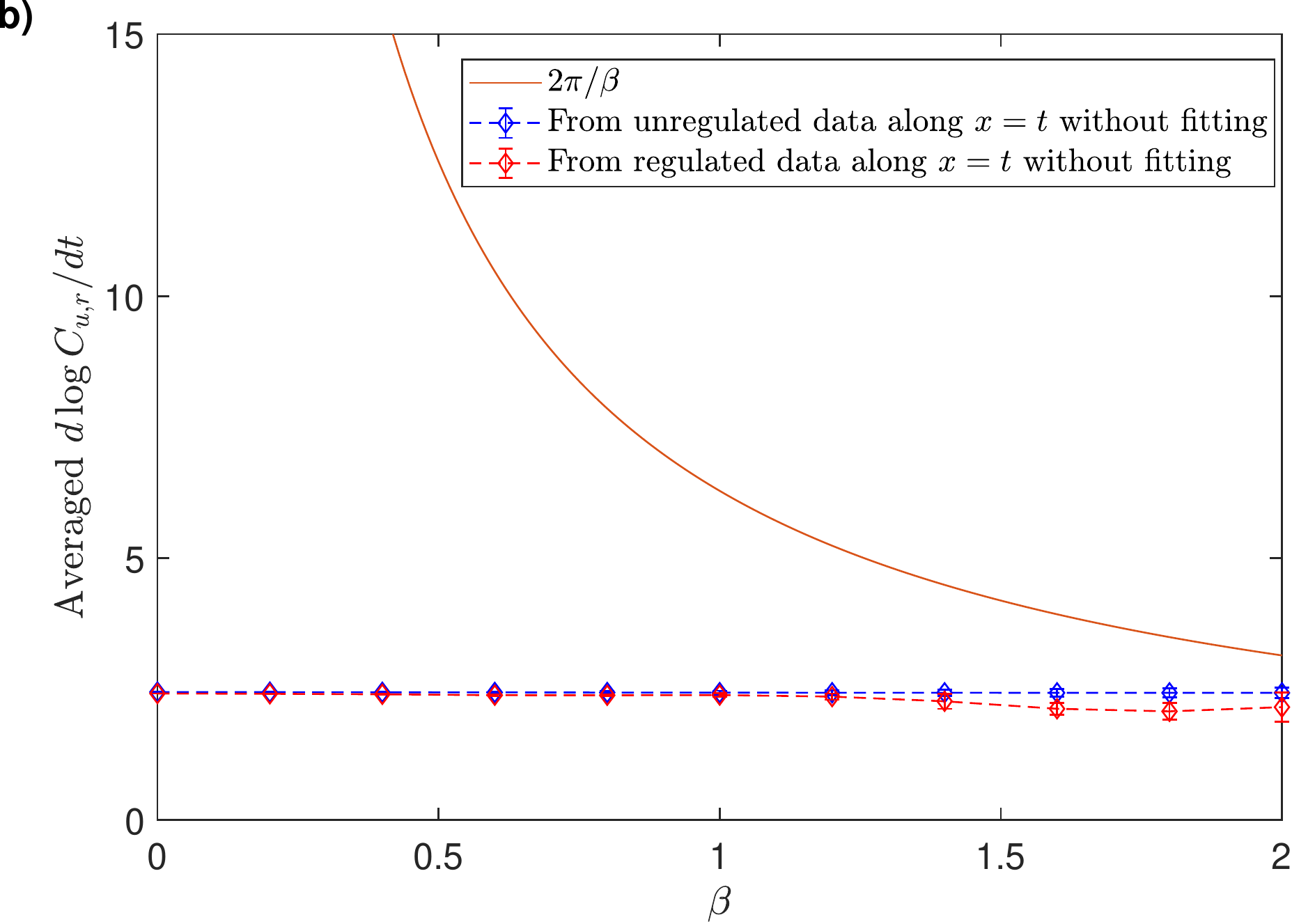}
	\caption{\textbf{a)} The data of the unregulated squared commutator for the data range $-50<\log C_{u}<-15$, is picked out along the `ray' $x=t$. $\partial_{t}C_{u}$ is evaluated in this domain, and the averaged $\partial_{t}C_{u}$ along $x=t$ is plotted as a function of $\beta$ in b). Similarly data for the unregulated case can be picked up. \textbf{b)} The averaged $\partial_{t}C_{u,r}$ along $x=t$ is plotted as a function of $\beta$.   }\label{fig:chaos_bound_data}
\end{figure}
\begin{figure}
	\includegraphics[width=0.48\columnwidth]{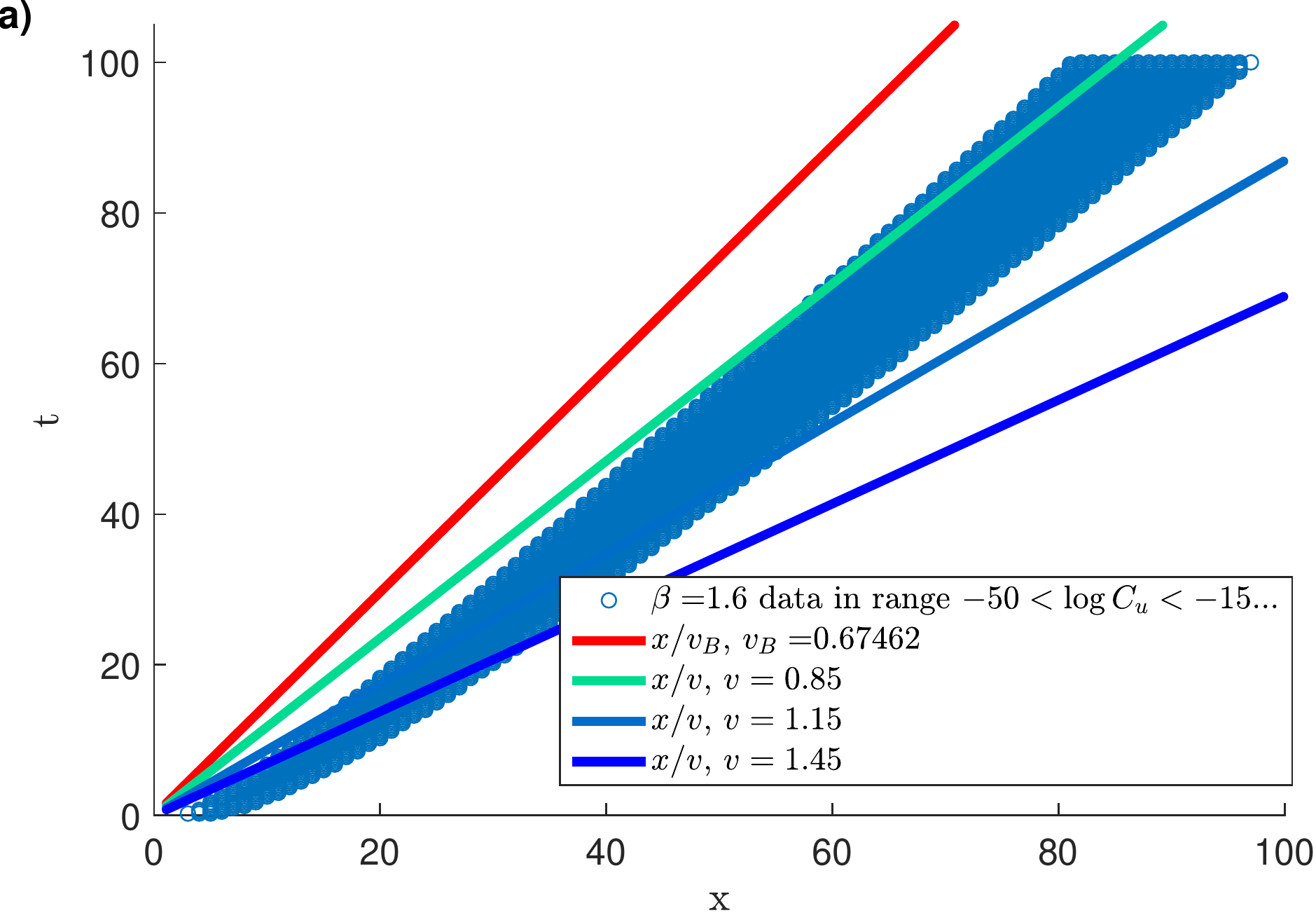}
	\includegraphics[width=0.48\columnwidth]{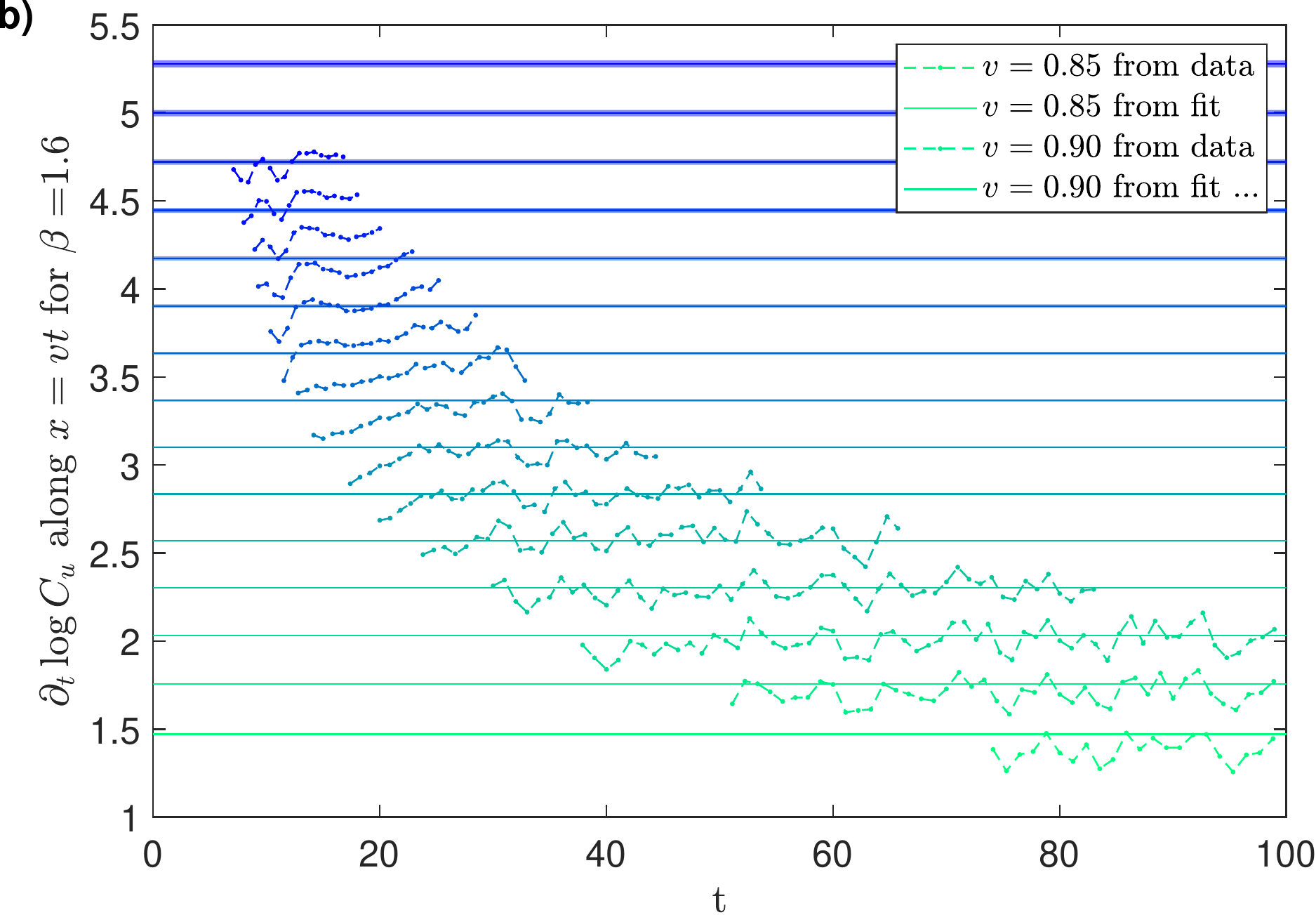}
	\caption{\textbf{a)} The data of the unregulated squared commutator for the data range $-50<\log C_{u}<-15$, is picked up along different`rays' $x=vt$. This procedure can be repeated for the regulated case. \textbf{b)} For different $v$-s, $\partial_{t}\log C_{u}$ is plotted as a function of $t$ (dots), and compared against the prediction from the near wavefront ansatz (constant lines whose thickness signify the confidence interval from the fitting to the ansatz). For lower $v$ (i.e.) closer to the butterfly velocity $v_{B}$, the near wavefront behavior and the numerical result are the same, but they deviate for high ray velocities. The constancy of $\partial_{t}\log C_{u}$ along rays allow us to study their time averages as a function of $\beta$. 
	 }\label{fig:chaos_bound_ray}
\end{figure}

We analyse in detail the contour dependence of $\partial_{t}C_{u,r}$, as was done in Sec. \ref{sec:chaosbound}. In Fig. \ref{fig:chaos_bound_data}, we sketch how $\partial_{t}C_{u,r}$ is found without numerical fitting. We first pick out data along a `ray' $x=t$, wherever the squared commutator has converged, and study $\partial_{t}C_{u,r}$ numerically. In Fig. \ref{fig:chaos_bound_data}b the averaged $\partial_{t}\log C_{u,r}$ along this ray is plotted as a function of $\beta$, and compared against the bound on chaos. The result is similar to Fig. \ref{Fig:lambda_reg_unreg}, which was obtained by fitting to the near wavefront ansatz. Given the constancy of the unregulated case, the chaos bound could be violated at lower temperatures. These results are for a particular ray $x=t$, and as a function of $\beta$. We can also study $\partial_{t}\log C_{u,r}$ as function of the ray velocity $v$, where $x = vt$, for a particular $\beta$. If the near wavefront scrambling ansatz (Eq. \ref{eq:scrambling_ansatz}) is satisfied, then $\partial_{t}\log C_{u,r}$ along a ray of velocity $v$ is given by $\lambda_{p}(v/v_{B}-1)^{p}(1+pv/v_{B})$. As $v$ is increased beyond the $v_{B}$, the near wavefront ansatz predicts that the chaos bound can be violated. We test this numerically in Fig. \ref{fig:chaos_bound_ray}, and we see that indeed $\partial_{t}\log C_{u,r}(t,vt)$ deviates from its near ansatz prediction at higher $v$.

\begin{figure}
	\includegraphics[width=0.48\columnwidth]{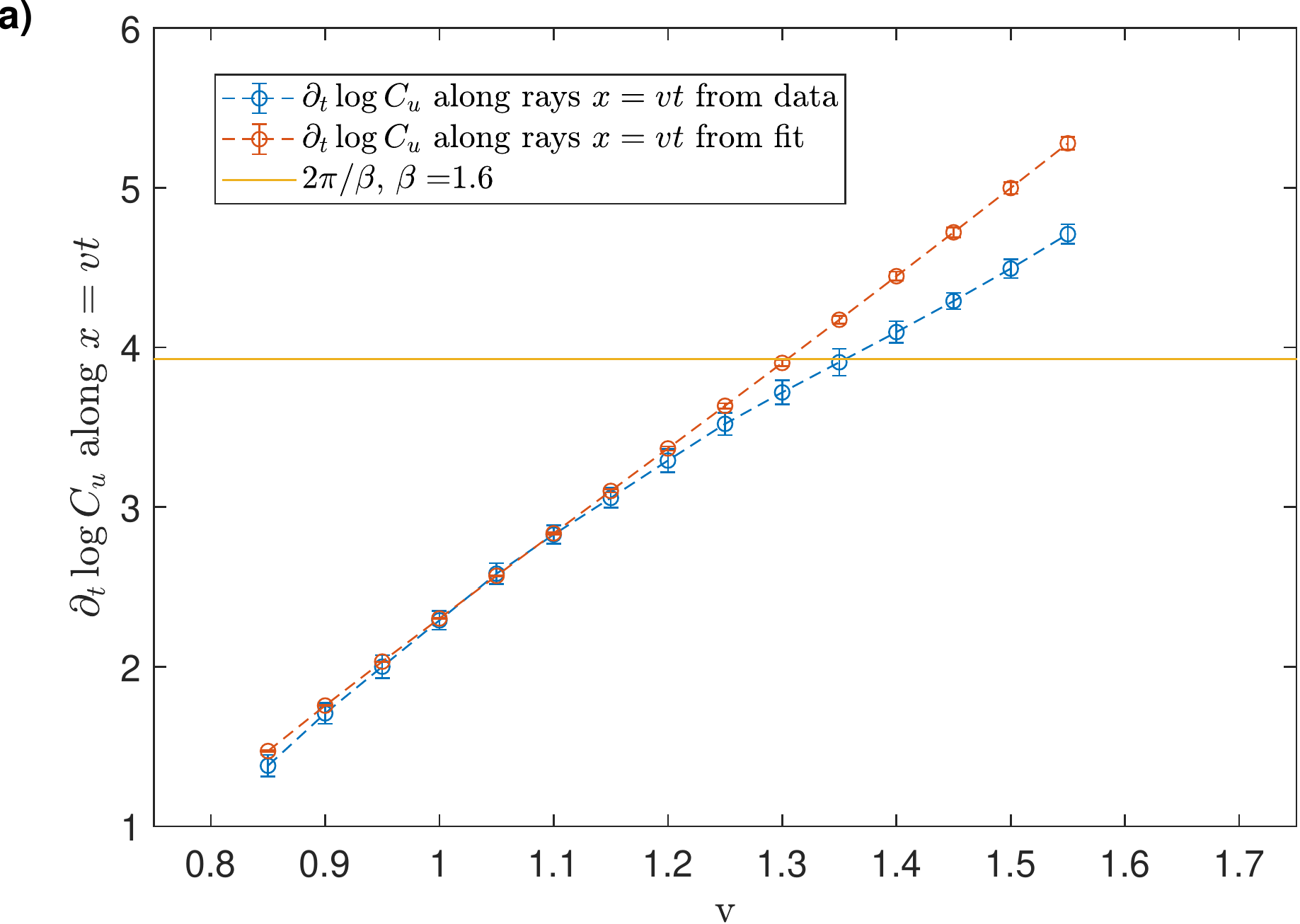}
	\includegraphics[width=0.48\columnwidth]{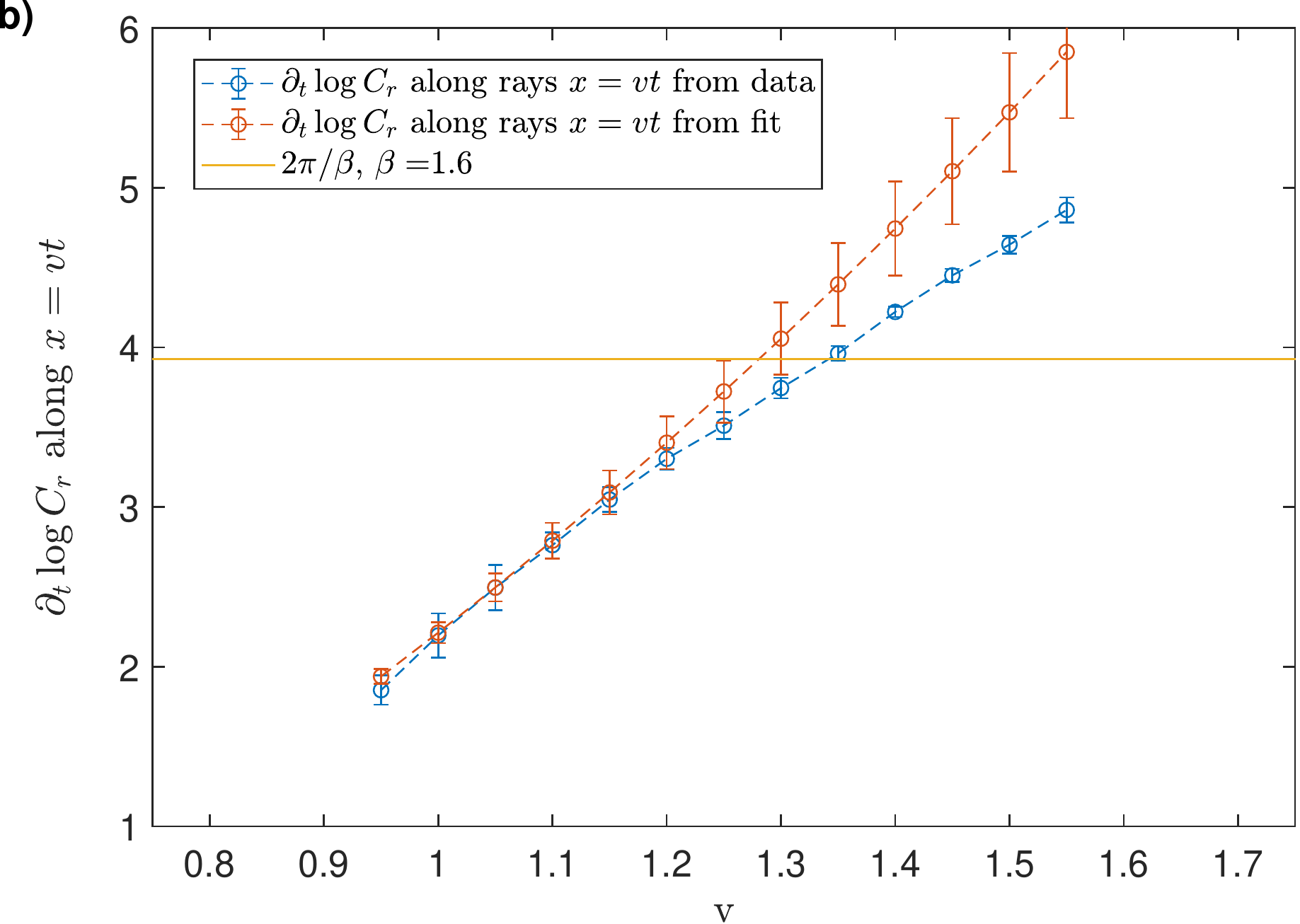}
	\caption{\textbf{a)} Time average of $\partial_{t}\log C_{u}(t,vt)$ is plotted for $\beta = 1.6$ as a function of ray velocity $v$ (blue dots), and compared against the prediction from the near wavefront ansatz (red dots). \textbf{b)} Same analysis is done for the regulated case. The yellow line in both case refer to the chaos bound at $\beta = 1.6$.
	 }\label{fig:chaos_bound_reg_unreg}
\end{figure}

We also compare $\partial_{t}\log C_{u,r}(t,vt)$ against the chaos bound as a function of ray velocity $v$ in Fig. \ref{fig:chaos_bound_reg_unreg}, and see that for high ray velocities, the bound is violated for both the regulated and unregulated cases. Note however that the analysis on the data is done only on the domain where the data has converged and also lies along the rays - severely restricting the domain on which numerical differentiation can be reliably done to obtain $\partial_{t}\log C_{u,r}(t,vt)$. 

\section{Spectral representation and the generalized Wightman function}\label{appsec:spec_wightman}
From the definition of the generalized Wightman function in Eq. \ref{eq:gen_wightman}, we go to the Fourier space, and expand in terms of many body eigenstates $|n\rangle$ with energy $E_{n}$ and momentum $P_{n}$,
\begin{equation}
    \mathcal{G}_{W}^{\alpha}(\omega,\mathbf{k}) = \frac{1}{(2\pi)^{3}}\int dt e^{i\omega t} \int d^{2}x \sum_{mn}\langle n |\rho^{\alpha}\phi(t,\mathbf{x})|m\rangle \langle m |\rho^{1-\alpha}\phi(0,\mathbf{0})|n\rangle. 
\end{equation}
In Heisenberg representation, $\phi(t,\mathbf{x}) = e^{-iPx}e^{iHt}\phi(0,\mathbf{0})e^{iPx}e^{-iHt}$. This allows us to write the spectral representation of the generalized Wightman function, 
\begin{equation}
    \mathcal{G}_{W}^{\alpha}(\omega,\mathbf{k}) = \frac{1}{Z} \sum_{mn}|\langle n |\phi|m\rangle|^{2} \delta\left(\omega - (E_{m}-E_{n})\right)\delta\left(\mathbf{k}-(P_{m}-P_{n})\right)e^{-\beta\left(\alpha E_{n}+(1-\alpha)E_{m}\right)}. 
\end{equation}
The spectral function can be similarly expanded in the spectral representation,
\begin{equation}
    A(\omega,\mathbf{k}) = \frac{1}{Z} \sum_{mn}|\langle n |\phi|m\rangle|^{2} \delta\left(\omega - (E_{m}-E_{n})\right)\delta\left(\mathbf{k}-(P_{m}-P_{n})\right)e^{-\beta E_{n}}\left(1-e^{-\beta \omega}\right). 
\end{equation}
Comparing the two spectral representations, we get the following relation,
\begin{equation}
    \mathcal{G}_{W}^{\alpha}(\omega,\mathbf{k}) = \frac{A(\omega,\mathbf{k})}{2\sinh{\frac{\beta \omega}{2}}}e^{\left(\alpha - \frac{1}{2}\right)\beta \omega}.
\end{equation}

\section{Polarization bubble calculation}\label{appsec:polarization}
\subsubsection*{T=0}
At $T=0$, the polarization bubble can be evaluated exactly, by changing the Matsubara sum to an integral,
\begin{align}
    \Pi^{T=0}\left(i\nu_{n},\mathbf{q}\right)=\frac{1}{2}\int_{\mathbf{k}}\int_{\mathbf{R}-(-\epsilon,\epsilon)}\frac{d\omega}{2\pi}\frac{1}{\left(\omega+\nu_{n}\right)^{2}+\left(\mathbf{k}+\mathbf{q}\right)^{2}+m^{2}}\frac{1}{\omega^{2}+\mathbf{k}^{2}+m^{2}}.
\end{align}
The retarded Polarization bubble is obtained by analytically continuing to real frequencies, $\Pi(\mathbf{q},i\nu_{n}\to\nu+i0^{+})$. The integral can be exactly evaluated, and we obtain,
\begin{align}
    \Pi_{R}^{T=0}\left(\nu,\mathbf{q}\right)&= \frac{1}{8\pi}\frac{1}{\sqrt{\mathbf{q}^{2}-\nu^{2}}}\arctan{\frac{\sqrt{\mathbf{q}^{2}-\nu^{2}}}{2m}}.
\end{align}
For $\nu^2\geq \mathbf{q}^{2}+4m^2$,
\begin{equation}
    \begin{split}
        Im[\Pi_{R}^{T=0}(\nu+i0^{+},\mathbf{q},)]&=-\frac{1}{16\sqrt{\nu^{2}-\mathbf{q}^{2}}} \\
        Re[\Pi_{R}^{T=0}(\nu+i0^{+},\mathbf{q})]&= \frac{1}{16\pi\sqrt{\nu^{2}-\mathbf{q}^{2}}}\log\left(\frac{\sqrt{\nu^{2}-\mathbf{q}^{2}}+2m}{\sqrt{\nu^{2}-\mathbf{q}^{2}}-2m}\right). 
    \end{split}
\end{equation}
For $\nu^{2}<\mathbf{q}^{2}+4m^{2}$, 
\begin{equation}\label{eq:re_pi_r_rel}
    \begin{split}
        Im[\Pi_{R}^{T=0}(\nu+i0^{+},\mathbf{q})]&= 0 \\
        Re[\Pi_{R}^{T=0}(\nu+i0^{+},\mathbf{q})]&= \frac{1}{8\pi}\frac{1}{\sqrt{\mathbf{q}^{2}-\nu^{2}}}\arctan{\frac{\sqrt{\mathbf{q}^{2}-\nu^{2}}}{2}}. 
    \end{split}
\end{equation}

\subsubsection*{Finite T}
Here, we obtain the low temperature correction to the $T=0$ polarization. 
At finite T, we introduce the function $b(z)=(e^{\beta z}-1)^{-1}$ and the $\phi$ polarization bubble can be calculated,  
\begin{equation}
    \begin{split}
        \Pi(i\nu_{n},\mathbf{q})&=\frac{T}{2}\sum_{i\omega_{n}}\int_{\mathbf{k}}^{\Lambda}\frac{1}{(\omega_{n}+\nu_{n})^{2}+\epsilon_{\mathbf{k}+\mathbf{q}}^{2}}\frac{1}{\omega_{n}^{2}+\epsilon_{\mathbf{k}}^{2}}\\
        &=\frac{1}{2}\int_{\mathbf{k}}^{\Lambda}\oint\frac{dz}{2\pi i}b(z)\frac{1}{(z+i\omega_{n})^{2}-\epsilon_{\mathbf{k}+\mathbf{q}}^{2}}\frac{1}{z^{2}-\epsilon_{\mathbf{k}}^{2}}\\
        &=-\frac{1}{2}\int_{\mathbf{k}}^{\Lambda}\frac{1}{4\epsilon_{\mathbf{k}}\epsilon_{\mathbf{k}+\mathbf{q}}}\left[\frac{b(\epsilon_{\mathbf{k}})-b(\epsilon_{\mathbf{k}+\mathbf{q}})}{\epsilon_{\mathbf{k}}-\epsilon_{\mathbf{k}+\mathbf{q}}+i\nu_{n}}-\frac{b(\epsilon_{\mathbf{k}})+b(\epsilon_{\mathbf{k}+\mathbf{q}})}{\epsilon_{\mathbf{k}}+\epsilon_{\mathbf{k}+\mathbf{q}}+i\nu_{n}}\right.\\&\left.- \frac{b(-\epsilon_{\mathbf{k}})+b(\epsilon_{\mathbf{k}+\mathbf{q}})}{\epsilon_{\mathbf{k}}+\epsilon_{\mathbf{k}+\mathbf{q}}-i\nu_{n}}-\frac{b(\epsilon_{\mathbf{k}})-b(\epsilon_{\mathbf{k}+\mathbf{q}})}{\epsilon_{\mathbf{k}+\mathbf{q}}-\epsilon_{\mathbf{k}}+i\nu_{n}}\right]
    \end{split}
\end{equation}
Using $b(-z)=-b(z)-1$ and for our hierarchy of scales, $b(\epsilon_{\mathbf{k}})\approx e^{-\beta \epsilon_{\mathbf{k}}}<<1$ for any $\mathbf{k}$, we can replace $b(-z)\to-1$. The retarded polarization bubble is obtained by analytically continuing from the imaginary Matsubara frequency to real frequency, $\Pi(i\nu_{n},\mathbf{q})\to\Pi_{R}(\nu+i0^{+},\mathbf{q})$. Using Cauchy imaginary value theorem, the imaginary part can be obtained to be (restricting to $\nu>0$)
\begin{equation}
    Im[\Pi_{R}(\nu+i0^{+},\mathbf{q})]=\frac{1}{2}\int_{\mathbf{k}}^{\Lambda}\frac{\pi}{4\epsilon_{\mathbf{k}}\epsilon_{\mathbf{k}+\mathbf{q}}}\left[\delta(\epsilon_{\mathbf{k}}+\epsilon_{\mathbf{k}+\mathbf{q}}-\nu)+2\left(e^{-\beta\epsilon_{\mathbf{k}+\mathbf{q}}}-e^{-\beta\epsilon_{\mathbf{k}}}\right)\delta(\epsilon_{\mathbf{k}+\mathbf{q}}-\epsilon_{\mathbf{k}}+\nu)\right].
\end{equation}
The first term is the $T=0$ result, which was also obtained in the previous paragraph. At finite $T$, the only modification is the second term, which we now evaluate.

In order to evaluate this integral, we need to impose the delta function condition. First we shift the $\mathbf{k}$ integral to $\mathbf{k}+\mathbf{q}/2$. We also change notation $\epsilon_{\pm}=\epsilon_{\mathbf{k}\pm\mathbf{q}/2}$. The delta function conditions are then, $\epsilon_{+}+s\epsilon_{-}=s\nu$, for $s= - 1$. Imposing the delta function condition, we get, $k^{*}=\frac{\nu}{2}\sqrt{\frac{\nu^2-q^2-4m^2}{\nu^2-q^2\cos^{2}\theta}}$, and $\epsilon^{*}_{\pm}=\mp\nu/2-k^{*}q\cos\theta /\nu$. For this to be consistent with the positivity of $\epsilon_{\pm}$, $\theta\in\left(\pi-\cos^{-1}\nu/q,\pi+\cos^{-1}\nu/q\right)$.

Now, by change of variable in the delta function,
\begin{align*}
    \delta\left(\nu+\epsilon_{+}-\epsilon_{-}\right)&=\left|\nabla f(\mathbf{k})\right|_{k=k^{*}}^{-1}\delta(k-k^{*}) \text{ where,}\\
    f(\mathbf{k})&=\sqrt{(\mathbf{k}+\mathbf{q}/2)^{2}+m^{2}}-\sqrt{(\mathbf{k}-\mathbf{q}/2)^{2}+m^{2}}+\nu.
\end{align*} 
We then do the radial $k$ integral, by setting $k\to k^{*}$. In order to do the $\theta$ integral, we can employ the Laplace method, as the integrand has the exponential factor,   $e^{\frac{\beta q\cos\theta\sqrt{q^2-\nu^2+4m^2}}{2\sqrt{q^{2}\cos^{2}\theta-\nu^2}}}$, and $\beta m >>1$. The exponent has a maxima at $\theta=\pi$, which lies in the allowed domain of $\theta$. Doing the integral, we get the full correction, for $\nu<q$,
\begin{equation}
Im \Pi_R(\nu ,\mathbf{q})=\frac{1}{8\pi}\sqrt{\frac{4 \pi }{\beta }} \sinh \left(\frac{\beta \nu }{2}\right) \left(\frac{1}{q^2 \left(4m^{2}-\nu ^2+q^2\right) \left(q^2-\nu ^2\right)}\right)^{1/4} e^{-\frac{\text{$\beta $q}}{2} \sqrt{\frac{4m^{2}-\nu ^2+q^2}{q^2-\nu ^2}}}
\end{equation}


\section{Self Energy calculation}\label{appsec:self_energy}
To study the temperature dependent relaxation time of the bosonic quasiparticles, we need to evaluate the self energy of $\phi$. The relevant diagrams are shown in Fig. \ref{fig:self_energy}. The imaginary part of the self energy has contribution only from the first diagram in Fig. \ref{fig:self_energy}, and can be evaluated to give,
\begin{equation}
    \begin{split}
        Im[\Sigma_{R}(\omega+i0^{+},\mathbf{q})]&=-\frac{1}{N}\int_{\mathbf{k}}\frac{\sinh{\beta\omega/2}}{4\epsilon_{\mathbf{k}}\sinh{\beta\epsilon_{\mathbf{k}}/2}}\times\\&\left[\frac{Im[\Pi_{R}^{-1}(\epsilon_{k}-\omega,\mathbf{k}-\mathbf{q})]}{\sinh[\beta(\epsilon_{k}-\omega)/2]}+\frac{Im[\Pi_{R}^{-1}(-\epsilon_{k}-\omega,\mathbf{k}-\mathbf{q})]}{\sinh[-\beta(\epsilon_{k}+\omega)/2]}\right].
    \end{split}
\end{equation}
Note, at low temperature, the second term in the imaginary part of the self-energy can be ignored. Recalling the definition of the Wightman function, we have, 
\begin{equation}
            Im[\Sigma_{R}(\omega+i0^{+},\mathbf{q})]\approx \frac{1}{N}\int_{\mathbf{k}}\frac{\sinh{\beta\omega/2}}{4\epsilon_{\mathbf{k}}\sinh{\beta\epsilon_{\mathbf{k}}/2}}G_{W,\lambda}^{(1/2)}\left(\epsilon_{k}-\omega,\mathbf{k}-\mathbf{q}\right).
\end{equation}
The inverse lifetime, or the relaxation rate of $\phi$ can be written in terms of the imaginary part of the self energy,
\begin{equation}\label{eq:inv_lifetime}
    \begin{split}
        \Gamma_{\mathbf{q}}=\frac{Im[\Sigma_{R}(\epsilon_{\mathbf{q}},\mathbf{q})]}{2\epsilon_{q}}&=\frac{1}{2N}\int_{\mathbf{k}}^{\Lambda}\frac{\sinh\left(\beta \epsilon_{\mathbf{q}}/2\right)}{\sinh \left(\beta \epsilon_{\mathbf{k}}/2\right)}\mathcal{R}^{(1/2)}_{1+}(\mathbf{k},\mathbf{q})\text{, where we have defined,}\\
        \mathcal{R}^{(1/2)}_{1+}(\mathbf{k},\mathbf{q})&=\frac{G_{W,\lambda}^{(1/2)}\left(\epsilon_{k}-\epsilon_{q},\mathbf{k}-\mathbf{q}\right)}{4\epsilon_{k}\epsilon_{q}}.
    \end{split}
\end{equation}
Note, $|\epsilon_{k}-\epsilon_{q}|<|\mathbf{k}-\mathbf{q}|$. The Wightman function $\mathcal{G}_{W,\lambda}^{(1/2)}(\epsilon_{k}-\epsilon_{q},\mathbf{k}-\mathbf{q})$ can be expressed as $$\frac{- Im[\mathcal{G}_{R}((\epsilon_{k}-\epsilon_{q},\mathbf{k}-\mathbf{q})]}{\sinh{\beta (\epsilon_{\mathbf{k}}-\epsilon_{\mathbf{q}})/2}},$$ where, $Im[\mathcal{G}_{R}]$ is given by $Im[\Pi_{R}^{-1}]=-\frac{Im[\Pi_{R}]}{Re[\Pi_{R}]^{2}+Im[\Pi_{R}]^{2}}$. From the calculations in Sec. \ref{appsec:polarization}, one can read off the expression for $Im[\Pi_{R}]$ which is exponentially suppressed in $\beta m$.  In the denominator, any temperature dependence can be ignored, because of the leading $T=0$ behavior of $Re[\Pi_{R}]$. Thus, we have the following approximation for $\mathcal{R}_{1}(\mathbf{k},\mathbf{q})$,
\begin{equation}\label{eq:r1_p}
    \begin{split}
        \mathcal{R}^{(1/2)}_{1+}(\mathbf{k},\mathbf{q}) \approx \frac{1}{8\pi}\sqrt{\frac{4\pi}{\beta m}}\frac{1}{4\epsilon_{\mathbf{k}}\epsilon_{\mathbf{q}}} \exp\left(-\frac{\beta|\mathbf{k}-\mathbf{q}|\sqrt{(\mathbf{k}-\mathbf{q})^{2}-\left(\epsilon_{\mathbf{k}}-\epsilon_{\mathbf{q}}\right)^{2}+4m^{2}}}{2\sqrt{(\mathbf{k}-\mathbf{q})^{2}-\left(\epsilon_{\mathbf{k}}-\epsilon_{\mathbf{q}}\right)^{2}}}\right)\times\\
        \left(\frac{\left(\left|\mathbf{k}-\mathbf{q}\right|^{2}-(\epsilon_{\mathbf{k}}-\epsilon_{\mathbf{q}})^{2}\right)^{3/4}}{\left|\mathbf{k}-\mathbf{q}\right|^{1/2}\left(4m^{2}-(\epsilon_{\mathbf{k}}-\epsilon_{\mathbf{q}})^{2}+\left|\mathbf{k}-\mathbf{q}\right|^{2}\right)^{1/4}}\right)\frac{64\pi^{2}}{\arctan^{2}\frac{\sqrt{\left(\left|\mathbf{k}-\mathbf{q}\right|^{2}-(\epsilon_{\mathbf{k}}-\epsilon_{\mathbf{q}})^{2}\right)}}{2}}.
    \end{split}
\end{equation}


At low temperature, the relaxation rate can be approximated by the Laplace method, since the integrand has a factor exponential in $\beta m$ (arising from both the prefactor $\sinh$ and $\mathcal{R}_{1}$ functions in Eq. \ref{eq:inv_lifetime}).

We define the phase coherence inverse time scale as, $\tau_{\phi}^{-1}=\Gamma_{\mathbf{q}=\mathbf{0}}$ \cite{Chubukov1994}, which can be evaluated,
\begin{align} 
    \Gamma_{0} = \frac{1}{\tau_{\phi}}\approx \frac{2\pi}{N\beta}e^{-\beta m}.
\end{align}

The momentum dependent $\Gamma_{\mathbf{q}}$ can be evaluated numerically, 
\begin{equation}
    \Gamma_{\mathbf{q}} \approx \frac{1}{2N}e^{\beta \epsilon_{\mathbf{q}}/2}\int_{\mathbf{k}}e^{-\beta \epsilon_{\mathbf{k}}/2}\mathcal{R}^{(1/2)}_{1+}(\mathbf{k},\mathbf{q}).
\end{equation}

\section{Ladder calculation in different contours}
The ladder calculation sets up a diagrammatic calculation of the squared commutator in terms of retarded Green functions and Wightman functions of the fields $\phi$ and $\lambda$. Here we give a sketch of how it works, following \cite{Chowdhury2017}, while also extending their results to the unregulated squared commutator.

Consider the generalized squared commutator, 
\begin{equation}
    C_{(\alpha)}(t,\mathbf{x}) = -\frac{1}{N^{2}}\sum_{ab}Tr\left(\rho^{\alpha}[\phi_{a,0}(t,\mathbf{x}),\phi_{b,0}(0,\mathbf{0})]\rho^{(1-\alpha)}[\phi_{a,0}(t,\mathbf{x}),\phi_{b,0}(0,\mathbf{0})]\right).
\end{equation}
To go to the interaction representation for the $\phi$ fields, we introduce time evolution operators in the interaction picture,
\begin{equation}
    U_{I} = \mathcal{T}\exp \left(\frac{i}{2\sqrt{N}}\sum_{a}\int_{0}^{t}ds\int_{\mathbf{x}}\lambda_{0}(s,\mathbf{x})\phi_{0}^{2}(s,\mathbf{x})\right),
\end{equation}
where the subscript $0$ indicates that the fields time evolve under the non-interacting part of the Hamiltonian. We further drop the factors of $N$ and the index structure to obtain,  
\begin{equation}\label{eq:int_c}
    C_{(\alpha)}(t,\mathbf{x}) \sim -Tr\left(\rho^{\alpha}[U_{I}^{\dagger}\phi_{0}(t,\mathbf{x})U_{I},\phi_{0}(0,\mathbf{0})]\rho^{(1-\alpha)}[U_{I}^{\dagger}\phi_{0}(t,\mathbf{x})U_{I},\phi_{0}(0,\mathbf{0})]\right).
\end{equation}
By expanding up to second order of $\lambda$, we get, 
\begin{equation}
\begin{split}
    U_{I}^{\dagger}\phi_{0}(t)U_{I} \approx \phi_{0}(t) + \frac{i}{2}\int_{0}^{t}ds\left[\phi_{0}(t),\lambda_{0}(s)\phi_{0}^{2}(s)\right]+\\\left(\frac{i}{2}\right)^{2}\int_{0}^{t}ds_{1}\int_{0}^{s_{1}}ds_{2}\left[\left[\phi_{0}(t),\lambda_{0}(s_{1})\phi_{0}^{2}(s_{1})\right],\phi_{0}(t),\lambda_{0}(s_{2})\phi_{0}^{2}(s_{2})\right]+...,
    \end{split}
\end{equation}
where we have suppressed the spatial dimension.

By combining fields from both `sides of the ladder' in the expanded expression Eq. \ref{eq:int_c}, we get the two distinct types of rungs - the contributions which are called the Type I and Type II rungs in Sec. \ref{sec:ladder_sum}. The contour dependence appears in the form of the contour dependence of the Wightman functions. For example, the Type I rung is a contour dependent $\lambda$-Wightman function, $Tr\left(\rho^{\alpha}\lambda_{0}(s)\rho^{1-\alpha}\lambda_{0}(s^{\prime})\right)$, or, $\mathcal{G}_{W,\lambda}^{(\alpha)}(s-s^{\prime})$. Similarly, for Type II, we get the corresponding contour dependent Wightman functions.

\section{Kernel functions at low temperature}\label{appsec:kernel}

\subsubsection{$\mathcal{R}^{(1/2)}_{1}$ kernel}
We already calculated the $\mathcal{R}^{(1/2)}_{1+}$ kernel in Sec. \ref{appsec:self_energy}, as given in Eq. \ref{eq:r1_p}. We now calculate $\mathcal{R}^{(1/2)}_{1-}$ at low temperatures,
\begin{equation}\label{eq:r1_m}
    \begin{split}
       \mathcal{R}^{(1/2)}_{1-}(\mathbf{p^{\prime}},\mathbf{p})&:= \frac{\mathcal{G}^{(1/2)}_{W,\lambda}(-\epsilon_{\mathbf{p}\prime}-\epsilon_{\mathbf{p}})}{4\epsilon_{\mathbf{p}\prime}\epsilon_{\mathbf{p}}}\\&\approx \frac{1}{2\epsilon_{\mathbf{p}\prime}\epsilon_{\mathbf{p}}}e^{-\frac{\beta(\epsilon_{\mathbf{p}}+\epsilon_{\mathbf{p}\prime})}{2}}\frac{Im \Pi_{R}^{T=0}(\epsilon_{\mathbf{p}\prime}+\epsilon_{\mathbf{p}},\mathbf{p}\prime-\mathbf{p})}{\left|\Pi_{R}(\epsilon_{\mathbf{p}\prime}+\epsilon_{\mathbf{p}},\mathbf{p}\prime-\mathbf{p})\right|^{2}}\\
       &=\frac{1}{32\epsilon_{\mathbf{p}\prime}\epsilon_{\mathbf{p}}}e^{-\frac{\beta(\epsilon_{\mathbf{p}}+\epsilon_{\mathbf{p}\prime})}{2}}\frac{1}{\sqrt{(\epsilon_{\mathbf{p}}+\epsilon_{\mathbf{p}\prime})^2-(\mathbf{p}\prime-\mathbf{p})^2}}\frac{1}{\left|\Pi_{R}^{T=0}(\epsilon_{\mathbf{p}\prime}+\epsilon_{\mathbf{p}},\mathbf{p}\prime-\mathbf{p})\right|^{2}}.
    \end{split}
\end{equation}
$\mathcal{R}^{(1/2)}_{1-}(\mathbf{p^{\prime}},\mathbf{p})$ is exponentially suppressed unless $p,p^{\prime}<<1$, while $\mathcal{R}^{(1/2)}_{1+}(\mathbf{p^{\prime}},\mathbf{p})$ is exponentially suppressed unless $|\mathbf{p}^{\prime}-\mathbf{p}|<<1$. Furthermore, even in the domain where both the exponents are comparable, it can be numerically verified that $\mathcal{R}^{(1/2)}_{1-}(\mathbf{p^{\prime}},\mathbf{p})$ is negligible compared to $\mathcal{R}^{(1/2)}_{1+}(\mathbf{p^{\prime}},\mathbf{p})$. Hence for the ladder calculation, we ignore $\mathcal{R}_{1-}$.

\subsubsection{$\mathcal{R}^{(1/2)}_{2}$ kernel}

In order to evaluate the $\mathcal{R}^{(1/2)}_{2}$ integration, we first need an expression for $\mathcal{G}^{(1/2)}_{\text{eff}}$ that was defined in Eq. \ref{eq:g_eff}. For results correct to the required order of $1/N$, it is enough to consider $\mathcal{G}^{(1/2)}_{W}(\omega)\sim \mathcal{Q}(\omega)A^{(0)}(\omega)$, where $A^{(0)}$ is the bare $\phi$ spectral function, given in Eq. \ref{eq:bare_phi_spec}. We have also defined the function, $\mathcal{Q}(\omega)=[2\sinh(\beta\omega/2)]^{-1}$. Inserting the spectral function in the expression for $\mathcal{G}^{(1/2)}_{W}(\omega^{\prime\prime}-\omega,\mathbf{p^{\prime\prime}}-\mathbf{p})\mathcal{G}_{W}(\omega^{\prime}-\omega^{\prime\prime},\mathbf{p^{\prime}}-\mathbf{p^{\prime\prime}})$ in Eq. \ref{eq:g_eff}, allows us to integrate over $\omega^{\prime\prime}$. We introduce notation $x = \mathbf{p^{\prime}}-\mathbf{p}$, $y = \frac{\mathbf{p^{\prime}}+\mathbf{p}}{2}$ and $\overline{\omega}=\omega^{\prime}-\omega$. We also denote $\epsilon_{x/2\pm\mathbf{p^{\prime\prime}}}=:\epsilon_{\pm}$. We now have the following expression for $\mathcal{G}^{(1/2)}_{\text{eff}}$,
\begin{equation}
    \begin{split}
        \mathcal{G}^{(1/2)}_{\text{eff}}(\omega^{\prime},\omega,\mathbf{p^{\prime}},\mathbf{p})=\frac{1}{2N}\int_{\mathbf{p^{\prime\prime}}}\frac{\pi}{\epsilon_{+}\epsilon_{-}}\bigg(\mathcal{Q}(\epsilon_{+})\mathcal{Q}(\overline{\omega}-\epsilon_{+})\mathcal{G}_{R,\lambda}(-\omega-\epsilon_{+},-\mathbf{p^{\prime\prime}}-y)\mathcal{G}_{R,\lambda}(\omega+\epsilon_{+},\mathbf{p^{\prime\prime}}+y)\\\left[\delta(\overline{\omega}-\epsilon_{+}-\epsilon_{-})-\delta(\overline{\omega}-\epsilon_{+}+\epsilon_{-})\right]\\-\quad\mathcal{Q}(-\epsilon_{+})\mathcal{Q}(\overline{\omega}+\epsilon_{+})\mathcal{G}_{R,\lambda}(-\omega+\epsilon_{+},-\mathbf{p^{\prime\prime}}-y)\mathcal{G}_{R,\lambda}(\omega-\epsilon_{+},\mathbf{p^{\prime\prime}}+y)\\\left[\delta(\overline{\omega}+\epsilon_{+}-\epsilon_{-})-\delta(\overline{\omega}+\epsilon_{+}+\epsilon_{-})\right]\bigg).
    \end{split}
\end{equation}
In this expression, because of the delta functions, one can replace the arguments of $\mathcal{Q}$ by $\pm \epsilon_{\pm}$. Note, at low temperature, $\mathcal{Q}(\epsilon_{\pm})\approx e^{-\beta\epsilon_{\pm}/2}$, and $\mathcal{Q}(-\epsilon_{\pm})\approx - e^{-\beta\epsilon_{\pm}/2}$. We can also use the fact that $\mathcal{G}_{R,\lambda}(\omega,-\mathbf{q})=\mathcal{G}_{R,\lambda}(\omega,\mathbf{q})$, and that the real and imaginary parts of $\mathcal{G}_{R,\lambda}(\omega,\mathbf{q})$ are even and odd functions of $\omega$ respectively. This allows for the following simplification,
\begin{equation}
    \begin{split}
        \mathcal{G}_{R,\lambda}(-\omega+\epsilon_{+},-\mathbf{p^{\prime\prime}}-y)&\mathcal{G}_{R,\lambda}(\omega-\epsilon_{+},\mathbf{p^{\prime\prime}}+y)\\&=\frac{1}{Re[\Pi_{R}(\omega-\epsilon_{+},\mathbf{p^{\prime\prime}}+y)]^{2}+Im[\Pi_{R}(\omega-\epsilon_{+},\mathbf{p^{\prime\prime}}+y)]^{2}}\\&\approx\frac{1}{|\Pi_{R}^{T=0}(\omega-\epsilon_{+},\mathbf{p^{\prime\prime}}+y)|^{2}}.
    \end{split}
\end{equation}
We finally arrive at a simple expression for $\mathcal{G}^{(1/2)}_{\text{eff}}$,
\begin{equation}
    \begin{split}
        \mathcal{G}^{(1/2)}_{\text{eff}}(\omega^{\prime},\omega,\mathbf{p^{\prime}},\mathbf{p})=\frac{1}{2N}\int_{\mathbf{p^{\prime\prime}}}\frac{\pi e^{-\frac{\beta(\epsilon_{+}+\epsilon_{-})}{2}}}{\epsilon_{+}\epsilon_{-}}\\\bigg(\left|\Pi_{R}^{T=0}(\omega+\epsilon_{+},\mathbf{p^{\prime\prime}}+y)\right|^{-2}\left[\delta(\overline{\omega}-\epsilon_{+}-\epsilon_{-})+\delta(\overline{\omega}-\epsilon_{+}+\epsilon_{-})\right]\\\left|\Pi_{R}^{T=0}(\omega-\epsilon_{+},\mathbf{p^{\prime\prime}}+y)\right|^{-2}\left[\delta(\overline{\omega}+\epsilon_{+}-\epsilon_{-})+\delta(\overline{\omega}+\epsilon_{+}+\epsilon_{-})\right]\bigg).
    \end{split}
\end{equation}

\subsubsection{$\mathcal{R}^{(1/2)}_{2+}$ kernel}
For $\mathcal{R}^{(1/2)}_{2+}$, the relevant function is $\mathcal{G}^{(1/2)}_{\text{eff}}(\epsilon_{\mathbf{p}^\prime},\epsilon_{\mathbf{p}},\mathbf{p}^\prime,\mathbf{p})$, where $\overline{\omega} = \epsilon_{\mathbf{p}^\prime}-\epsilon_{\mathbf{p}}$, and $\mathbf{x} = \mathbf{p^\prime}-\mathbf{p}$. The only delta functions in the equation above that can be satisfied are $\delta(\overline{\omega}+\epsilon_{+}-\epsilon_{-})$ and $\delta(\overline{\omega}-\epsilon_{+}+\epsilon_{-})$. We can impose the delta function to do the $p\prime\prime$ radial integration, which fixes the radial component at $p^{\prime\prime}_{*}(\theta) = \frac{\overline{\omega}}{2}\sqrt{\frac{\overline{\omega}^2-x^2-4m^2}{\overline{\omega}^2-x^2\cos^{2}{\theta}}}$, where $\theta$ is the angle with $\mathbf{x}$. This can be followed by the angular integration approximated by the Laplace method, since there is an exponential factor with large $\beta m$ in the exponent. The calculation closely follows the evaluation of $Im \Pi_{R}$ at finite $T$ in Appendix \ref{appsec:polarization}. The final expression for $\mathcal{R}^{(1/2)}_{2+}$ is,
\begin{equation}\label{eq:r2_p}
    \begin{split}
        \mathcal{R}^{(1/2)}_{2+}(\mathbf{p}^{\prime},\mathbf{p})\approx\frac{1}{8\pi}\sqrt{\frac{4\pi}{\beta m}}\frac{1}{4\epsilon_{\mathbf{p}^{\prime}}\epsilon_{\mathbf{p}}} \exp\left(-\frac{\beta |\mathbf{p}^{\prime}-\mathbf{p}|\sqrt{(\mathbf{p}^{\prime}-\mathbf{p})^{2}-\left(\epsilon_{\mathbf{p}^{\prime}}-\epsilon_{\mathbf{p}}\right)^{2}+4m^{2}}}{2\sqrt{(\mathbf{p}^{\prime}-\mathbf{p})^{2}-\left(\epsilon_{\mathbf{p}^{\prime}}-\epsilon_{\mathbf{p}}\right)^{2}}}\right)\times\\
        \left(\frac{1}{\left|\mathbf{p}^{\prime}-\mathbf{p}\right|^{1/2}\left(4m^{2}-(\epsilon_{\mathbf{p}^{\prime}}-\epsilon_{\mathbf{p}})^{2}+\left|\mathbf{p}^{\prime}-\mathbf{p}\right|^{2}\right)^{1/4}\left(\left|\mathbf{p}^{\prime}-\mathbf{p}\right|^{2}-(\epsilon_{\mathbf{p}^{\prime}}-\epsilon_{\mathbf{p}})^{2}\right)^{1/4}}\right)\times\\
        \left(\left|\Pi_{R}^{T=0}\left(\frac{\epsilon_{\mathbf{p}^{\prime}}+\epsilon_{\mathbf{p}}}{2}-\frac{x p^{\prime\prime}_{*}(\theta = \pi)}{\overline{\omega}},\frac{\mathbf{p}^{\prime}+\mathbf{p}}{2}+p^{\prime\prime}_{*}(\theta = \pi)\right)\right|^{-2}\right.\\
        \\
        +\left.\left|\Pi_{R}^{T=0}\left(\frac{\epsilon_{\mathbf{p}^{\prime}}+\epsilon_{\mathbf{p}}}{2}+\frac{x p^{\prime\prime}_{*}(\theta = 0)}{\overline{\omega}},\frac{\mathbf{p}^{\prime}+\mathbf{p}}{2}+p^{\prime\prime}_{*}(\theta = 0)\right)\right|^{-2}\right).
    \end{split}
\end{equation}
\subsubsection{$\mathcal{R}^{(1/2)}_{2-}$ kernel}
We can similarly evaluate the $\mathcal{R}^{(1/2)}_{2-}$, for which the relevant function is $\mathcal{G}_{\text{eff}}(-\epsilon_{\mathbf{p}^\prime},\epsilon_{\mathbf{p}},\mathbf{p}^\prime,\mathbf{p})$. We further define, $\overline{\omega} = \epsilon_{\mathbf{p}^\prime}+\epsilon_{\mathbf{p}}$, and $\mathbf{x} = \mathbf{p^\prime}-\mathbf{p}$. The only delta function in the equation above that can be satisfied is $\delta(\overline{\omega}-\epsilon_{+}-\epsilon_{-})$. We can impose the delta function to do the $p\prime\prime$ radial integration, which fixes the radial component at $p^{\prime\prime}_{*}(\theta) = \frac{\overline{\omega}}{2}\sqrt{\frac{\overline{\omega}^2-x^2-4m^2}{\overline{\omega}^2-x^2\cos^{2}{\theta}}}$. This brings an exponential factor of $e^{-\frac{\beta(\epsilon_{\mathbf{p}^{\prime}}+\epsilon_{\mathbf{p}})}{2}}$ to the expression for $\mathcal{R}^{(1/2)}_{2-}$, and hence $\mathcal{R}^{(1/2)}_{2-}(\mathbf{p}^{\prime},\mathbf{p})$ is substantial only at $p,p^{\prime}<<1$. The approximate expression (after the angular integration) is,

\begin{equation}\label{eq:r2_m}
    \begin{split}
        \mathcal{R}^{(1/2)}_{2-}(\mathbf{p}^{\prime},\mathbf{p}) \approx \frac{1}{8\epsilon_{\mathbf{p}^{\prime}}\epsilon_{\mathbf{p}}}e^{-\frac{\beta(\epsilon_{\mathbf{p}^{\prime}}+\epsilon_{\mathbf{p}})}{2}}\frac{\sqrt{\overline{\omega}^{2}-4m^2}}{\overline{\omega}^{2}}\left|\Pi_{R}^{T=0}\left(\frac{\epsilon_{\mathbf{p}}-\epsilon_{\mathbf{p}^{\prime}}}{2},\frac{\sqrt{\overline{\omega}^{2}-4m^2}}{2}\right)\right|^{-2}.
    \end{split}
\end{equation}
Numerically, it can be verified that $\mathcal{R}^{(1/2)}_{2-}(\mathbf{p}^{\prime},\mathbf{p})$ can always be ignored with respect to $\mathcal{R}^{(1/2)}_{2+}(\mathbf{p}^{\prime},\mathbf{p})$, for similar reasons as $\mathcal{R}_{1}$. Hence, for the ladder calculation, we can ignore $\mathcal{R}_{2-}(\mathbf{p}^{\prime},\mathbf{p})$.

\section{Details of numerics of ladder calculation}\label{appsec:numerics_ladder}

\begin{figure}
	\includegraphics[width=0.7\columnwidth]{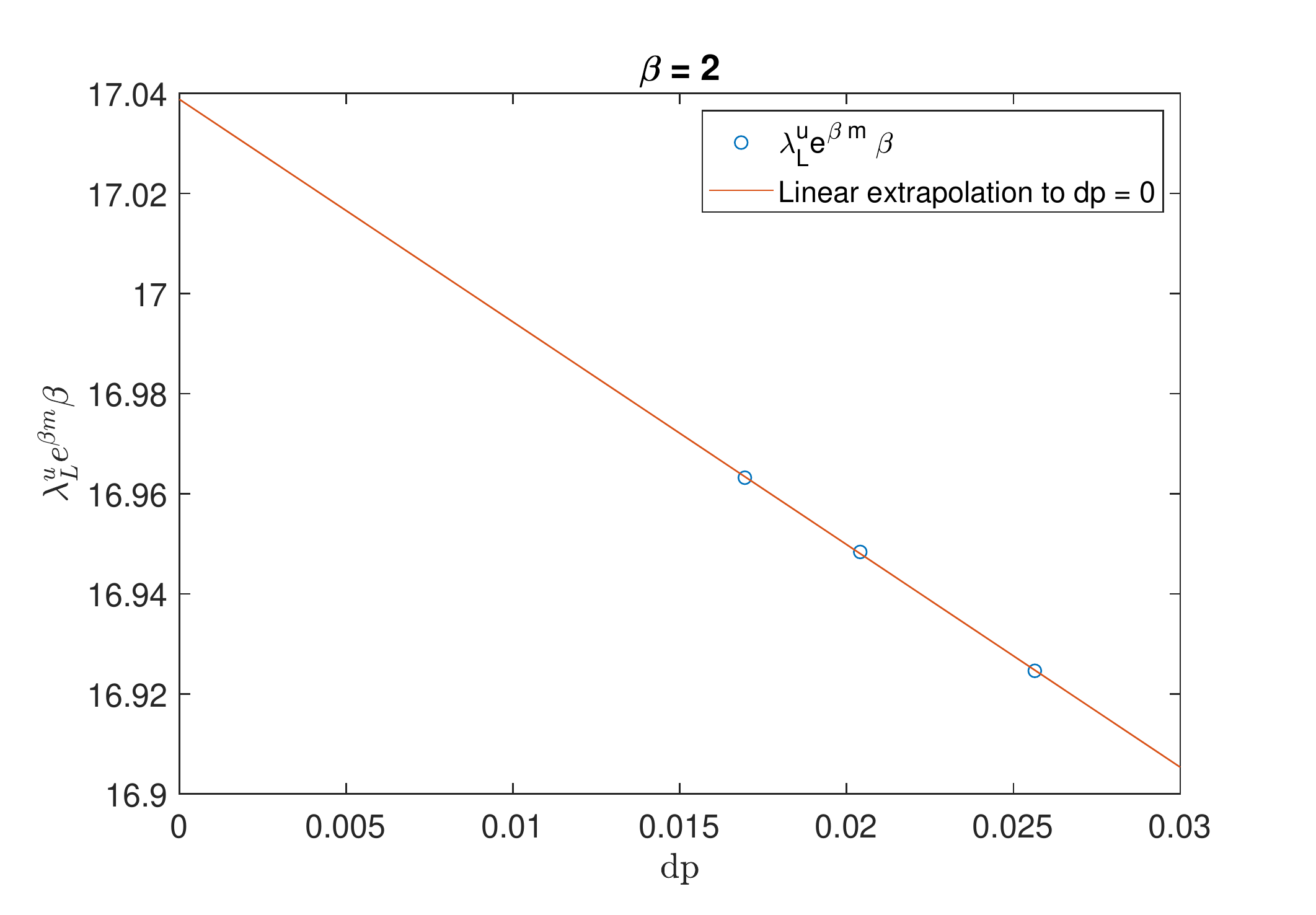}
	\caption{The maximum eigenvalue $\lambda_{L}e^{\beta m}\beta$ is determined by taking the linear extrapolation of $\lambda_{L}e^{\beta m}\beta$ at each grid interval $dp$ to $dp\to 0$. The error is determined as the uncertainty in the extrapolation from its $95\%$ confidence interval. The graph here is shown for the unregulated calculation at $\beta = 2$.} \label{Fig:extrapolate}
\end{figure}

Here we provide some details of the numerical computation of the ladder sum. We fix the mass as $m=1$, and do all the calculation in these units. Having determined the approximate values of the kernel functions $\mathcal{R}_{1,2}$, we need to discretize the 2D momentum space to set up the matrix form of the kernel integration. For that purpose, we set up a hard momentum cut-off of $|\mathbf{p}_{x}|,|\mathbf{p}_{y}|\leq 1$. The choice is justified for the kernel in rescaled momenta, which is exponentially suppressed - $\exp\left(-|\mathbf{p}-\mathbf{p^{\prime}}|^{2}/8\right)$. Next, we create 2D grid of momenta, with the momentum interval $dp$ determined by the number of points that we consider - 40 by 40, 50 by 50 and 60 by 60 grids. Next, we set up the matrix form of the kernel, $\hat{K}_{\mathbf{p}^{\prime}\mathbf{p}} = dp^2 \hat{\mathcal{K}}(\mathbf{p}^{\prime},\mathbf{p})$, given in Eq. \ref{eq:bs_eq_mat}. The matrices are of sizes, 1600 by 1600, 2500 by 2500, and 3600 by 3600, respectively. In constructing the matrix, we need to evaluate $\Gamma_{\mathbf{p}}$ by performing a 2D integration (in Eq. \ref{eq:gamma_q}) within the grid area ($|\mathbf{p}_{x}|,|\mathbf{p}_{y}|\leq 1$). We find the maximum magnitude eigenvalue of the matrix, and find that the largest magnitude eigenvalue has a positive real part, thereby resulting in exponential growth. The eigenvalues are then extrapolated to the $dp\to0$ limit by a linear extrapolation. Errors in the estimation are denoted as the errorbars for this eigenvalue (see Fig. \ref{Fig:extrapolate}).

\begin{figure}
	\includegraphics[width=0.48\columnwidth]{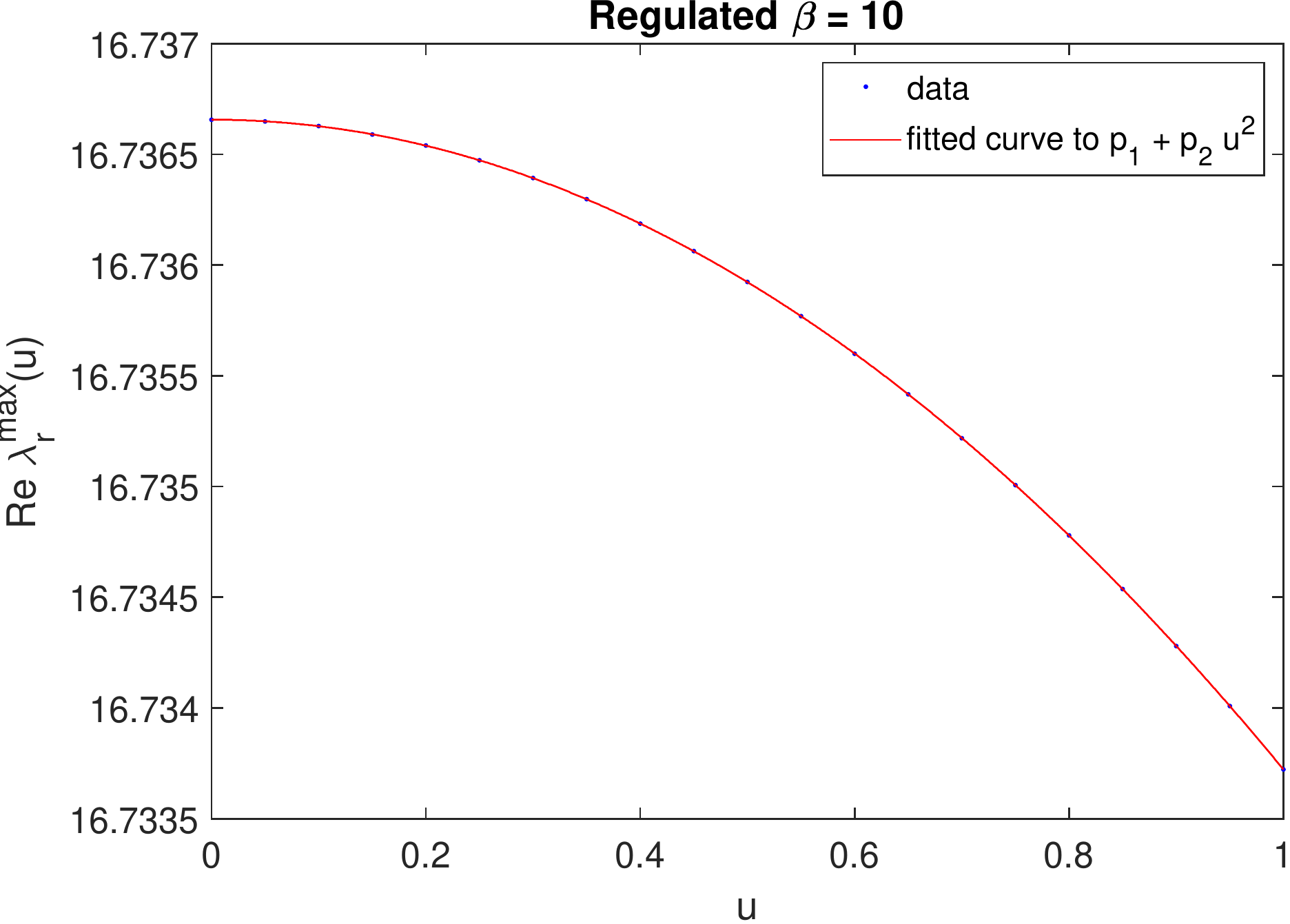}
	\includegraphics[width=0.48\columnwidth]{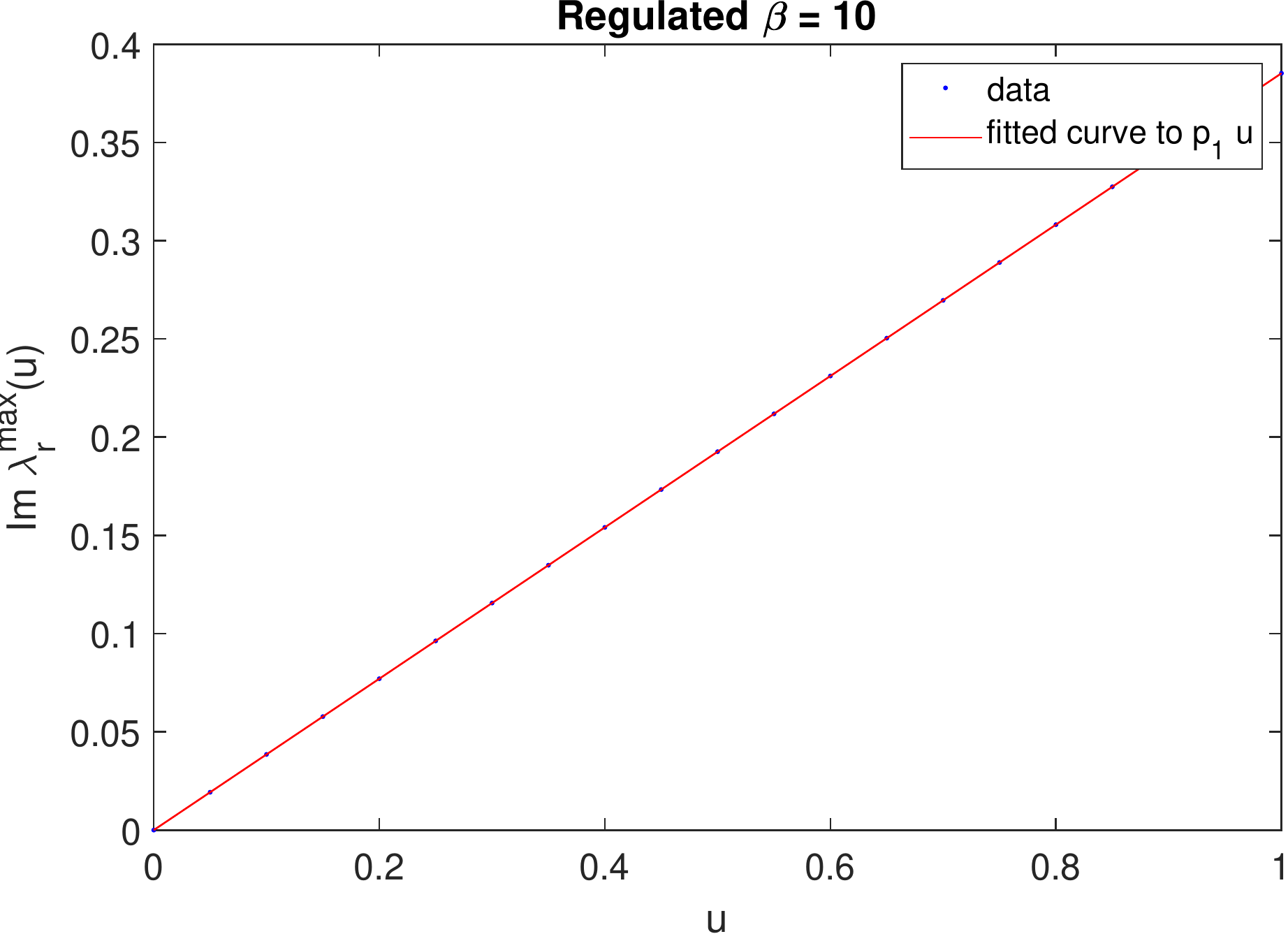}
	\caption{A sample fit of the numerically obtained $\lambda_{L}^{r}(u)$, at $\beta = 10$. $m$ is chosen to be 1. The real part is fit to $f(u) = \lambda_{0} - \lambda_{2} u^2$, while the imaginary part is fit to $f(u) = \lambda_{i}u$, and the fit works very well even at quite large u. }\label{Fig:knon0_quadfit}
\end{figure}

In Fig. \ref{Fig:knon0_quadfit}, we study the external momentum dependence of the largest magnitude eigenvalue of the kernel equation $\lambda_{L}(u)$ at non zero external momentum $u$. The real part of $\lambda_{L}(u)$ shows a quadratically decreasing behavior, $\lambda_{0}-\lambda_{2}u^{2}$ even at significantly high $u$, while the the imaginary part shows a linear behavior, $\lambda_{i}u$. At $u=0$, the eigenvalue is real and positive.

\section{Bounds on temperature dependence of butterfly velocity}\label{appsec:scrambling_bound}



Locality in gapped quantum spin chains can lead to microcausality and short ranged correlation \cite{Hastings2006}. Can we use similar techniques to bound the behavior of butterfly velocity?

In this Appendix we discuss state dependent bounds on butterfly velocity in local gapped systems which were introduced in \cite{Han2019}. The general definition of squared commutator in Eq. \ref{eq:alpha_square_norm} can be rewritten as $\mathcal{C}_{\alpha}(t,\mathbf{x},\rho)=-Tr\left(\rho^{\alpha}O\rho^{1-\alpha}O\right)$, where, $O=i\left[W_{\mathbf{0}}(t),V_{\mathbf{x}}\right]$. By restricting to $x=vt$, one can define the velocity dependent Lyapunov exponents, \begin{equation}
    \lambda(v,\rho)=\lim_{t\to\infty}\frac{1}{t}\ln C(vt,t,\rho).
\end{equation}
The butterfly velocity can be defined as the largest velocity for which the Lyapunov exponent is positive,
\begin{equation}
    v_{B}(\rho)=\sup\left\{v:\lambda(v,\rho)\geq 0\right\}.
\end{equation}
We define the support of the commutator, $O$ as a region S of diameter $2R(v,t)$, around a point $\mathbf{0}$. The scrambling velocity is defined as the rate of increase of this support,
\begin{equation}\label{scrvel_eq}
    v_{S}(\rho)=\lim_{t\to\infty}\frac{R(v,t)}{t}.
\end{equation}

We consider the Hamiltonian $H$ to be defined on a lattice, composed of geometrically local terms, and such that it has a finite gap. We introduce the shifted zero expectation-value Hamiltonian,  $\Tilde{H}=H-Tr(\rho H)$. We can divide the shifted Hamiltonian into terms supported inside and outside $S$,
\begin{equation}
    \Tilde{H}=\sum_{i\in S}\Tilde{h}_{i}+\sum_{j\in \Lambda-S}\Tilde{h}_{j}.
\end{equation}
Let us consider the near wavefront ansatz,
\begin{equation}
    \lambda(v,\rho)=-\lambda\left(\frac{v}{v_{B}}-1\right)^{1+p}.
\end{equation}
In \cite{Han2019}, it was shown that for the unregulated squared commutator, the rate of change of butterfly velocity with temperature, $\partial_{\beta} v_{B}$ can be bounded, 
\begin{equation}
    \lambda (\Delta v)^{p}(\Delta v +1)|\partial_{\beta} \ln v_{B}| \leq 2 h\left(v_{S}(\rho)-\xi\lambda(v,\rho)\right),
\end{equation}
where $\Delta v = v/v_{B}-1$, $\xi>0$ is the finite correlation length, and $h$ is given by,
\begin{equation}\label{eq:h_def2_eq}
    h=\text{sup}_{i}\frac{\left|Tr\left(\sqrt{\rho}\Tilde{h}_{i}\sqrt{\rho} O O\right)\right|}{Tr\left(\rho OO\right)}.
\end{equation}
At low temperature, $\beta\to\infty$, $\rho\sim|0\rangle\langle0|$. From Eq. \ref{eq:h_def2_eq}, $h\propto \langle 0|\Tilde{h}_{i}|0\rangle$, and hence 0, which implies,
\begin{equation}\label{eq:scrambling_bound_low_T}
    \partial_{\beta}\ln v_{B}\to 0 \text{ as } \beta\to\infty.
\end{equation}
We first review the proof for the unregulated case due to \cite{Han2019} and then also extend the bound to the butterfly velocity obtained from the regulated squared commutator, and show that the same low temperature behavior as in Eq. \ref{eq:scrambling_bound_low_T} holds in that case as well. However, we note that the bound can't differentiate between a power-law vanishing butterfly velocity at low temperature and a constant butterfly velocity. Low temperature behaviors of both the regulated and unregulated cases which were obtained in Sec.  \ref{sec:mpo_spins}, i.e., $v_{B}\sim\beta^{-1/2}$ and $v_{B}\sim \text{constant}$ respectively, are consistent with Eq. \ref{eq:scrambling_bound_low_T}. 

We first discuss the bound on butterfly velocity obtained from the unregulated squared commutator as given in \cite{Han2019}. We differentiate $C_{u}$ with respect to the inverse temperature $\beta$ to obtain,
\begin{equation}
    \partial_{\beta}C_{u}=-Tr\left(\Tilde{H}\rho OO)\right).
\end{equation}

We want to upper bound $|\partial_{\beta}C_{u}|$. By separating out the contributing terms to two parts - inside and outside a ball of radius $R+\delta$ around the point $x_{0}$ (a region we call $S^{\prime}$), we have, 
\begin{equation} \label{bound_eq}
    |\partial_{\beta}C_{u}|\leq\sum_{i\in S^{\prime}}\left|Tr\left(\rho O O \Tilde{h}_{i}\right)\right|+\sum_{j\in \Lambda-S^{\prime}}\left|Tr\left(\rho O O \Tilde{h}_{j}\right)\right|.
\end{equation}
For the terms outside the ball $S^{\prime}$, we invoke the Exponential Clustering Theorem, which states, for two operators $W_{1}$ and $W_{2}$ supported on non-overlapping regions $A$ and $B$ on a lattice system with a gapped Hamiltonian, there exist, $\xi$ and $\mathcal{N}$, such that,
\begin{equation}
    \left|Tr\left(\rho W_{1}W_{2}\right)-Tr\left(\rho W_{1}\right)Tr\left(\rho W_{2}\right)\right|\leq \mathcal{N} min\{|\partial A|,|\partial B|\}\lVert W_{1}\rVert \lVert W_{2}\rVert e^{-|A-B|/\xi}, 
\end{equation}
where, $|A-B|$ is the minimum distance between the regions $A$ and $B$. Here, $\xi$ is the correlation length, which is finite because of the presence of the gap. The Exponential Clustering Theorem can be proved using Lieb Robinson bound techniques \cite{Hastings2006}. Now, $Tr\left(\rho \Tilde{h}_{i}\right)=0$. Thus the sum of `outside' terms in the RHS of Eq. \ref{bound_eq}, can bounded in the following way - 
\begin{equation}
\begin{split}
    \sum_{j}... &\leq 2\mathcal{N} min\{|\partial A|,|\partial B|\}\lVert O\rVert^{2}\sum_{j=\delta}^{\infty}e^{-j/\xi}\\
    &=\mathcal{M}\int_{\delta}^{\infty}dx e^{-x/\xi}\text{ where $\mathcal{M}$ is suitably defined,}\\
    & = \mathcal{M}\xi e^{-\delta/\xi}.
\end{split}
\end{equation}
The `inside' terms in the RHS of Eq. \ref{bound_eq}, can be bounded in the following way,
\begin{equation}
    \begin{split}
        \sum_{i} ... &\leq h\sum_{i\in S^{\prime}} |Tr\left(\rho O O\right)|\\
        &= V_{R+\delta} C_{u}(t,vt,\rho),
    \end{split}
\end{equation}
where, $h$ is a maximum over the different terms of the shifted Hamiltonian, and $V_{r}$ is the size of the region of radius $r$, i.e., $V_{r}=2r+1$. Two convenient choices of $h$ are,
\begin{align}
    h&=2 \text{ sup}_{i} \lVert h_{i} \rVert \text{ or,} \\h&=\text{sup}_{i}\frac{\left|Tr\left(\sqrt{\rho}\Tilde{h}_{i}\sqrt{\rho} O O\right)\right|}{Tr\left(\rho OO\right)}.
\end{align}

Combining both the contributions, we get, 
\begin{equation}\label{bound2_eq}
    |\partial_{\beta}C_{u}|\leq V_{R+\delta} h C_{u}(t,vt,\rho) + \mathcal{M} \xi e^{-\delta/\xi}
\end{equation}
Usually at late times, $C(t\to\infty)=e^{\lambda(v,\rho)t}$. For $v>v_{B}$, $\lambda(v,\rho)<0$. We can choose $\delta=(-\xi\lambda(v,\rho)+\epsilon)t$ for some positive $\epsilon$, which makes the second term in Eq. \ref{bound2_eq} subleading compared to the first term, and hence can be dropped. Essentially, the contribution to the bound from sufficiently outside the support of the operator $O$ can be dropped. 

Now, using the ansatz $C_{u}=e^{\lambda(v,\rho)t}$, we obtain the following bound for the rate of change of the Lyapunov exponent, 
\begin{equation}
    \begin{split}
        |\partial_{\beta}\lambda|&\leq h \lim_{t\to\infty}\frac{V_{R-\xi\lambda(v,\rho)t}}{t}\\
        &=2 h \left(\lim_{t\to\infty}\frac{R}{t}-\xi\lambda(v,\rho)\right)\\
        &= 2 h \left(v_{S}(\rho)-\xi\lambda(v,\rho)\right) \text{ from the definition of the scrambling velocity from Eq. \ref{scrvel_eq}.}
    \end{split}
\end{equation}
We can further analyze this scrambling bound by using the near wavefront ansatz,
\begin{equation}
    \lambda(v,\rho)=-\lambda\left(\frac{v}{v_{B}}-1\right)^{1+p}.
\end{equation}
Let's introduce the short hand $\Delta v = v/v_{B}-1$. For this ansatz, we have, 
\begin{equation}
    \partial_{\beta}\lambda(v,\rho)=\lambda(\Delta v)^{1+p}\left[\partial_{\beta}\ln \lambda +\ln(\Delta v) \partial_{\beta}p-(1+p)\frac{v/v_{B}}{\Delta v}\partial_{\beta}\ln v_{B}\right]
\end{equation}
Close to the Butterfly velocity, i.e., when $v\gtrsim v_{B}$, the last term is the leading term. Thus for $\Delta v=0^{+}$, we have the bound on rate of change of butterfly velocity, 
\begin{equation}
    \lambda (\Delta v)^{p}(\Delta v +1)|\partial_{\beta} \ln v_{B}| \leq 2 h\left(v_{S}(\rho)-\xi\lambda(v,\rho)\right)
\end{equation}
Now, say $\beta\to\infty$. For the gapped system, $\rho=|0\rangle\langle0|$. We can estimate $h$ using the definition, in Eq. \ref{eq:h_def2_eq}. For this $\rho$, $h\propto \langle 0|\Tilde{h_{i}}|0\rangle$, and hence 0, which implies,
\begin{equation}
    \partial_{\beta}\ln v_{B}\to 0 \text{ as } \beta\to\infty
\end{equation}
Note, however, unlike the assertion in \cite{Han2019}, this doesn't imply a freezing out of the Butterfly Velocity at temperatures below the gap. In fact, even power-law ansatz, $v_{B}\sim \beta^{-a}$ for $a>0$, satisfies the above bound, and our observation $v_{B}\sim \beta^{-1/2}$ is certainly admissable. 

\section{Scrambling bounds for regulated squared commutator}
We can extend the bounds to the butterfly velocity from regulated squared commutator, $C_{r} =-Tr\left(\sqrt{\rho}O\sqrt{\rho}O\right)$, as well. Differentiating with $\beta$, we obtain,
\begin{equation}
    \begin{split}
        \partial_{\beta}C_{r}&=-Tr\left(\Tilde{H}\sqrt{\rho}O\sqrt{\rho}O\right)\\
        &=-Tr\left(\Tilde{H}\rho O\rho^{1/2}O\rho^{-1/2}\right).
    \end{split}
\end{equation}
Now, we invoke the Araki bound \cite{Araki1969}, which states, in 1 dimensional quantum lattice systems with a gap, for any finitely supported operator $A$ with support $R$, the operator $\rho^{s}A\rho^{-s}$ is also supported, upto exponential correction, on a ball of support $R+l(\beta s)$, where $l(x)$ is and entire function not larger than exponential in $x$. Thus, the support of $\rho^{1/2}O\rho^{-1/2}$, and hence of $O\rho^{1/2}O\rho^{-1/2}$ has radius $\sim R+\mathcal{A}e^{\mathcal{B}\beta}$, for appropriately defined numbers $\mathcal{A}, \mathcal{B}$. Hence, the entire argument of the previous section follows by replacing $R\to R+l(\beta/2)$, and we can bound the rate of change of Lyapunov exponent and Butterfly velocity obtained from the regulated squared commutator as well. In particular, in deriving these bounds, the effect of this thermal broadening can be ignored, since, $l(\beta)/t\to 0$, as $t\to\infty$. Hence, all the scrambling bounds derived for the unregulated case also follow naturally for the regulated case.

\section{Carbon cost of simulations}
Here we quote the approximate carbon cost of the numerical simulations. The template is from \href{https://scientific-conduct.github.io}{scientific-conduct.github.io}. This provides a lower bound of the carbon cost.
\begin{center}
\begin{tabular}[b]{l c}
\hline
\textbf{Numerical simulations} & \\
\hline
Total Kernel Hours [$\mathrm{h}$] (approx)& 3000\\
Thermal Design Power Per Kernel [$\mathrm{W}$]& 11.5\\
Total Energy Consumption Simulations [$\mathrm{kWh}$] & 34.5\\
Average Emission Of CO$_2$ In Maryland (2017) [$\mathrm{kg/kWh}$]& 0.39\\
Total CO$_2$-Emission For Numerical Simulations [$\mathrm{kg}$] & 13.5\\
Were The Emissions Offset? & \textbf{No}\\
\hline
\hline
\end{tabular}
\end{center}


\end{appendices}

\end{document}